%% file: thesis.tex
\documentclass[a4paper,oneside,12pt]{book}
\usepackage{psfig}

\begin{document}
\setlength{\baselineskip}{1.5\baselineskip}

\input{titlepage.tex}

\input{abstract.tex}

\input{ack.tex}

\pagestyle{plain}
\pagenumbering{roman}

\tableofcontents

\pagenumbering{arabic}
\input{chapter1.tex}

\input{chapter2.tex} 
\input{chapter3.tex}

\input{chapter4.tex}

\input{chapter5.tex}

\end{document}

%% file: titlepage.tex
\thispagestyle{empty}
\begin{titlepage}
\begin{center}
{\LARGE\bf String Theory and Black Holes \\ }
\end{center}

\vskip 1.5in

\centerline{\bf A thesis submitted to the}
\centerline{\bf University of Mumbai}
\centerline{\bf for the}
\centerline{\bf Ph.D. Degree in Physics}

\vskip 1.5in

\centerline{\bf by}
\centerline{\bf Justin Raj David}
\centerline{\bf Department of Theoretical Physics}
\centerline{\bf Tata Institute of Fundamental Research}
\centerline{\bf Mumbai 400 005, India.}

\vskip 1.5in

\centerline{\Large\bf 1999}

\end{titlepage}

%% file: abstract.tex
\chapter*{Abstract}
\thispagestyle{empty}

This thesis aims to make precise the microscopic understanding of
Hawking radiation from the D1/D5 black hole. We present an explict
construction of all the shortmultiplets of the ${\cal N}=(4,4)$ SCFT
on the symmetric product $\tilde{T}^4/S(Q_1Q_5)$. An investigation of
the symmerties of this SCFT enables us to make a one-to-one
correspondence beween the supergravity moduli and the marginal
opeerators of the SCFT. We analyse the gauge theory dynamics of the
splitting of the D1/D5 system into subsystems and show that it agrees
with supergravity. We have shown that the fixed scalars of the D1/D5
system couple only to $(2,2)$ operators thus removing earlier
discrepancies between D-brane calculations and semiclassical
calculations. The absorption cross-section of the minimal scalars is 
determined from first
principles upto a propotionality constant. We show that the
absorption cross-section of the minimal scalars computed in
supergravity and the SCFT is independent of the 
moduli.

%% file: ack.tex
\chapter*{Acknowledgements}
\thispagestyle{empty}

Firstly, I would like to thank Spenta Wadia, my thesis advisor for
inspiration, support and encouragement throughout my five years at TIFR.
It is from him I learnt how to approach problems in physics. 
He has emphasized to me the importance of thinking from first principles and
attacking physically relevant problems. During my early years in TIFR,
Spenta's guidance helped me overcome the maze of superfluous literature and to
concentrate on the basics. He emphasized to me the importance of independent
thinking in physics. I thank him for sharing with me his wonderful ideas. His
inspiring confidence helped me overcome moments of despair when all means of
attacking a physics problem seemed to fail.

During the course of my five years I worked very closely with Gautam Mandal
and Avinash Dhar. I have learnt (not completely) from Gautam the art of
reducing a physics problem to its very essentials. It is amazing to see how
simple a seemingly complicated problem is, when reformulated by Gautam. Avinash
helped me to realize the importance of being persistent and pursuing a
physics problem to its entire logical conclusion.
Avinash and I were
closely working on a problem which would not have yielded results had it not
been for the constant encouragement and persistence of Avinash.

I owe most of my understanding of gauge theories to the wonderful course 
that Sumit
Das took great pains to teach. The course stretched for the entire year of
1995. I thank the string community in TIFR for being active in fostering the
nascent enthusiasm for the subject.

There are other members of the Theory group 
working in different fields who provided
inspiration, help and guidance indirectly and unknowingly. 
They did it through their 
lectures, seminars and discussions. I have benefitted in this way from 
Deepak Dhar and Mustansir Barma. 

\newpage

\thispagestyle{empty}
The students of the Theory group during the past five years contributed a lot
to help me imbibe the spirit of search and discovery. I would like to mention
Abhishek Dhar whose uncompromising attitude towards physics continues to
be a source of inspiration.

I would like to acknowledge the help from the efficient and friendly staff of
the Department of Theoretical Physics.

Finally I would like to thank my family and friends for their support in
helping me pursue physics as a career.

\newpage

%% file: chapter1.tex
\chapter{Introduction}
\pagenumbering{arabic}
\def\be{\begin{equation}}
\def\ee{\end{equation}}
\def\ba{\begin{array}{l}}
\def\ea{\end{array}}
\def\bea{\begin{eqnarray}}
\def\eea{\end{eqnarray}}
\def\eq#1{(\ref{#1})}
\def\del{\partial}
\def\bibi#1{\item\label{#1}}
\def\cit#1{[\ref{#1}]}
\markright{Chapter 1. Introduction}
\vspace{-1cm}

A unified theory of fundamental interactions should include within its
framework a quantum theory of gravity. The quantization of
Einstein's theory of gravity using conventional methods
poses a host of problems. Einstein's theory of gravity is a
non-renormalizable field theory. Therefore one can not extract
meaningful answers from quantum perturbation theory. The existence of
black holes as solutions in Einstein's theory of gravity is another
stumbling block. Arguments
involving physics of black holes within the framework of quantum mechanics 
seem to lead  to the conclusion that quantum mechanical evolution of 
black hole is not unitary. This contradicts the basic rules of quantum
mechanics.

At present string theory is the leading candidate for a unified theory
of fundamental interactions. The massless spectrum of string theory includes
 the graviton. The low energy effective action of string
theory includes the Einstein action. String
theory is also consistent with the rules of quantum mechanics.
Thus the natural generalization of Einstein's theory within the
framework of quantum mechanics is string theory.
Furthermore string theory is  perturbatively finite. This cures the
non-renormalizability of gravity. 
The  understanding of  non-perturbative spectra of string theory like
D-branes has paved the way for addressing the problems of black hole
physics. In this thesis we will attempt to make precise the 
description of black hole physics within the framework of string
theory. 

\section{Black hole physics}

Let us briefly review some general properties of black holes.
\cit{chp1:Wald}
Black holes are objects which result as end points of gravitational
collapse of matter. For masses greater than 3.6 solar mass, the
gravitational force overcomes all other forces  and the matter
generically collapses into a black hole (in some exceptional cases a
naked singularity might result). 
This would suggest that to specify a black hole it is necessary to
give in detail the initial conditions of the collapse. As we will see
below a black hole is completely specified by a few parameters only.

To introduce various concepts related to black hole we will discuss
two examples of black holes. First, let us consider the Schwarzschild
black hole in $3+1$ dimensions. 
It is a time independent, spherically symmetric
solution of pure Einstein gravity. Its metric is given by
\be
ds^2= -\left(1 - \frac{2 G_N  M}{r}\right) dt^2 + 
\left(1-\frac{2 G_N M}{r}\right)^{-1} dr^2 + r^2 d\Omega ^2
\ee
where $t$ refers to time, $r$ the radial distance, $\Omega$
the solid angle in $3$ dimensions and $G_N$ the Newton's constant. 
We have chosen units so that the velocity of light, $c=1$.
The surface $r=2 G_N M$ is called
the event horizon. It is a co-ordinate singularity but not a curvature
singularity. 
Light-like geodesics and time-like geodesics
starting at $r<2 G_N M$ end up at $r=0$ (the curvature singularity) in
finite proper time. 
This means that classically the black hole is
truly black, it cannot emit anything. Note that the solution is completely
specified by only one parameter $M$, the mass of the black hole. 

Next we consider the Reissner-N\"{o}rdstrom black hole. 
It is a time independent, spherically symmetric solution of Einstein
gravity coupled to the electromagnetic field. The solution is given
by the following backgrounds.
\bea
ds^2&=& -\left( 1-\frac{2G_N M}{r} + \frac{G_NQ^2}{r^2} \right) dt^2
+ \left( 1- \frac{2G_N M}{r} + \frac{G_NQ^2}{r^2}\right)^{-1} dr^2 +
r^2 d\Omega^2 ,\\  \nonumber
A_0 &=&\frac{Q}{r}.
\eea
where $A_0$ is the time component of the vector potential. This
solution carries a charge $Q$. There are two co-ordinate singularities
at
\be
r_+ = G_N M + \sqrt{G_N^2 M^2 -G_N Q^2} 
\ee
and
\be
r_- = G_N M - \sqrt{G_N^2 M^2 -G_N Q^2}
\ee
The event horizon is at $r=r_+$. When $M=
|Q|/\sqrt{G_N}$, the outer horizon at $r=r_+$
and the inner horizon at $r=r_-$
coincide. A black hole with coincident inner and outer horizons 
is called  an extremal black hole. 
Note that in this case the black hole is completely specified by its
mass $M$ and the charge $Q$. 

In general, collapsing matter results in black holes which are
completely specified by the
mass $M$, the $U(1)$ charges $Q_i$ 
and the angular momentum $J$. This is called
the no hair theorem. Whatever other information 
(for example, multipole moments) present decays
exponentially fast during the collapse. Thus, all detailed information carried
by the collapsing matter is completely lost.

So far we have discussed the black holes only classically. In the
seventies, the works of Bekenstein, Hawking and others furthered
the understanding of black holes within the framework of quantum
mechanics. It was found by Hawking that the Schwarzschild black
hole is not  truly black. A semi-classical calculation by
Hawking showed that it emits radiation with the spectrum of a black
body at a temperature T given by
\be
T= \frac{\hbar}{8\pi G_N M}
\ee
The quantum nature of this effect is clearly evident from the fact
that the temperature is proportional to $\hbar$. For the 
Reissner-N\"{o}rdstrom black hole the temperature of 
Hawking radiation is 
\be
T= \frac{(r_+ - r_-)\hbar}{4\pi r_+^2}
\ee
One notes that the extremal Reissner-N\"{o}rdstrom black hole does
not Hawking radiate.
In general the Hawking temperature turns out to be function of mass, 
the charges and the angular momentum alone. Thus even semi-classical effects 
do not provide further information of the black hole. The works in the
seventies culminated in the following laws of black holes which are
analogous to the laws of thermodynamics.\\
i. First Law: Two neighboring black hole equilibrium states are related
by
\be
dM= Td\left[\frac{A}{4G_N\hbar}\right] + \Phi_i dQ_i + \Omega dJ
\ee
where $A$ is the area of the event horizon, $\Phi$ the electric surface
potential and $\Omega$ the angular velocity. For the special case of the
Reissner-N\"{o}rdstrom black hole the first law reduces to
\be
dM= Td\left[\frac{A}{4G_N\hbar}\right] + \Phi dQ.
\ee
where $\Phi= Q/r_+$. \\
ii. Second Law: Black holes have entropy $S$ given by
\be
S= \frac{A}{4G_N\hbar}
\ee
In any irreversible 
process, the sum of the entropy of the black hole and the
surroundings always increases.

The above facts taken together lead to the following puzzle. Consider
a black hole formed out of the collapse of a pure state. It eventually
evaporates through Hawking radiation. The radiation is purely thermal.
The final state is a mixed state. Thus we have a situation in which a
pure state has evolved into a mixed state. This contradicts the
unitary evolution of quantum mechanics. 

To emphasize where we lack in understanding let us contrast the above
situation with that of radiation from a black body. A black body
made up of a pure state radiates thermally. 
This is because we have averaged over the
various micro-states of the black body. The radiation would 
contain sufficient information to reconstruct the pure state of the
black body. In this case we do not have a contradiction with quantum
mechanics. 

The difficulty with the understanding of 
black holes within the framework of quantum mechanics is because we do
not know what are the  microstates which contribute to the entropy of
the black hole. Any serious candidate for the quantum theory of gravity
should enable us to understand black holes from a microscopic point
of view. It should provide us with a microscopic theory analogous to
the way statistical mechanics is the microscopic theory for
thermodynamics. String theory has demonstrated the potential to
fulfill this requirement.

\section{Black holes in string theory}

The low energy effective action of string theory   
has black holes as  solutions of its equation of motion. 
Based on earlier suggestions 
[\ref{chp1:Hawking},\,\ref{chp1:Salam},\,\ref{chp1:Hooft},\,
\ref{chp1:HolWil}] 
an attempt was made to understand 
the entropy of black holes in string theory. 
[\ref{chp1:Susskind},\,\ref{chp1:Sen_bh}] showed that the microscopic
states contributing to the entropy of certain black holes in string
theory arises due to the degeneracy of the
perturbative string spectrum. 
For bosonic string theory it was seen \cit{chp1:Susskind} that a
very massive string with mass M at strong string coupling $g_s$
has a Schwarzschild radius, 
$r_{ {\rm Schwarzschild} } \sim g_s^2 M $ 
larger than the string length $\sqrt{\alpha '}$ 
and therefore can be treated as a black hole. The entropy  of
states of the perturbative string for a given mass level can be
computed as follows. One evaluates the degeneracy of the string
states at a given mass. The logarithm of this degeneracy is the entropy
of the string states at that mass level. For a perturbative string of mass $M$, the
entropy is proportional to $M$. 
But it is well known that the entropy of a Schwarzschild black hole is proportional to
$M^2$. It was hoped that  strong coupling
effects would renormalize the mass of the string so 
that the entropy of the string could agree with that of the
corresponding black hole. If one has a situation where mass does not get renormalized then the entropy evaluated at weak coupling should agree with that calculated at strong coupling. 
In superstring theory there are states whose
masses are not renormalized. These are the BPS states. It was shown in
\cit{chp1:Sen_bh} that a massive BPS string of heterotic string
theory compactified on $T^6$ has an entropy due to  perturbative
string states which is proportional to the area of the stretched 
horizon. The supergravity configuration corresponding to this BPS
string has a horizon of zero area. Thus it was necessary to find a black
hole solution in string theory which is BPS and has finite area of
horizon. Such black holes are known, but they carry Ramond-Ramond
charges. They are solitonic 
solutions of the supergravity equations of motion.
It was not until
the discovery of the carriers of these charges known as D-branes
\cit{chp1:Polchinski} and their understanding in terms of
perturbative string theory that a microscopic understanding of black
holes developed. Strominger and Vafa in \cit{chp1:StrVaf} calculated
the entropy of an extremal and BPS black hole of type IIB string theory 
compactified on $K3 \times S^1$. This extremal black hole was obtained
by wrapping D5-branes along the 4-cycle of $K3$ and introducing
momentum along the $S^1$. Strominger and Vafa showed that the entropy
calculated as the logarithm of the degeneracy of states in the
microscopic theory corresponding to the extremal black hole agrees
precisely with that computed from supergravity.

\section{The D1/D5 black hole}

A black hole solution which is BPS and has a finite area of horizon
was also studied by Callan and Maldacena \cit{chp1:CalMal}.  
It is a solution of type IIB string theory compactified on $T^4\times
S^1$ to five dimensions. This solution avoids the complication of
having to deal with $K3$.  This black hole will be the 
working model in this thesis \footnote{In the rest of the thesis, we set 
$\hbar =1.$}. We will discuss the solution in detail
below.

The relevant terms in the supergravity action for the solution are 
\be
\frac{-1}{128 \pi^7 \alpha '^4 g_s^2 }\int d^{10} x \sqrt -g \left[
e^{-2 \phi} ( R - 4(d\phi )^2 + \frac{1}{12}(dB)^2 ) \right]
\ee
$B$ denotes the Ramond-Ramond two form potential, $\phi $ the dilaton and $g$ the
metric. 
To construct the supergravity solution, we take $Q_5$ D5-branes and align
them along the $5, 6,7,8, 9$ directions and $Q_1$ D1-branes along the
$x_5$ direction. We then compactify the $6, 7,8,9$ directions on a
torus $T^4$ of volume $V_4$ and the $x_5$ direction on a circle of
radius $R_5$. To obtain a black hole with nonzero area for its event
horizon we must introduce $N$ units of momentum along the $x_5$
direction.
The supergravity solution in $10$ dimensions for this configuration
is obtained by setting all the other
fields of type IIB supergravity to zero and setting the fields in the
action to the following values.
\bea
\label{d1_d5}
ds^2 &=& f_1^{-\frac{1}{2}} f_5^{-\frac{1}{2}} (-dt^2 + dx_5^2
+ k(dt - dx_5)^2 ) 
+ f_1^{\frac{1}{2}} f_5^{\frac{1}{2}} (dx_1^2 + \cdots + dx_4^2) 
\\ \nonumber 
    & & + f_1^{\frac{1}{2}} f_5^{-\frac{1}{2}} 
(dx_6^2 + \cdots + dx_9^2),
\\ \nonumber
e^{-2 \phi} &=&\frac{1}{g_s^2} f_5 f_1^{-1} , \\ \nonumber
B_{05} &=& \frac{1}{2} (f_1^{-1} -1), \\ \nonumber
H_{abc} &=& (dB)_{abc}
=\frac{1}{2}\epsilon_{abcd}\partial_{d} f_5, \;\;\;\;
 a, b, c, d = 1, 2, 3, 4 
\eea
where $f_1$, $f_5$ and $k$ are given by
\be
f_1 = 1 + \frac{16 \pi ^4 g_s \alpha '^3 Q_1}{V_4 r^2}, \;
f_5= 1+ \frac{ g_s \alpha' Q_5}{r^2},\;
k= \frac{16 \pi ^4 g_s ^2 \alpha '^3 N}{V_4 R_5^2 r^2}
\ee
Here $r^2 = x_1^2 + x_2^2 + x_3^2 + x_4^2$ denotes the distance
measured in the transverse direction to all the D-branes. 
On compactifying the above solution to 5 dimensions using the
Kaluza-Klein ansatz one obtains
an extremal black hole with the horizon at $r=0$ but
with a finite area. The entropy of the 5 dimensional black hole is given
by
\be
\label{chp1:entropy}
S= 2\pi \sqrt{Q_1 Q_5 N}
\ee
This solution preserves 4 supersymmetries out of the 32
supersymmetries of type IIB theory. It is a BPS configuration. 

A non-extremal version of the above black hole 
solution can be obtained by
having Kaluza-Klein momentum $N$ which is distributed  along
the left and right directions of $x_5$. This breaks all the 32
supersymmetries. Such a solution in ten dimensions is given by the
following background. 
\bea
\label{chp1:nonextremal}
e^{-2 \phi} &=& \frac{1}{g_s^2}
\left(1 + \frac{r_5^2 }{r^2} \right) 
\left( 1 + \frac{r_1^2}{r^2} \right)^{-1} 
,\\  \nonumber
H &=& \frac{2r_5^2}{g_s}\epsilon_3 +
2g_se^{-2\phi}r_1^2 *_6\epsilon_3 , \\ \nonumber
ds^2 &=& 
\left( 1 + \frac{r_1^2}{r^2} \right)^{-1/2} 
\left(1 + \frac{r_5^2}{r^2} \right) ^{-1/2} 
\left[ -dt^2 + dx_9^2 \right.  \\  \nonumber
&+& \frac{r_0^2}{r^2} (\cosh\sigma dt + \sinh\sigma dx_9)^2 +
\left.
\left( 1 + \frac{r_1^2}{r^2} \right) 
(dx_5^2 + \ldots dx_8^2) \right] \\ \nonumber
&+&\left( 1 + \frac{r_1^2}{r^2} \right)^{1/2} 
\left(1 + \frac{r_5^2}{r^2} \right) ^{1/2} 
\left[ \left( 1- \frac{r_0^2}{r^2} \right)^{-1} dr^2 + r^2 d\Omega_3^2
\right],
\eea
where $*_6$ is the Hodge dual in the six dimensions $x_0, \ldots ,x_5$
and $\epsilon_3$ is the volume form on the unit three-sphere. $x_5$ is
periodically identified with period $2\pi R_5$ and directions $x_6,
\ldots ,x_9$ are compactified on a torus $T^4$ of volume $V_4$.
$\Omega_3$ is the volume of the unit three-sphere in the transverse
directions. This solution 
is parameterized by six independent quantities --
$r_1, r_5, r_0, \sigma,  R_5$ and $V_4$. These are related to the 
number of D1-branes , D5-branes and Kaluza-Klein momentum on $x_5$  
as follows.
\bea
\label{fields}
Q_1 &=&\frac{V_4}{64\pi^6 g_s^2\alpha^{\prime 3}}\int e^{2\phi}*_6 H 
= \frac{V_4 r_1^2}{16\pi^4\alpha^{\prime 3} g_s} ,
\\ \nonumber
Q_5 &=&\frac{1}{4\pi^2 \alpha^{\prime}} \int H  =
\frac{r_5^2}{g_s\alpha^{\prime}} , \\ \nonumber
N &=& \frac{R_5^2 V_4 r_0^2}{32\pi^4\alpha^{\prime 4} g_s^2} \sinh
2\sigma .
\eea
On compactifying this solution to five dimensions using the
Kaluza-Klein ansatz one obtains a five-dimensional black hole with a
horizon at $r=r_0$. 
 The entropy and the mass of this black hole is given
by
\bea
\label{mass_entropy}
S&=& \frac{A}{4G_5}= \frac{2\pi^2 r_1 r_5 r_0 \cosh 2\sigma}{4G_5}, \\
\nonumber 
M&=& \frac{\pi}{4 G_5}(r_1^2 + r_5^2 +\frac{r_0^2\cosh 2 \sigma}{2} ),
\eea
where the five-dimensional Newton's constant is 
\be
G_5= \frac{4\pi^5\alpha^{\prime 4} g_s^2}{V_4 R_5}
\ee

We will now discuss the various limits in which physics of this black
hole solution can be understood \cit{chp1:MalStr96}. 
As the supergravity action has a $1/g_s^2$ prefactor the classical
limit can be obtained by  setting $g_s\rightarrow 0$. The fields are
held fixed in the classical limit. From \eq{chp1:nonextremal} and
\eq{fields} we see that this is equivalent to
\bea
\label{class_limit}
g_s\rightarrow 0, \\  \nonumber
\mbox{with} \; g_sQ_1, \; g_sQ_5, \; g_s^2N \; \mbox{fixed}.
\eea
The formulae in \eq{fields} indicate that this is also equivalent to
\bea
g_s\rightarrow 0, \\ \nonumber
\mbox{with} \; r_1,\;  r_5, \; r_N \; \mbox{fixed}.
\eea
where $r_N=r_0 \sinh \sigma$.
Hawking's semi-classical analysis of the radiation emitted by the black hole 
is valid in this limit. 

The D1/D5 black hole is a solution of the 
equations of motion of type
IIB supergravity. Type IIB supergravity is the low energy effective
theory of type IIB string theory. Therefore the classical limit 
\eq{class_limit} is described by genus zero closed string theory. In
string theory for every genus there is an $\alpha^{\prime}$ 
expansion. Closed
string perturbation is valid when the typical length scales involved
are large compared to the string length $\sqrt{\alpha^{\prime}}$. 
From \eq{chp1:nonextremal} we find the typical length scales
involved are $r_1, r_5, r_N$. Thus the $\alpha^{\prime}$ expansion is
valid when $r_1, r_5, r_N$ are large compared to string length. From
\eq{fields} this is equivalent to
\be
\label{l_1}
g_sQ_1>>1, \; g_sQ_5>>1, \; g_s^2N >>1
\ee

The D1/D5 black hole can also be described by the constituent
D-branes.  D-brane perturbation theory involves  open string
loops. Open string loops have factors of $g_sQ_1$ or $g_sQ_5$,
corresponding to the fact that the open string loops can end on any of
the D-branes. Hence the classical limit \eq{class_limit} corresponds to
a large $N$ limit of the open string field theory. 
Perturbation theory of the large $N$ limit of open string theory is
good if
\be
\label{l_2}
g_sQ_1<<1 , \; g_sQ_5 <<1 ,\; g_s^2N <<1.
\ee
In the limit \eq{l_2} the classical solution is still a black hole, though 
the typical size of this solution in the limit \eq{l_2} is smaller than the 
string length. This is because in the classical limit \eq{class_limit} the 
entropy S in \eq{mass_entropy} diverges.It costs an infinite amount of entropy 
for the black hole to loose any finite fraction of its mass in outgoing 
radiation. The second law thus prohibits radiation from escaping and black 
holes are black in the classical limit \eq{l_2}, independent of their size.
The microscopic
understanding of the D1/D5 black hole involves understanding the D1/D5
black hole in terms of the constituent D-branes. Thus any calculation
of a macroscopic property of the black hole like entropy 
or Hawking radiation in terms of the D-branes is valid
only in the small black hole region. 
Surprisingly many of the calculations
done in the limit \eq{l_2} agree with those done in the limit \eq{l_1}.
The reason is generally ascribed to some non-renormalization theorems. 

The entropy of the extremal and the near extremal black hole 
discussed above was derived  in \cit{chp1:MalSus}
using a  phenomenologically motivated microscopic model. It
consists of a string wound along the $x_5$ direction. 
The string was assumed to carry the effective degrees of freedom of
the D1/D5 black hole.  The length of
the string is $Q_1 Q_5 R_5$. 
The string is assumed to have oscillation only along the directions
of the torus $T^4$. It has four bosonic and four fermionic degrees of
freedom.
For an extremal black hole with $N$ units
of Kaluza-Klein momentum along $x_5$ the string is excited to a level
$NQ_1Q_5$. This is necessary because of the quantization of the
Kaluza-Klein momentum in units of $1/R_5$. The value of
$L_0-\bar{L}_0$ of the string is identified with the Kaluza-Klein
momentum of the black hole. Thus we have
\be
L_0-\bar{L}_0 = \frac{NQ_1Q_5}{Q_1Q_5R_5} = \frac{N}{R_5}
\ee
Using Cardy's formula, 
\be
S_{\mbox{string}}=2\pi \sqrt{NQ_1Q_5}
\ee
This agrees with 
\eq{chp1:entropy} obtained from the supergravity solution.
We will call this phenomenological model the `long string'
model.

Hawking radiation of scalars 
from the near-extremal black hole was understood
microscopically from the long string model in
[\ref{chp1:DhaManWad},\,\ref{chp1:DasMat}]. 
To derive Hawking radiation it was
necessary to couple the microscopic model to various fields of
supergravity. In [\ref{chp1:DhaManWad},\,\ref{chp1:DasMat}] 
this was done using
couplings derived from a Dirac-Born-Infeld (DBI) action of the
long string. This method gave results which agreed with supergravity
for minimal scalars. These scalars are 
massless in the black hole geometry. A detailed calculation by
\cit{chp1:MalStr96} showed that using the long string as the
microscopic theory for the D1/D5 black hole reproduces the Hawking rate
taking into account of the grey body factors of the supergravity. The
Hawking rate with grey body factor and the decay rate using the long
string is given by
\be
\Gamma_H= \Gamma_{\mbox{string}}= 2\pi^2r_1^2r_5^2 \frac{\pi\omega}{2}
\frac{1}{e^{\frac{\omega}{2T_L}} -1}
\frac{1}{e^{\frac{\omega}{2T_R}} -1}
\frac{d^4k}{(2\pi)^4}
\ee
where 
\bea
\label{temp}
T_L = \frac{1}{\pi}\frac{r_o e^{\sigma}}{2r_1r_5} = \frac{1}{\pi R_5
Q_1 Q_5} \sqrt{N_L} \\ \nonumber
T_R = \frac{1}{\pi}\frac{r_o e^{-\sigma}}{2r_1r_5} = \frac{1}{\pi R_5
Q_1 Q_5} \sqrt{N_R} 
\eea
The right movers of the string are excited to a level $N_R$ and
the left movers of the string are excited to a level $N_L$.

Inspite of these achievements 
the long string model had many short commings.
As we will see the long string model could explain Hawking radiation
of those minimal scalars which correspond to the metric fluctuation
of the torus $T^4$. In the supergravity background of the D1/D5 black
hole there are four additional minimal scalars discussed in Chapter 2
and Chapter 3. The long string model has no way of explaining Hawking
radiation of these additional minimal scalars.
The long string model
failed to reproduce the Hawking rate expected
in supergravity for the fixed scalars \cit{chp1:GubKleTsy}. 
These  scalars  are massive in the black hole geometry.
This suggests that the long string model 
along with the DBI couplings could
not be a microscopic model for the D1/D5 black hole. 
In some ways the model has an ad-hoc nature. It does not expain why
the string has to be of length $Q_1Q_5R_5$. 
The DBI action is valid only for a single set of branes. Its use
for the case of multiple D1, D5-branes is at best a crude
approximation.

The first attempt
to obtain a microscopic model for the D1/D5 black hole based on the low
energy effective theory of the D1/D5 system was made in 
\cit{chp1:HasWad}.
It was found that the low energy effective degrees of freedom turn out
to be that of a effective string. The fact that the Kaluza-Klein
momenta has to be quantized in units of $1/(R_5Q_1Q_5)$ was explained
by constructing the effective string from the
infrared gauge invariant degrees of
freedom of the D1/D5 gauge theory.
However using the methods of 
\cit{chp1:HasWad} it was
difficult to obtain the couplings of the effective theory with the
supergravity.

\section{Outline of the Thesis}

In this thesis we attempt to make precise the microscopic 
derivation of Hawking radiation for the D1/D5 black hole. 
Abstracting from the discussion in the last paragraphs of the previous
section, 
the requirements for understanding Hawking radiation are: 1.
Discovering the microscopic degrees of freedom responsible for the
black hole entropy. 2. Deriving  how these degrees of freedom
couple with the fields present in the bulk of the space-time. 3.
Explaining the process of the interaction of the bulk fields with the
microscopic degrees of freedom to produce Hawking radiation within a
unitary framework.

The outline of this thesis is as follows. In Chapter 2, we discuss
the microscopic degrees of freedom relevant for the  D1/D5 black hole. 
The microscopic degrees of freedom are best described by first
isolating the degrees of freedom for a system closely related to the
D1/D5 black hole. This system is obtained by type IIB theory compactified on
$T^4$ with $Q_5$ D5-branes wrapped on $T^4$. The compact directions
are $x_6,x_7,x_8,x_9$.
The single non-compact
direction of the D5-branes is chosen to be along the $x_5$ direction.
The $Q_1$ D1-branes are also aligned along the  $x_5$ direction.
The difference between the D1/D5 black hole and this system is that
the $x_5$ direction is not compact in the latter. 
The supergravity solution is that
of a black string in six dimensions. This solution lifted to ten
dimensions is given by setting $N=0$ in \eq{d1_d5}. We will 
call this system the D1/D5 system.

We will study the D1/D5 system from the D-brane point of view.
As we have seen before this is valid in the limit \eq{l_2}.
size. For understanding Hawking radiation, it is sufficient to study the 
low energy effective theory of the
D1/D5 system.
The low energy theory of the D1/D5 system is a $1+1$ dimensional
supersymmetric gauge theory with gauge group $U(Q_1)\times U(Q_5)$. It
has eight supersymmetries. The matter content of this theory consists of
hypermultiplets transforming in the adjoint representation of each of
the gauge groups and hypermultiplets transforming as bifundamentals
of $U(Q_1)\times \overline{U(Q_5)}$. We will review in detail the field content of
this theory and its symmetries. The bound state of the D1 and
D5-branes is described by the Higgs branch of this theory. The Higgs
branch in the infrared will flow to a certain ${\cal N}=(4,4)$
superconformal field theory (SCFT). 
The D1-branes bound to the D5-branes can be considered as 
instanton strings of the $U(Q_5)$ gauge theory. 
From this point of view the  target space 
of the SCFT is the moduli space of $Q_1$ instantons of 
a $U(Q_5)$ gauge theory on $T^4$. This moduli space 
is known to be a resolution of the orbifold
$(\tilde{T}^4)^{Q_1Q_5}/S(Q_1Q_5)$ which we shall denote by 
${\cal M}$. $\tilde{T}^4$ can be distinct from the compactification
torus $T^4$. The evidence for this is mainly topological
and is related to dualities which map the black string corresponding to
D1/D5 system to a perturbative string of Type IIB theory with $Q_1$
units of momentum and $Q_5$ units of winding along the $x_5$
direction. We will discuss in 
detail a realization of this orbifold SCFT as a free field theory with
identifications. The symmetries of this SCFT including a new $SO(4)$
algebra will also be discussed. 

In this thesis we will use the 
Higgs branch of the D1/D5 system, realized as a ${\cal N}=(4,4)$ SCFT on a
resolution of the orbifold ${\cal M}$, to provide a microscopic
understanding of Hawking radiation from the D1/D5 black hole. 
The point of view adopted in this thesis is to attempt at
a first priciple derivation
of Hawking radiation from the Higgs branch of the D1/D5 system. 
The long string model
used by \cit{chp1:MalSus} is at best a
phenomologically motivated theory.

The complete specification of the D1/D5 system involves the
specification of the various supergravity moduli. The supergravity
solution \eq{d1_d5} is the solution with no moduli. We would like to
understand what the SCFT is corresponding to the D1/D5 system at
generic values of the moduli. To this end we construct all the marginal
operators of the ${\cal N}=(4,4)$ SCFT on ${\cal M}$ including
operators involving twist field which correspond to blowing up modes. 
We classify the marginal operators according to the short multiplets of
the global part of the ${\cal N}= (4,4)$ superalgebra and a new 
$SO(4)$ algebra.  It is known from supergravity that the D1/D5 bound
state $(Q_1,Q_5)$
with no moduli turned on is marginally stable with respect to decay to
subsystems $(Q_1', Q_5')$ and $(Q_1^{\prime\prime},
Q_5^{\prime\prime})$. Analyzing the gauge theory of the D1/D5 system
we find that the dynamics of this splitting is described by an
effective $(4,4)$ $U(1)$ theory coupled to $Q'_1 Q''_5 + Q''_1
Q'_5$ hypermultiplets. We show, by an analysis of the D-term equations
and the potential, that the splitting is possible only when the
Fayet-Iliopoulos terms and the theta term of the effective gauge
theory are zero \cit{chp1:DavManWad2}.

In Chapter 3, we take the next step towards the 
microscopic understanding of Hawking
radiation. We find the precise coupling of the fields of the
supergravity to the microscopic SCFT. 
This is given by a specific SCFT
operator ${\cal O}(z, \bar{z})$, which couples to the supergravity
field $\phi$ in the form of  an interaction
\be
\label{interaction}
S_{\mbox{int}}= \mu\int d^2 z \;\phi_ (z, \bar{z}) {\cal O}(z, \bar{z})
\ee
where $\mu$ is the strength of the coupling.
As the ${\cal N}=(4,4)$ SCFT on ${\cal M}$ 
is an ``effective'' theory of the D1/D5 system, it is difficult to fix the
coupling of this theory to the supergravity fields. 
Traditionally couplings in effective theories have been fixed using
the method of symmetries. For example, the pion-nucleon coupling in
the pion model is fixed by symmetries. The pion $\Pi^i$ transforms as a
${\bf 3}$ of the $SU(2)$ isospin symmetry. The nucleon $N$ transforms
as a ${\bf 2}$ under this $SU(2)$. Therefore the pion-nucleon coupling
consistent with this symmetry is 
\be
{\cal L}_{\mbox{int}}= g_{\rm eff} \bar{N}\sigma^i\Pi^i\gamma_5N
\ee
where $g_{\rm eff}$ is the strength of the pion-nucleon coupling.
The operator $\bar{N}\sigma^i \gamma_5 N$ transforms as ${\bf 3}$ under
the isospin $SU(2)$. The $\gamma_5$ occurs because the pion is a pseudoscalar.
The strength of the pion-nucleon coupling $g_{eff}$ can be
fixed only be appealing to the QCD, the microscopic theory relevant
for the pion-nucleon system.
In this thesis we fix the operator in the 
SCFT corresponding to the 
supergravity field using the method of symmetries. 
It is seen that in the near horizon limit 
the D1/D5 system exhibits enhanced symmetries. 
This is a special case of the AdS/CFT correspondence. 
The near horizon geometry of the D1/D5 system reduces to that of
$AdS_3\times S^3\times T^4$. 
The AdS/CFT correspondence \cit{chp1:Mal} states
that string theory on $AdS_3\times S^3\times T^4$ is dual to the $1+1$
dimensional SCFT describing the Higgs branch of the D1/D5 system on
$T^4$. 
The radius of $S^3$ is $2\pi
{\alpha'} (g_s^2 Q_1 Q_5/V_4)^{1/4}$. 
The Kaluza-Klein modes on $T^4$ is
of the order of the string length. Thus for large $Q_1 Q_5$, type IIB
string theory on $AdS_3\times S^3\times T^4$ passes over to 6
dimensional $(2,2)$ supergravity on $AdS_3\times S^3$. We will work
in the supergravity limit. The evidence for this correspondence comes
from symmetries.
The  isometries of $AdS_3$ correspond to the global part
of the Virasoro group of the SCFT. The R-symmetry of the SCFT is
identified with the isometry of the $S^3$. The number of
supersymmetries of the bulk get enhanced to 16 from 8. These
correspond to the global supersymmetries of the ${\cal N} =(4,4)$
superalgebra of the SCFT. Thus the global part of the superalgebra
of the SCFT is identified with the $AdS_3\times S^3$ supergroup,
$SU(1,1|2)\times SU(1,1|2)$. 

Therefore a viable strategy to fix the coupling is to
classify both the bulk fields and the SCFT operators according to the
symmetries. The question then would be if this procedure can fix the
SCFT operator required for analysing Hawking radiation. We will review the
classification of the entire set of Kaluza-Klein modes of the
six-dimensional supergravity on $S^3$ as short multiplets of
$SU(1,1|2)\times SU(1,1|2)$ \cit{chp1:Deboer}. We use symmetries,
including a new global $SO(4)$ algebra, to identify the marginal
operators constructed in Chapter 2  with their corresponding
supergravity fields. This enables us to identify the operators
corresponding to the minimal scalars. 
We show that the four of the minimal scalars which are not metric
fluctuations of $T^4$ correspond to the blow up modes of the orbifold
$(\tilde{T}^4)^{Q_1Q_5}/S(Q_1Q_5)$. The long
string model has no operators corresponding to these minimal scalars
because it is not based on an orbifold. Therefore it cannot be the
effective theory for the D1/D5 system. 
We also find the representation
of the operators of the fixed scalars under the shortmultiplet of the
supergroup $SU(1,1|2)\times SU(1,1|2)$
[\ref{chp1:DavManWad2},\,\ref{chp1:DavManWad1}]. 
We show that the SCFT operators corresponding  to the all the
fixed scalars have conformal dimension $(h, \bar{h})= (2,2)$. 
The long string along with the DBI action predicts that the fixed
scalars also couple to operators with dimension $(1,3)$ and $(3,1)$.
Hawking radiation computed from the long string model using $(1,3)$
and $(3,1)$ operators gave results which did not agree with that
computed from supergravity. Operators of dimension $(2,2)$ gave
results which agreed with supergravity.
Thus this method of fixing the coupling
removes the discrepancy found
earlier for the fixed scalars using the long string model. 
In Chapter 4 we will see that the Hawking rate/absorption cross-section
is determined upto a proportionality constant by the dimension of the
SCFT operator corresponding to the supergravity field. Thus for the
case of the fixed scalars we have detemined the Hawking rate upto a
normalization constant. This constant can be determined by the same
methods used for the case of the 
minimal scalars discussed in Chapter 4. 
We compare the description of the Higgs branch of the effective $U(1)$
gauge theory describing the splitting of $(Q_1, Q_5)$ system into
subsystems in terms of Coulomb variables \cit{chp1:AhaBer} with the
supergravity description of the splitting. We find that in both cases
the splitting is described by a linear dilaton theory with the same
background charge.

In Chapter 4, we finally address the problem of Hawking radiation.
We discuss the Hawking radiation of various scalars from the D1/D5
black hole. The near horizon geometry of the D1/D5 black hole is
$BTZ\times S^3\times T^4$ \cit{chp1:MalStr96}
 where $BTZ$ refers to a a black hole in
three-dimensional Anti-de Sitter space discovered by \cit{chp1:Btz}.
The mass $M$and the angular momentum $J$
of the $BTZ$ black hole is given by 
\bea
\frac{M}{2} = L_0 + \bar{L}_0=\frac{N_R +N_L}{Q_1Q_5} \\ \nonumber
\frac{J}{2R} = L_0-\bar{L}_0=\frac{N_L -N_R}{Q_1Q_5} 
\eea
where $R= 2\pi{\alpha'} (g_s^2Q_1Q_5/V_4)^{1/4}$, $N_L$ and $N_R$
are as defined in \eq{temp}. 
The metric of the zero mass
BTZ black hole is almost identical to that of $AdS_3$ except for the
identification of $x_5$ with $x_5+ 2\pi R_5$. This difference leads to
killing spinors which are periodic in the $x_5$ coordinate for the
$BTZ$ black hole and anti-periodic for the $AdS_3$. Thus the SCFT dual
to the BTZ is the Ramond sector of the SCFT of the D1/D5 system
whereas
$AdS_3$ corresponds to the Neveu-Schwarz sector.
The microscopic degrees of freedom for the D1/D5 extremal black hole
are the states with $L_0 =N$ and $\bar{L}_0=0$
in the Ramond sector of this SCFT. The
number of such states agree precisely with the entropy formula
\eq{chp1:entropy}
We argue that the identification of the operator
${\cal O}$ corresponding to the minimal and fixed scalars carried out in
the Neveu-Schwarz sector remains valid in the Ramond sector of the
SCFT.Only non-extremal
black holes can Hawking radiate. The general D1/D5 black hole 
corresponds to a SCFT with $L_0\neq 0$ and $\bar{L}_0\neq0$ over the
Ramond vacuum. 

We calculate Hawking radiation of minimal scalars from the D1/D5
black hole using the ${\cal N}=(4,4)$  SCFT on ${\cal M}$. This is
done by relating the absorption cross-section of a quanta to the thermal
Green's function of the SCFT
operator ${\cal O}$ which couples to the minimal scalars.
The thermal Green's function is 
determined upto a normalization constant 
by the conformal dimensions $(h, \bar{h})$ of the operator
${\cal O}$ corresponding to the minimal scalars. This method of symmetries
makes it possible to derive
Hawking radiation of the minimal scalars corresponding to the 
blow up modes of the orbifold ${\cal M}$. In the long string model 
there is no way of understanding the Hawking radiation of these
minimal scalars. This is because these operators are not present in
the action of the long string model. The method of using thermal
Green's function of the operator ${\cal O}$ also makes it clear that
upto a normalization constant one can understand Hawking radiation
using symmetries.

The precise matching of the Hawking rate computed from
the microscopic theory and that computed from supergravity occurs only
on fixing the strength of the coupling $\mu$ in \eq{interaction}. 
Since the microscopic
theory is an effective theory, the precise value of
the coupling $\mu$ is difficult to fix. We have no first principle
derivation of the strength of the coupling. (Note that even in the
pion-nucleon model the strength of the coupling $g_{\rm eff}$ 
could not be fixed by symmetries, one had to appeal to QCD, the
microscopic theory relevant for the pion-nucleon model.)
We show that consistency with the
AdS/CFT correspondence fixes the strength so that the Hawking rate
computed from the  SCFT based on the orbifold ${\cal M}$ 
and supergravity match exactly. A first
principle derivation of the strength of this coupling from the
microscopic theory of the D1/D5 system is still an open problem. To
this extent a complete understanding of Hawking radiation is still
elusive.

We also investigate the Hawking 
radiation of minimal scalars for the
D1/D5 black hole with moduli turned on. 
We argue that the entropy of the D1/D5 black holes is independent of
the moduli.
We show that both in the
supergravity and in the SCFT the moduli does not modify the
Hawking rate \cit{chp1:DavManWad2}. 
Recently the supergravity solutions of the D1/D5 black
hole with moduli were constructed \cit{chp1:GASY}. 
It would be interesting to verify this
prediction for Hawking  rate of minimal scalars using the explicit
supergravity solution. 

It is clear that the method used in this thesis relies 
only on the symmetry
properties of the SCFT and the near horizon supergravity. 
The effective SCFT captures the
essential symmetries required to reproduce the supergravity result.
So it is not a surprise that  the 
SCFT calculations of Hawking radiation done in the
limit \eq{l_2} should match  with the supergravity calculations done
in the limit \eq{l_1}.

\section*{References}

\begin{enumerate}
\bibi{chp1:Wald} For a review of black hole physics and 
a list of relevant references, see ``General
Relativity,'' by Robert M. Wald, 
The University of Chicago Press, 1984.
\bibi{chp1:Hawking} S. Hawking, ``Gravitational collapsed objects of
very low mass,'' Monthly Notices Roy. Astron. Soc. {\bf
152} (1971) 75.
\bibi{chp1:Salam} A. Salam, ``Impact of quantum gravity theory on 
particle physics,'' in {\em Quantum Gravity: an Oxford Symposium}, eds.
C.J. Isham, R. Penrose and D.W. Sciama, Oxford University Press (1975).
\bibi{chp1:Hooft} G. 't Hooft, ``The black hole interpretation of 
string theory,'' Nucl. Phys. {\bf B335} (1990) 138.
\bibi{chp1:HolWil} C. Holzhey and F. Wilczek, 
``Black holes as elementary particles,'' Nucl. Phys. {\bf B335}
(199) 138, hep-th/9202014.
\bibi{chp1:Susskind}  L. Susskind, ``Some speculations about black hole
entropy in string theory,'' hep-th/9309145.
\bibi{chp1:Sen_bh} A. Sen, ``Extremal black holes and elementary
string states,'' Mod. Phys. Lett. {\bf A10} (1995) 2081,
hep-th/9504147.
\bibi{chp1:Polchinski} J. Polchinski, ``Dirichlet branes and Ramond-Ramond charges,'' Phys. Rev. Lett. {\bf 75} (1995) 4724, hep-th/9510017.
\bibi{chp1:StrVaf} A. Strominger and C. Vafa, ``Microscopic origin
of the Bekenstein-Hawking entropy,'' Phys. Lett. {\bf B379}
(1996) 99, hep-th/9601029.
\bibi{chp1:CalMal}C. G. Callan and J. Maldacena, ``D-brane
 approach to black hole quantum mechanics,'' Nucl. Phys. {\bf
B472} (1996) 591, hep-th/9602043.
\bibi{chp1:AhaBer} O. Aharony and M. Berkooz, ``IR dynamics of $d=2,
{\cal N}=(4,4)$ gauge theories and DLCQ of `Little String Theories',''
hep-th/9909101.
\bibi{chp1:MalStr96}J. Maldacena and A. Strominger, ``Black hole
greybody factors and D-brane spectroscopy,'' Phys. Rev. {\bf D55}
(1997) 861, hep-th/9609026.
\bibi{chp1:MalSus}J. Maldacena and L. Susskind, ``D-branes and fat
black holes,'' Nucl. Phys. {\bf B475} (1996) 679, hep-th/9604042.
\bibi{chp1:DhaManWad}A. Dhar, G. Mandal and S.R. Wadia, ``Absorption vs
decay of black holes in string theory and T-symmetry,'' Phys.
Lett. {\bf B388} (1996) 51, hep-th/9605234.
\bibi{chp1:DasMat}S. R. Das, and S.D. Mathur, ``Comparing decay rates
for black holes and D-branes,'' Nucl. Phys. {\bf B478} (1996)
561, hep-th/9606185; ``Interactions involving D-branes,'' Nucl.
Phys. {\bf B482} (1996) 153, hep-th/9607149.
\bibi{chp1:GubKleTsy}C. G. Callan, S.S. Gubser, I.R.
Klebanov and A. A. Tseytlin, ``Absorption of fixed scalars and the
D-brane approach to black holes,'' Nucl. Phys. {\bf B489} (1997)
65, hep-th/9610172.
\bibi{chp1:HasWad}S. F. Hassan and  S.R. Wadia, ``Gauge theory
description of D-brane black holes: emergence of the effective
SCFT and Hawking radiation,'' Nucl. Phys. {\bf B526} (1998) 311,
hep-th/9712213.
\bibi{chp1:DavManWad2} J.R. David, G. Mandal and S.R. Wadia, ``D1/D5
moduli in SCFT and gauge theory and Hawking radiation,''
hep-th/9907075.
\bibi{chp1:Mal} J. Maldacena, ``The large N limit of superconformal
field theories and supergravity,'' Adv. Theor. Math. Phys. {\bf 2}
(1997) 231, hep-th/9711200.
\bibi{chp1:Deboer}J. de Boer, ``Six-dimensional supergravity on
$S^3\times AdS_3$ and 2d conformal field theory,''
hep-th/9806104.
\bibi{chp1:DavManWad1} J.R. David, G. Mandal and S.R. Wadia, 
``Absorption and Hawking radiation of minimal and fixed
scalars, and $AdS/CFT$ correspondence,''  
Nucl. Phys. {\bf B544} (1999) 590, hep-th/9808168.
\bibi{chp1:Btz} M. Banados, C. Teitelboim and J. Zanelli,
``Black hole in
Three-dimensional spacetime,'' Phys. Rev. Lett {\bf 69} (1992) 1849.
\bibi{chp1:GASY} A. Dhar, G. Mandal, S.R. Wadia and K.P. Yogendran,
``D1/D5 system with B-field, noncommutative geometry and the CFT of
the Higgs branch,'' hep-th/9910194.
\end{enumerate}

%% file: chapter2.tex
\chapter{ The Microscopic Theory of the D1/D5 system}
\markright{Chapter 2. The Microscopic Theory of the D1/D5 system}
\vspace{-1cm}
\def\re#1{{\bf #1}}

In this chapter, we will focus on the microscopic description of the
D1/D5 system. 
We will describe the general D1/D5 system with moduli
in terms of the ${\cal N}=(4,4)$ SCFT on the resolved 
orbifold ${\cal M}$. We construct all the marginal operators of this
SCFT including the blow up modes. 
We will provide an explicit construction of all short multiplets of
single particle states of the ${\cal N}=(4,4)$ SCFT on the 
orbifold ${\cal M}$.
We also derive an effective $U(1)$
gauge theory describing the dynamics of the splitting of the D1/D5
system into subsystem \cit{chp2:DavManWad2}. We start by reviewing the gauge theory of the
D1/D5 system in detail. 

\section{The gauge theory of the D1/D5 system}

The gauge theory relevant for understanding the low energy degrees of
freedom of the D1/D5 system
consists of the  massless open string excitations
attached to the various D-branes 
[\ref{chp2:Malthesis},\,\ref{chp2:HasWad}].
Consider the $Q_1$ D1-branes and
the $Q_5$ D5-branes of type IIB string theory in ten-dimensions
arranged as follows. The D5-branes are
extended along the directions 
$ x_5, x_6, x_7, x_8, x_9$ and the D1-branes are
aligned along the $x_5$ direction. 
Let us call the open strings joining the D1-branes among themselves as
$(1,1)$ strings and the strings joining the D5-branes among themselves
as $(5,5)$ strings. 
The strings joining the D1-branes
and the D5-branes are called $(1,5)$ strings or $(5,1)$ strings
according to their orientations.
The boundary conditions of the various open strings 
at both ends along the various co-ordinates 
can be read off from the following table.
\be
\label{chp2:bc}
\begin{array}{ccccccccccc}
\mbox{co-ordinate} & x_0& x_1& x_2& x_3 &x_4&x_5&x_6&x_7&x_8&x_9 \\
\hline
Q_1\; \mbox{D1-branes} & N&D&D&D&D&N&D&D&D&D \\
Q_5\; \mbox{D5-branes} & N&D&D&D&D&N&N&N&N&N 
\end{array}
\ee
Here  $D$ stands for Dirichlet boundary conditions and $N$ stands
 for Neumann boundary conditions. The directions $x_6,x_7,x_8,x_9$ are
 compactified on a torus $T^4$.

To find the low energy degrees of freedom let us first 
us look at the symmetries
preserved by this configuration of D-branes.
 The $SO(1,9)$
 symmetry of ten-dimensions is broken down 
 to $SO(1,1)\times SO(4)_E \times SO(4)_I$ by the boundary conditions.
 The $SO(4)_E$ stands for rotations of the $1,2,3,4$ directions and
 the $SO(4)_I$ stands for rotations of the $6,7,8,9$ directions. 
 As the $6,7,8,9$  directions are compactified on the torus $T^4$, 
 the $SO(4)_I$ symmetry is also broken. But we can still use the
 $SO(4)_I$ algebra to classify states and organize fields.
 This
 configuration preserves eight supersymmetries out of the 
 32 supersymmetries of the type IIB string theory. Therefore the
 massless excitations of the various open strings can be 
 organized into fields of $N=2$ supersymmetry in four-dimensions.
Since the  size of the $T^4$  is of the order of string length 
 we can ignore the massive Kaluza-Klein modes on the $T^4$.
 Therefore the low energy effective theory of the D1/D5 system
is a $1+1$ field theory with eight
 supersymmetries. We will discuss the field content of this theory as
 obtained from the various strings joining the D-branes.
 \\
 
 \noindent
{\bf $(1,1)$ strings} 

The fields corresponding to the massless 
excitations of the $(1,1)$ strings can be obtained from the 
dimensional reduction of $U(Q_1)$
$N=1$ super Yang-Mills in ten-dimensions 
to two-dimensions. The bosonic fields of this theory can be 
organized
into the vector multiplet and the hypermultiplet of $N=2$ theory in
four-dimensions as
\bea
\mbox{Vector multiplet:} \; A_0^{(1)}, A_5^{(1)}, \phi_2^{(1)},
\phi_3^{(1)}, \phi_4^{(1)} \\   \nonumber
\mbox{Hypermultiplet:} \; Y_6^{(1)}, Y_7^{(1)}, Y_8^{(1)}, Y_9^{(1)}
\eea
The $A_0^{(1)}, A_5^{(1)}$ are the $U(Q_1)$ gauge fields in the
non-compact directions. The $Y^{(1)}$'s and $\phi^{(1)}$'s 
are gauge fields in the compact directions of the $N=1$
super Yang-Mills in ten-dimensions. 
They are hermitian $Q_1\times Q_1$ matrices transforming as
 adjoints of
$U(Q_1)$. The hypermultiplet of $N=2$ supersymmetry are
doublets of the $SU(2)_R$ symmetry of the theory. 
The $Y^{(1)}$'s can be arranged as 
doublets under $SU(2)_R$ as
\be
N_{a\bar{a}}^{(1)} = 
\left(
\begin{array}{c}
N_{1(a\bar{a})}^{(1)} \\ \nonumber
N_{2(a\bar{a})}^{(1)\dagger}
\end{array}
\right)
=\left(
\begin{array}{c}
Y_{9 (a\bar{a})}^{(1)} + i 
Y_{8 (a\bar{a})}^{(1)}  \\
Y_{7 (a\bar{a})}^{(1)} - i 
Y_{6 (a\bar{a})}^{(1)} 
\end{array}
\right)
\ee
where $a, \bar{a}$ runs from $1,\ldots ,Q_1$. 
\\

\noindent
{\bf $(5,5)$ strings} 

The field content of these massless open 
strings is similar to the the $(1,1)$ strings except for 
the fact that the gauge group is 
$U(Q_5)$ instead of $U(Q_1)$.
Normally one would have expected
the gauge theory of the $(5,5)$ strings to be a 
dimensional reduction of $N=1$ $U(Q_5)$ super Yang-Mills to
$5+1$ dimensions. Since we are ignoring the Kaluza-Klein modes 
on $T^4$ this is effectively a theory in $1+1$ dimensions. The vector
multiplets and the hypermultiplets are given by
\bea
\mbox{Vector multiplet:} \; A_0^{(5)}, A_5^{(5)}, \phi_2^{(5)},
\phi_3^{(5)}, \phi_4^{(5)} \\   \nonumber
\mbox{Hypermultiplet:} \; Y_6^{(5)}, Y_7^{(5)}, Y_8^{(5)}, Y_9^{(5)}
\eea
The $A_0^{(5)}, A_0^{(5)}$ are the $U(Q_5)$ gauge fields in the
non-compact directions. The $Y^{(5)}$'s and $\phi^{(5)}$'s 
are gauge fields in the compact directions of the $N=1$
super Yang-Mills in ten-dimensions. They 
are hermitian $Q_5\times Q_5$ matrices transforming as adjoints of
$U(Q_5)$. The hypermultiplets $Y^{(5)}$'s can be 
arranged as doublets under
$SU(2)_R$ as
\be
N_{b\bar{b}}^{(5)} = 
\left(
\begin{array}{c}
N_{1(b\bar{b})}^{(5)} \\ \nonumber
N_{2(b\bar{b})}^{(5)\dagger}
\end{array}
\right)
=
\left(
\begin{array}{c}
Y_{9 (b\bar{b})}^{(5)} + i 
Y_{8 (b\bar{b})}^{(5)}  \\
Y_{7 (b\bar{b})}^{(5)} - i 
Y_{6 (b\bar{b})}^{(5)} 
\end{array}
\right)
\ee
where $b, \bar{b}$ runs from $1,\ldots ,Q_5$. 
\\

\noindent
{\bf $(1,5)$ and $(5,1)$ strings} 

From \eq{chp2:bc} we find that these
strings have $ND$ boundary conditions along $6,7,8,9$ directions. These
boundary conditions imply that the world sheet bosons $x^6, x^7, x^8,
x^9$ of the open string have anti-periodic boundary conditions and
the world sheet fermions $\psi^6, \psi^7, \psi^8,\psi^9$ are periodic in the
Neveu-Schwarz sector. 
It can be shown that the net zero point energy of the Neveu-Schwarz 
sector vanishes. Thus the massless mode is a boson 
transforming as a spinor of $SO(4)_I$. This gives $2^2$ number of
bosons.  The GSO projection projects out half of these which
reduces the number of bosons to $2$.  
The two bosons of the $(1,5)$ strings and the $(5,1)$ strings combine
to form a complex doublet transforming under the diagonal $SU(2)$ of
the $SO(4)_I$. 
As the hypermultiplets of $N=2$ theory transform as doublets under
$SU(2)_R$, the diagonal $SU(2)$ of $SO(4)_I$ can be identified with
the $SU(2)_R$ of the gauge theory.
The Chan-Paton factors show  that 
they transform as bi-fundamentals of $U(Q_1)$ and $\overline{U(Q_5)}$. 
We arrange these hypermultiplets as doublets of the $SU(2)_R$ symmetry
of the theory as
\be
\chi_{a\bar{b}} = 
\left(
\begin{array}{c}
A_{a\bar{b}} \\
B^{\dagger}_{a\bar{b}}
\end{array}
\right)
\ee
As an aside we note that the fermionic superpartners of these
hypermultiplets which arise from the Ramond sector of the massless
excitations of
$(1,5)$ and $(5,1)$ strings carry spinorial indices under $SO(4)_E$
and they are singlets under $SO(4)_I$.

Therefore the gauge theory of the D1/D5 system is a $1+1$ dimensional 
$(4,4)$ supersymmetric gauge theory with gauge group $U(Q_1)\times
U(Q_5)$. The matter content of this theory consists of hypermultiplets
$Y^{(1)}$'s, $Y^{(5)}$'s transforming as adjoints of $U(Q_1)$ and
$U(Q_5)$ respectively. It also has the hypermultiplets $\chi$'s which
transform as bi-fundamentals of $U(Q_1)\times \overline{U(Q_5)}$. 

We will now show that this gauge theory has the required degrees of
freedom to describe the entropy of the extremal D1/D5 black hole. The
D1/D5 bound state is  described by the Higgs branch of this gauge
theory. The Higgs branch  is obtained by giving 
expectation values to the hypers. This makes the vector multiplets
massive. For a supersymmetric vacuum the hypers take values over the
surface which is given by setting the superpotential of the gauge
theory to zero. 
Setting the superpotential to zero imposes two sets of  D-flatness
conditions corresponding to each of the gauge groups $U(Q_1)$ and
$U(Q_5)$. The D-terms for the gauge group $U(Q_1)$ is given by
\bea
\label{chp2:dtermq1}
A_{a\bar{b}}A^{*}_{a'\bar{b}} -B_{b\bar{a}'}B^{*}_{b\bar{a}} 
+ [N_1^{(1)}, N_{1}^{(1)\dagger}]_{a\bar{a}'}
- [N_2^{(1)}, N_2^{(1)\dagger}]_{a\bar{a}'} =0 \\ \nonumber
A_{a\bar{b}}B_{b\bar{a}'} + [N_1^{(1)},
N_2^{(1)}]_{a\bar{a}'} = 0 \\ \nonumber
\eea
While the D-terms of the gauge group $U(Q_5)$ are
\bea
\label{chp2:dtermq5}
A_{a\bar{b}'}A^{*}_{a\bar{b}} -B_{b\bar{a}}B^{*}_{b'\bar{a}} 
+ [N_1^{(5)}, N_{1}^{(5)\dagger}]_{b\bar{b}'}
- [N_2^{(5)}, N_2^{(5)\dagger}]_{b\bar{b}'} =0 \\ \nonumber
A_{a\bar{b}'}B_{b\bar{a}} + [N_1^{(5)},
N_2^{(5)}]_{b\bar{b}'} = 0 
\eea
where $a'$ runs from $1,\ldots ,Q_1$ and $b'$ runs from $1, \ldots ,
Q_5$.

The total number of bosonic degrees of 
freedom from all the hypermultiplets is
\be
4Q_1^2 + 4Q_5^2 + 4Q_1Q_5
\ee
The first equation in \eq{chp2:dtermq1} is real and the second
equation in \eq{chp2:dtermq1} is complex. The total number of
constraints imposed by \eq{chp2:dtermq1} is $3Q_1^2$. Similarly the
set of D-term equations in \eq{chp2:dtermq5} imposes $3Q_5^2$
constraints. Equations \eq{chp2:dtermq1} and \eq{chp2:dtermq5} have
the same trace parts corresponding to the vainishing of $U(1)$
D-terms, namely,
\bea
A_{a\bar{b}}A^{*}_{a\bar{b}} -B_{b\bar{a}}B^{*}_{b\bar{a}} &=& 0 \\
\nonumber
A_{a\bar{b}}B_{a\bar{b}} &=& 0
\eea
which are three real equations. Therefore, the vanishing of D-terms
imposes $3Q_1^2 + 3Q_5^2 -3$ constraints on the fields. 
One can use the gauge symmetry $U(Q_1)$ and $U(Q_5)$ to
remove another $Q_1^2 + Q_5^2 -1$ degrees of freedom. The number of
degrees of freedom fixed by the gauge symmetry is 
less than the number of
generators of $U(Q_1)$ and $U(Q_5)$ by one. This is because all the
hypermultiplets are invariant under the diagonal $U(1)$ of
$U(Q_1)\times U(Q_5)$. After gauge fixing, the number of gauge
invariant bosonic degrees of freedom to parameterize the moduli space
is $4(Q_1Q_5 +1)$. 

We are interested in low energy black hole processes so it is
sufficient to study the SCFT of the Higgs branch. The SCFT will
have ${\cal N}= (4,4)$ SUSY with central charge $6(Q_1Q_5+1)$
on a target space ${\cal M}$.
To find the microstates corresponding to the extremal D1/D5 black hole
we look for states with $L_0 =N$ and $\bar{L}_0 =0$. 
The assymptotic number of
distinct states of this SCFT given by Cardy's formula
\be
\Omega = \exp (2\pi \sqrt{Q_1Q_5 N} )
\ee
From the Boltzmann formula one obtains
\be
S=2\pi \sqrt{Q_1Q_5 N}
\ee
This exactly reproduces the Bekenstein-Hawking entropy
\eq{chp1:entropy} of the extremal D1/D5 black hole.

\section{The instanton moduli space}
\label{chp4:instanton}

In the previous section we found that the Higgs branch of the 
gauge theory of the D1/D5 system flows 
in the infrared to  ${\cal N} =(4,4)$ SCFT 
on a target space ${\cal M}$ with central charge $6Q_1Q_5$. 
For black hole processes like Hawking radiation it is important to
know the target space ${\cal M}$. 
In this section we review the arguments which show that the
target space ${\cal M}$ is a resolution of the orbifold
$(\tilde{T}^4)^{Q_1 Q_5}/S(Q_1Q_5)$. ($\tilde{T}^4$ can 
be different from the 
compactification torus $T^4$.)

The $Q_1$ D1-branes can be thought of as $Q_1$ instantons in the $5+1$
dimensional $U(Q_5)$ super Yang-Mills theory of the $Q_5$ D5-branes 
\cit{chp2:Vafa1}.
To see this note that the DBI action of the D5-branes have a coupling
\be
\int d^6x \;\; C^{(2)}  \wedge \mbox{Tr} [F^{(5)} \wedge F^{(5)}]
\ee
The non-trivial gauge configurations which are independent of
 $x^0, x^5$ and have zero values of $A_0^{(5)}$ and  $A_5^{(5)}$ 
but non-zero values of 
$\mbox{Tr} [F^{(5)} \wedge F^{(5)}]$ act as sources of the Ramond-Ramond
two-form $C^{(2)}_{05}$. If these gauge field configurations 
have to preserve half the supersymmetries of the D5-brane action 
they should be self dual. Thus they are instanton solutions of 
four-dimensional Euclidean Yang-Mills of the $6,7,8,9$ directions. 

Additional evidence for this comes from the fact that the
integral property of $\mbox{Tr} [F^{(5)}\wedge F^{(5)}]$ corresponds to the
quantization of the D1-branes charge. The action for a $Q_1$ instanton
solution is $Q_1/g_{YM}^2$. This agrees with the tension of $Q_1$
D1-branes, namely $Q_1/g_s$. If one is dealing with non-compact
D5-branes and D1-branes it is seen that the D-flatness conditions of
the D1-brane theory is identical to the ADHM construction of
$Q_1$ instantons of  $U(Q_5)$ gauge theory \cit{chp2:Douglas}.

{}From the discussion in the preceding paragraphs we conclude that ${\cal M}$ the target
space of SCFT can be thought of as the moduli space of $Q_1$ 
instantons of a $U(Q_5)$ gauge theory on $T^4$. 
This moduli space is known to be the Hilbert scheme of $Q_1Q_5$ points
on $\tilde{T}^4$ \cit{chp2:Dijkgraaf}. $\tilde{T}^4$ can be different from the
compactification torus $T^4$.
This is a smooth resolution of the singular orbifold
$(\tilde{T}^4)^{Q_1Q_5}/S(Q_1Q_5)$. We will 
provide physically motivated 
evidence for the fact that the moduli space of $Q_1$
instantons of a $U(Q_5)$ gauge theory on $T^4$ is a smooth resolution of
the orbifold $(\tilde{T}^4)^{Q_1Q_5}/S(Q_1Q_5)$ using string
dualities. The evidence is topological and it 
comes from realising that the cohomology of
${\cal M}$ is  the degeneracy of the ground states of the D1/D5 gauge
theory. We can calculate this degeneracy in two ways. One is by
explicitly counting the cohomology of 
$(\tilde{T}^4)^{Q_1 Q_5}/S(Q_1 Q_5)$ \cit{chp2:Vafa2}.
The second method is to use string dualities as discussed below. 
Both these methods give identical answers. Thus at least at the 
level of cohomology we are able to verify that the moduli space is
$(\tilde{T}^4)^{Q_1 Q_5}/S(Q_1 Q_5)$. 

Consider type IIB string theory compactified on 
$S^1\times T^4$ with a fundamental string having $Q_5$ units of
winding along $x^6$ and $Q_1$ units of momentum along $x^6$. On
performing the sequence of dualities $ST_{6789}ST_{56}$ we can map the
fundamental string to the D1/D5 system 
(we can de-compactify the $x^5$ direction finally)
with $Q_1$ D1-branes along
$x^5$ and $Q_5$ D5-branes along $x^5, x^6, x^7, x^8, x^9$ \cit{chp2:Sen}. 
Therefore using this U-duality sequence the BPS states of this
fundamental string (that is, states with either purely left moving or
right moving oscillators) maps to ground states of the D1/D5 system. 
The number of ground states of the D1/D5 system is given by the 
dimension of the cohomology of ${\cal M}$. From the 
perturbative string degeneracy counting the generating function of BPS
states with left moving oscillator number $N_L$ is given by
\be
\sum_{N_L=o}^\infty d(N_L)q^{N_L} = 256\times \prod_{n=1}^\infty
\left(\frac{1+ q^n}{1-q^n} \right)^8
\ee
where $d(N_L)$ refers to the degeneracy of states with left moving
oscillator number $N_L$. The D1/D5 system is U-dual to the perturbative
string with $N_L=Q_1Q_5$.  

Explicit counting of the cohomology of ${\cal M}$ gives
$d(Q_1Q_5)/256$. The factor $256$ comes from quantization of the
center of mass coordinate along $1,2,3,4$ directions and the $6,7,8,9$
directions. The center of mass coordinate is represented by the
relative $U(1)$ of $U(Q_1)\times U(Q_5)$. Therefore the low energy
theory of the bound D1/D5 system is a SCFT on the target space
\be
R^4\times T^4\times (\tilde{T^4})^{Q_1 Q_5}/S(Q_1 Q_5) 
\ee
From now on we will suppress the center of mass coordinate and
${\cal M}$ will denote 
\be
(\tilde{T}^4)^{Q_1 Q_5}/S(Q_1 Q_5)
\ee

\section{The SCFT on the orbifold ${\cal M}$}


The ${\cal N}= (4,4)$ SCFT on ${\cal M}$ 
is described by the free Lagrangian
\be
\label{free}
S = \frac{1}{2} \int d^2 z\; \left[\del
x^i_A \bar\del x_{i,A} + 
\psi_A^i(z) \bar\del \psi^i_A(z) + 
\widetilde\psi^i_A(\bar z) \del \widetilde \psi^i_A(\bar z) 
 \right]
\ee
Here $i$ runs over the $\widetilde{ T^4}$ coordinates
1,2,3,4 and $A=1,2,\ldots,Q_1Q_5$ labels various copies
of the four-torus. The symmetric group $S(Q_1Q_5)$
acts by permuting the copy indices. It introduces various twisted
sectors which we will discuss later. The free field realization of
this SCFT has ${\cal N}=(4,4)$ superconformal symmetry. 
To set up our notations and conventions we review the
${\cal N}=4$ superconformal algebra.

\subsection{The ${\cal N}=4$ superconformal algebra}
%

The algebra is generated by the stress energy
tensor, four supersymmetry currents, and a local $SU(2)$ $R$ symmetry
current. The operator product expansions(OPE) of the algebra 
with central charge $c$ are given by (See for example \cit{chp2:Yu}.)
\bea
\label{chp2:scft_algebra}
T(z)T(w) &=& \frac{\del T(w)}{z-w} + \frac{2 T(w)}{(z-w)^2} +
\frac{c}{2 (z-w)^4},  \\  \nonumber
G^a(z)G^{b\dagger }(w) &=& 
\frac{2 T(w)\delta_{ab}}{z-w} + \frac{2 \bar{\sigma}^i_{ab} \del J^i}
{z-w} + \frac{ 4 \bar{\sigma}^i_{ab} J^i}{(z-w)^2} + 
\frac{2c\delta_{ab}}{3(z-w)^3}, \\
\nonumber
J^i(z) J^j(w) &=& \frac{i\epsilon^{ijk} J^k}{z-w} + \frac{c}{12 (z-w)^2}
, \\  \nonumber
T(z)G^a(w) &=& \frac{\del G^a (w)}{z-w} + \frac{3 G^a (z)}{2 (z-w)^2},
\\   \nonumber
T(z) G^{a\dagger }(w) &=& \frac{\del G^{a\dagger} (w)}{z-w} + 
\frac{3 G^{a\dagger} (z)}{2 (z-w)^2}, \\   \nonumber
T(z) J^i(w) &=& \frac{\del J^i (w)}{z-w} + \frac{J^i}{(z-w)^2}, \\
\nonumber
J^i(z) G^a (w) &=& \frac{G^b(z) (\sigma^i)^{ba}}{2 (z-w)}, \\
\nonumber
J^i(z) G^{a\dagger}(w) &=& -\frac{(\sigma^i)^{ab} G^{b\dagger
}(w)}{2(z-w)} 
\eea
Here $T(z)$ is the stress energy tensor, $G^a(z), G^{b\dagger}(z)$ 
the  $SU(2)$
doublet of supersymmetry generators and  $J^i(z)$ the $SU(2)$ $R$
symmetry current. The $\sigma$'s stand for Pauli matrices and the
$\bar{\sigma}$'s stand for the complex conjugates of Pauli matrices.
In the free field realization described below,
the above holomorphic currents occur together with
their antiholomorphic counterparts, which we will
denote by $\widetilde J(\bar z), \widetilde G(\bar z)$
and $\widetilde T(\bar z)$. In
particular, the R-parity group will be denoted
by $SU(2)_R \times \widetilde{SU(2)}_R$.

\subsection{Free field realization of ${\cal N}=(4,4)$ SCFT on the
orbifold ${\cal M}$}
A free field realization of the ${\cal N}=4$ superconformal 
algebra with $c=6Q_1Q_5$
can be constructed out of $Q_1Q_5$ copies 
of four real fermions and bosons. 
The generators are given by
\bea
T(z) &=& \del X_A (z)\del X^\dagger_A(z) 
+ \frac{1}{2}\Psi_A (z)\del \Psi^{\dagger}_A (z) 
- \frac{1}{2}\del\Psi_A (z) \Psi^{\dagger}_A (z) 
\\   \nonumber
G^a(z) &=&
\left(
\begin{array}{c}
G^1(z)\\
G^2(z)
\end{array}
\right) =
\sqrt{2} \left(
\begin{array}{c}
\Psi^1_A (z) \\
\Psi^2_A (z) \end{array}  \right)  \del X^2_A (z)  +
\sqrt{2} \left( \begin{array}{c}
-\Psi^{2\dagger}_A (z) \\
\Psi^{1\dagger}_A (z)  \end{array}  \right)  \del X^1_A (z) 
\\    \nonumber
J^i_R(z) &=& \frac{1}{2} \Psi_A(z)\sigma^i\Psi^\dagger_A (z)\\
\nonumber
\eea
We will use the following notation for the zero mode
of the R-parity current:
\be
J^i_R = \frac{1}{2}\int\frac{dz}{2\pi i} 
\Psi_A(z)\sigma^i\Psi^\dagger_A(z)
\ee
In the above the summation over $A$ which runs from $1$ to $Q_1Q_5$ is
implied.
The bosons $X$ and the fermions $\Psi$ are 
\bea
\label{defn}
X_A(z) &=& (X^1_A(z), X^2_A(z)) = \sqrt{1/2} (x^1_A(z) + i x^2_A(z),
x^3_A(z) + i x^4_A(z)),   \\  \nonumber
\Psi_A (z) &=& (\Psi^1_A(z), \Psi^2_A(z)) = \sqrt{1/2} (\psi^1_A(z) +
i\psi^2_A(z), \psi^3_A(z) + i\psi^4_A(z)) \\   \nonumber
X_A^\dagger (z) &=& 
\left(
\begin{array}{c}
X_A^{1\dagger} (z) \\
X_A^{2\dagger} (z)
\end{array}
\right)   = \sqrt{\frac{1}{2}}
\left(
\begin{array}{c}
x^1_A(z)-ix^2_A (z)\\
x^2_A(z)-ix^2_A(z)
\end{array}
\right) \\   \nonumber
\Psi_A^\dagger (z) &=&
\left(
\begin{array}{c}
\Psi_A^{1\dagger} (z) \\
\Psi^{2_A \dagger} (z)
\end{array}
\right)     =\sqrt{\frac{1}{2}} 
\left(  
\begin{array}{c}
\psi^1_A(z) - i\psi^2_A(z)\\
\psi^3_A(z) - i\psi^4_A(z)
\end{array}
\right)
\eea

\subsection{The $SO(4)$ algebra}

In addition to the local $R$ symmetry the free field realization of
the ${\cal N}=4$ superconformal algebra has additional global
symmetries which can be used to classify the states. There are $2$
global $SU(2)$ symmetries which correspond to the $SO(4)$ rotations of
the $4$ bosons $x^i$. The corresponding charges are given by 
\bea
I_1^i &=& 
\frac{1}{4}\int\frac{dz}{2\pi i} X_A \sigma^i \del X_A^\dagger 
-\frac{1}{4}\int\frac{dz}{2\pi i} \del X_A \sigma^i X_A^\dagger 
+ \frac{1}{2}\int\frac{dz}{2\pi i}
\Phi_A \sigma^i \Phi_A^\dagger  \\  \nonumber
I_2^i &=& 
\frac{1}{4} 
\int\frac{dz}{2\pi i}{\cal X}_A\sigma^i\del{\cal X}_A^\dagger
-\frac{1}{4} 
\int\frac{dz}{2\pi i}\del{\cal X}_A\sigma^i{\cal X}_A^\dagger
\eea
Here 
\bea
{\cal X }_A = (X^1_A, -X^{2\dagger}_A) \;&\;&\;\;
{\cal X}^\dagger  =
\left(
\begin{array}{c}
X^{1\dagger}_A \\
-X^2_A 
\end{array}  \right) \nonumber \\
\Phi_A = (\Psi^1_A, \Psi^{2\dagger}_A ) \;&\;&\;\;
\Phi_A^\dagger =
\left(
\begin{array}{c}
\Psi^{1\dagger}_A  \\
\Psi^2_A 
\end{array} \right).
\eea
These charges are generators of $SU(2)\times SU(2)$ algebra:
\bea
[I_1^i, I_1^j] = i\epsilon^{ijk} I_1^k \;&\;&\;\;
[I_2^i, I_2^j] = i\epsilon^{ijk} I_2^k 
\\  \nonumber
[I_1^i, I_2^j] &=&0
\eea
The commutation relation of these new global charges with the various
local charges are given below
\bea
\label{so4_on_g}
[I_1^i, G^a(z)] =0 \;&\;&\;\;
[I_1^i, G^{a\dagger}(z)] =0 \\ \nonumber 
[I_1^i, T(z)] =0 \;&\;&\;\;
[I_1^i, J(z)]=0 \\ \nonumber 
[I_2^i, {\cal G}^a(z)] = 
\frac{1}{2}{\cal G}^{b} (z)\sigma^i_{ba} \;&\;&\;\;
[I_2^i, {\cal G}^{a\dagger} (z) ]
= - \frac{1}{2}\sigma^i_{ab}{\cal G}^{b\dagger}(z) \\ \nonumber
[I_2^i, T(z)] =0  \;&\;&\;\;
[I_2^i, J(z)]=0
\eea
where
\bea
{\cal G} = ( G^1, G^{2\dagger}) \;&\;&\;\;
{\cal G}^\dagger=
\left(
\begin{array}{c}
G^{1\dagger} \\
G^2
\end{array}
\right)
\eea

The following commutations relation show that 
the bosons transform as $(\bf 2, \bf 2)$ under $SU(2)_{I_1}\times
SU(2)_{I_2}$
\bea
\label{boson}
[I_1^i, X^a_A] = \frac{1}{2} X^b_A\sigma^i_{ba} \;&\;&\;\;
[I_1^i, X^{a\dagger}_A] = -\frac{1}{2}\sigma^i_{ab}X^{b\dagger}_A 
\\   \nonumber
[I_2^i, {\cal X}^a_A ] =
\frac{1}{2} {\cal X}^b_A \sigma^i_{ba}  \;&\;&\;\;
[I_2^i, {\cal X}^{a\dagger}_A] =
-\frac{1}{2}\sigma^i_{ab}{\cal X}^{b\dagger}_A 
\eea
The fermions transform as $(\bf 2, \bf 1)$ under $SU(2)_{I_1}\times 
SU(2)_{I_2}$ as can be seen from the commutations relations
given below.
\bea
[I_1^i, \Phi^a_A] = \frac{1}{2}\Phi^b_A \sigma^i_{ba} \;&\;&\;\;
[I_1^i, \Phi^{a\dagger}_A] =
-\frac{1}{2}\sigma^i_{ab}\Phi^{b\dagger}_A  \\   \nonumber
[I_2^i, \Psi^a] =0  \;&\;&\;\;
[I_2^i, \bar{\Psi}^a]=0
\eea
We are interested in  studying the states of the ${\cal N}=(4,4)$ SCFT
on ${\cal M}$. The classification of the states and their symmetry
properties can be analyzed by studying the states of a free field
realization of a ${\cal N}=(4,4)$ SCFT on $R^{4Q_1Q_5}/S(Q_1Q_5)$. 
This
is realized by considering the holomorphic and the anti-holomorphic
${\cal N}=4$ superconformal algebra with $c=\bar{c}=6Q_1Q_5$
constructed out of $Q_1Q_5$ copies of four real fermions and bosons. 
So we have an anti-holomorphic component for each field, generator and
charges discussed above. These are labelled by the same symbols used
for the holomorphic components but distinguished by a tilde.

The charges $I_1, I_2$ constructed
above  generate $SO(4)$
transformations only on  the {\em holomorphic} bosons $X_A(z)$. 
Similarly, we can construct charges
$\widetilde{I_1}, \widetilde{I_2}$ which generate $SO(4)$
transformations only on  the {\em antiholomorphic} 
bosons $\widetilde{X_A}(\bar z)$. 
Normally one would expect these
charges to give rise to a global $SO(4)_{hol}
\times SO(4)_{antihol}$ symmetry. However, 
the kinetic term of the bosons in the
free field realization is not invariant under independent holomorphic
and antiholomorphic $SO(4)$ rotations. It is
easy to see, for example by using the Noether
procedure, that there is a residual $SO(4)$ symmetry
generated by the charges 
\bea
J_I= I_1  + \widetilde{I}_1 \;\;&\;&\;
\widetilde{J}_I = I_2 + \widetilde{I}_2
\eea
We will denote this symmetry as $SO(4)_I =
SU(2)_I\times \widetilde{SU(2)}_I$, where the
$SU(2)$ factors are generated by $J_I,
\widetilde{J}_I$. These charges satisfy the
property that (a) they 
correspond to $SO(4)$ transformations of the
bosons $X_A(z,
\bar z)= X_A(z) 
+ \widetilde{X_A}(\bar z)$ and (b) they fall into representations of the
${\cal{N}}=(4,4)$ algebra (as can 
be proved by using the commutation relations \eq{boson}
of the $I$'s). The bosons $X(z,\bar{z})$
transform as $(\bf 2 , \bf 2)$ under $SU(2)_I\times \widetilde{SU(2)}_I$. 

\subsection{The supergroup $SU(1,1|2)$}

The global part of the ${\cal N}=4$ superconformal algebra forms the
supergroup $SU(1,1|2)$. 
Let $L_{\pm,0} ,J^{(1),(2),(3)}_R$  be
the global charges of the currents
$T(z)$ and $J^{(i)}_R(z)$ and  $G^a_{1/2,-1/2} $  the
global charges of the supersymmetry currents $G^a(z)$ 
in the Neveu-Schwarz sector. From the OPE's \eq{chp2:scft_algebra}
we obtain the following commutation relations for the global charges.
\bea
[L_0, L_{\pm}] = \mp L_{\pm} \;\;&\;&\;\; [L_{1} , L_{-1}] = 2L_{0} \\ \nonumber
\{ G^a_{1/2} , G^{b\dagger}_{-1/2} \} &=& 2\delta^{ab}L_0 + 2
\sigma^i_{ab} J^{(i)}_{R} \\  \nonumber
\{ G^a_{-1/2} , G^{b\dagger}_{1/2} \} &=& 2\delta^{ab}L_0 - 2
\sigma^i_{ab} J^{(i)}_{R} \\  \nonumber
[J^{(i)}_R, J^{(j)}_R] &=& i\epsilon^{ijk}J^{(k)}_R \\ \nonumber
[L_0, G^a_{\pm 1/2}] = \mp\frac{1}{2} G^a_{\pm 1/2} \;\;&\;&\;\;[L_0, G^{a\dagger}_{\pm 1/2}] = \mp\frac{1}{2} G^{a\dagger}_{\pm 1/2} 
\\ \nonumber
[L_+ , G^a_{1/2}] = 0 \;\;&\;&\;\; [L_- , G^a_{-1/2}] = 0 \\ \nonumber
[L_- , G^a_{1/2}] = -G^a_{-1/2} \;\;&\;&\;\; [L_+ , G^a_{-1/2}] = G^a_{1/2} \\ \nonumber
[L_+ , G^{a\dagger}_{1/2}] = 0 \;\;&\;&\;\; [L_- , G^{a\dagger}_{1/2}] = 0 \\ \nonumber
[L_- , G^{a\dagger}_{1/2}] = -G^a_{-1/2} \;\;&\;&\;\; [L_+ , G^{a\dagger}_{-1/2}] = G^a_{1/2} \\ \nonumber
[J^{(i)}_R, G^a_{\pm 1/2} ] = \frac{1}{2} G^{b}_{\pm 1/2} (\sigma^i)^{ba}
\;\;&\;&\;\; [J^{(i)}_R, G^{a\dagger}_{\pm 1/2} ] = -\frac{1}{2}
 (\sigma^i)^{ba}G^{b\dagger}_{\pm 1/2} \\  \nonumber
\eea
The above commutation relations form the algebra of the
supergroup $SU(1,1|2)$. The global part of the ${\cal N}= (4,4)$
superconformal algebra form the super group $SU(1,1|2)\times
SU(1,1|2)$.

\section{Short multiplets of $SU(1,1|2)$}
\label{chp2:short_multiplets}

The representations of the supergroup $SU(1,1|2)$
are classified according to the conformal weight 
and $SU(2)_R$ quantum number. The highest weight states
$ |\mbox{hw}\rangle = 
|h,{\bf j}_R,j_R^3 =j_R \rangle $ satisfy the following
properties
\bea
L_1 |\mbox{hw}\rangle = 0 \;\;\; 
L_0 |\mbox{hw}\rangle = h|\mbox{hw}\rangle \\ \nonumber
J^{(+)}_{R}|\mbox{hw}\rangle =0  \;\;\; 
J_R^{(3)}|\mbox{hw}\rangle = j_R|\mbox{hw}\rangle\\  \nonumber
G_{1/2}^a|\mbox{hw}\rangle =0 \;\;\; G_{1/2}^{a\dagger}
|\mbox{hw}\rangle =0
\eea
where $J^+_R = J^{(1)}_R + i J^{(2)}_R$.
Highest weight states which satisfy 
$
G^{2\dagger }_{-1/2}|\mbox{hw}\rangle =0 ,\;\;\;
G^1_{-1/2}|\mbox{hw}\rangle =0
$
are chiral primaries. They satisfy $h=j$. We will denote 
these states as $|\mbox{hw}\rangle _{S}$. Short multiplets are
generated from the chiral primaries through the action of the raising
operators $J_{-}, G^{1\dagger }_{-1/2}$ and $G^2_{-1/2}$. The structure
of the short multiplet is given below
\bea
\label{short}
\begin{array}{cccc}
\mbox{States} & j & L_0 & \mbox{Degeneracy} \\
|\mbox{hw}\rangle_{S} & h & h& 2h+1 \\
G^{1\dagger }_{-1/2}|\mbox{hw}\rangle_{S}, 
G^2_{-1/2}|\mbox{hw}\rangle_{S} &h-1/2& h+1/2 & 2h + 2h = 4h \\
G^{1\dagger }_{-1/2} G^2_{-1/2}|\mbox{hw}\rangle_{S} & h-1& h+1& 2h-1
\end{array}
\eea
The short multiplets of the supergroup $SU(1,1|2)\times SU(1,1|2)$ are
obtained by the tensor product of the above multiplet. We denote the
short multiplet of  $SU(1,1|2)\times SU(1,1|2)$ as
$(\bf{2h +1}, \bf{2h'+1})_S$. These stand for the degeneracy of the
bottom component, the top row in  \eq{short}. The top component of
the short multiplet are the states belonging to  the last row in
\eq{short}. The short multiplet $(\bf{2}, \bf{2})_S$ is special, it
terminates at the middle row of \eq{short}. For this case, the top
component is the middle row. These states have $h=\bar{h}=1$
and transform as $(\bf{1}, \bf{1})$ of $SU(2)_R\times
\widetilde{SU(2)}_R$. There are $4$ such states for each $(\bf{2},
\bf{2})_S$.

\section{The resolutions of the symmetric product} 
\label{chp2:res}

We would like to discuss D1/D5 systems with various moduli turned on.
To do this we will construct marginal operators of the ${\cal
N}=(4,4)$ SCFT on the symmetric product orbifold ${\cal M}$. We will
find the four operators which correspond to resolution of the orbifold
singularity.

\subsection{The untwisted sector}
Let us first focus on the operators constructed from the
untwisted sector. The operators of lowest conformal weight 
are
\bea 
\label{chiral}
\Psi^1_A(z) \widetilde{\Psi}^1_A(\bar{z})  \;&\;&\;  
\Psi^1_A(z)\widetilde{\Psi}^{2\dagger}_A(\bar z) \\   \nonumber
\Psi^{2\dagger}_A(z)\widetilde{\Psi}^1_A(\bar z )  \;&\;&\;
\Psi^{2\dagger}_A(z)\widetilde{\Psi}^{2\dagger}_A(\bar z) 
\eea
where summation over $A$ is implied. These four operators have conformal
dimension $(h, \bar{h})=(1/2, 1/2)$ and
$(j_R^3, \widetilde{j}_R^3)= (1/2, 1/2)$ under
the R-symmetry  $SU(2)_R\times \widetilde{SU(2)}_R$.
Since $(h, \bar{h})=(j_R^3, \widetilde{j}_R^3)$,
 these operators are chiral primaries and
have non-singular operator product expansions (OPE) with the
supersymmetry currents 
$G^1(z), G^{2\dagger}(z), \widetilde{G}^1(\bar z),
\widetilde{G}^{2\dagger}(\bar z)$. 
These properties indicate
that they belong to the bottom component of the short multiplet $(\bf
2, \bf 2)_S$.  Each of the four chiral
primaries gives rise to four top components of the short multiplet
$(\bf 2, \bf 2)_S$. They are given by the leading pole
($(z-w)^{-1} (\bar z - \bar w)^{-1}$) in the
OPE's
\bea
\label{OPE}
G^2(z)\widetilde{G}^2(\bar z){\cal P} (w, \bar w)
\; &\;&\; 
G^2(z)\widetilde{G}^{1\dagger}(\bar z){\cal P} (w, \bar w)
\\ \nonumber
G^{1\dagger}(z) \widetilde{G}^2(\bar z){\cal P}(w, \bar w) \; &\;& \;
G^{1\dagger}(z)\widetilde{G}^{1\dagger} 
(\bar z){\cal P} (w, \bar w)
\eea
where ${\cal P}$ stands for any of the four chiral primaries in
\eq{chiral}. From the superconformal algebra it is easily seen that
the top components constructed above have weights $(1,1)$ and
transform as $(\bf 1, \bf 1 )$ under $SU(2)_R\times
\widetilde{SU(2)}_R$.  The OPE's \eq{OPE}\ can be easily evaluated. We
find that the $16$ top components of the $4 (\bf 2, \bf 2)_S$ short
multiplets are $\del x_A^i \bar{\del} x_A^j$.

We classify the above operators belonging to the top component
according to representations of (a) the $SO(4)_I$ rotational symmetry
of the $\widetilde{T}^4$, (The four torus $\tilde{T}^4$ breaks this
symmetry but we assume the target space is $R^4$ for the
classification of states)
(b) $R$ symmetry of the SCFT and (c) the
conformal weights. As all of these operators belong to the top
component of $(\bf 2, \bf 2 )_{\bf S}$ the only property which
distinguishes them is the representation under $SO(4)_I$. 
The quantum
numbers of these operators under the various symmetries are
\bea
\label{untwist_operator}
\begin{array}{lccc}
\mbox{Operator}&SU(2)_I\times \widetilde{SU(2)}_I&
SU(2)_R\times\widetilde{SU(2)}_R& (h, \bar{h}) \\
\del x^{ \{ i }_A(z) \bar{\del}x^{ j\} }_A (\bar z) -
\frac{1}{4}\delta^{ij}
\del x^k_A(z) \bar{\del}x^k_A (\bar z) &(\bf 3, \bf 3) & (\bf 1,\bf 1) 
& (1, 1) \\  
\del x^{[i}_A(z) \bar{\del}x^{j]}_A (\bar z) & (\bf 3, \bf 1) +
(\bf 1, \bf 3) & (\bf 1, \bf 1)& (1,1) \\ 
\del x^i_A(z) \bar{\del}x^i_A (\bar z) &(\bf 1, \bf 1)& (\bf 1, \bf 1)
& (1,1)
\end{array}
\eea
Therefore we have $16$ marginal operators from the untwisted sector.
As these are top components they can be added to the free SCFT as
perturbations without violating the ${\cal N}=(4,4)$ supersymmetry.

\subsection{$Z_2$ twists.}
\label{chp2:z_2twists}

We now construct the marginal operators from the various twisted
sectors of the orbifold SCFT.  The twist fields of the SCFT on the
orbifold ${\cal M}$ are labeled by the conjugacy classes of the
symmetric group $S(Q_1 Q_5)$ 
[\ref{chp2:VafWit},\,\ref{chp2:DijMooVerVer}]. The conjugacy
classes consist of cyclic groups of various lengths. The various
conjugacy classes and the multiplicity in which they occur in
$S(Q_1Q_5)$ can be found from the solutions of the equation
\be
\sum nN_n = Q_1 Q_5
\ee
where $n$ is the length of the cycle and $N_n$ is the multiplicity of
the cycle. Consider the simplest nontrivial conjugacy class which is
given by $N_1 = Q_1 Q_5 -2, N_2 = 1$ and the rest of $N_n =0$.  
A representative element of this class is 
\be
\label{group_element}
(X_1\rightarrow X_2, \; X_2\rightarrow X_1), 
\; X_3\rightarrow X_3 , \; \ldots ,
\; X_{Q_1Q_5} \rightarrow X_{Q_1Q_5}
\ee
Here the $X_A$'s are related to the $x_A$'s appearing in 
the action \eq{free} by \eq{defn} in.

To exhibit the singularity of this group action we go over to the
following new coordinates
\be
X_{cm} = X_1+ X_2 \;\; \mbox{and}\;\; \phi = X_1-X_2
\ee
Under the group action \eq{group_element}\ 
$X_{cm}$ is invariant and $\phi
\rightarrow -\phi$. Thus the singularity is {\em locally} of the type
$R^4/Z_2$. The bosonic twist operators for this orbifold singularity
are given by following OPE's
\cit{chp2:DixFriMarShe}
\bea
\del \phi^1 (z) \sigma^1(w, \bar{w} ) = 
\frac{ \tau^1(w, \bar w ) }{ (z-w)^{1/2} }  \; &\;& \;
\del {\phi}^{1\dagger} (z) \sigma^1(w, \bar{w} ) = 
\frac{ \tau'^1(w, \bar w ) }{ (z-w)^{1/2} }  \\   \nonumber
\del \phi^2 (z) \sigma^2(w, \bar{w} ) =
\frac{ \tau^2(w, \bar w ) }{ (z-w)^{1/2} }  \; &\;& \;
\del {\phi}^{2\dagger} (z) \sigma^2(w, \bar{w} ) =
\frac{ \tau'^2(w, \bar w ) }{ (z-w)^{1/2} }  \\  \nonumber
\bar{\del} \widetilde{\phi}^1 (\bar{z}) \sigma^1(w, \bar{w} ) = 
\frac{ \widetilde\tau'^1(w, \bar w ) }{ (\bar z-\bar w)^{1/2} }  \; &\;&
\;
\bar{\del} \widetilde{\phi}^{1\dagger} (\bar z) \sigma^1(w, \bar{w} ) = 
\frac{ \widetilde\tau^1(w \bar w ) }{ (\bar z-\bar w)^{1/2} } 
\\   \nonumber
\bar{\del} \widetilde{\phi}^2 (\bar z) \sigma^2(w, \bar{w} ) =
\frac{ \widetilde\tau'^2(w, \bar w ) }{ (\bar z-\bar w)^{1/2} }  \; &\;&
\;
\bar{\del} \widetilde{\phi}^{2\dagger} (\bar z ) \sigma^2(w, \bar{w} ) =
\frac{ \widetilde\tau^2(w, \bar w ) }{ (\bar z-\bar w)^{1/2} }  
\eea
The $\tau$'s are excited twist operators.
The fermionic twists are constructed from bosonized currents defined
by
\bea
\chi^1(z) = e^{iH^1(z)} \; &\;& \; \chi^{1\dagger}(z) = e^{-iH^1(z)} \\
\nonumber
\chi^2(z) = e^{iH^2(z)} \; &\;& \; \chi^{2\dagger}(z) = e^{-iH^2(z)} \\
\nonumber
\eea
Where the $\chi$'s, defined as
$\Psi_1 - \Psi_2$, are the superpartners of the bosons $\phi$.

{}From the above we construct the supersymmetric twist fields which act
both on fermions and bosons as follows:
\bea
\label{defSigma}
\Sigma^{(\frac{1}{2}, \, \frac{1}{2})}_{(12)} = \sigma^1(z,\bar z)
\sigma^2 (z,\bar z)
e^{iH^1(z)/2} e^{-iH^2(z)/2} 
e^{i\widetilde{H}^1(\bar z )/2} e^{-i\widetilde{H}^2(\bar z )/2}
\\  \nonumber
\Sigma^{(\frac{1}{2},\,  -\frac{1}{2})}_{(12)} = \sigma^1(z,\bar z)
\sigma^2 (z,\bar z)
e^{iH^1(z)/2} e^{-iH^2(z)/2} 
e^{- i\widetilde{H}^1(\bar z )/2} e^{i\widetilde{H}^2(\bar z )/2} 
\\  \nonumber
\Sigma^{(-\frac{1}{2}, \,  \frac{1}{2})}_{(12)} = \sigma^1(z,\bar z)
\sigma^2 (z,\bar z)
e^{-iH^1(z)/2} e^{+iH^2(z)/2} 
e^{i\widetilde{H}^1(\bar z )/2} e^{-i\widetilde{H}^2(\bar z )/2} 
\\   \nonumber
\Sigma^{(-\frac{1}{2}, \,  -\frac{1}{2})}_{(12)} = 
\sigma^1(z,\bar z) \sigma^2 (z,\bar z)
e^{-iH^1(z)/2} e^{+iH^2(z)/2} 
e^{-i\widetilde{H}^1(\bar z )/2} e^{+i\widetilde{H}^2(\bar z )/2} 
\\  \nonumber
\eea
The subscript $(12)$ refers to the fact that these twist operators were
constructed for the representative
group element \eq{group_element}\ which exchanges the $1$ and
$2$ labels of the coordinates of $\widetilde{T}^4$. 
The superscript stands
for the $(j^3_R, \widetilde{j}^3_R)$ quantum numbers.
The twist operators for the orbifold ${\cal M}$ belonging to the
conjugacy class under consideration is obtained by summing over these
$Z_2$ twist operators for all representative elements of this class.
\be
\Sigma^{(\frac{1}{2}, \,  \frac{1}{2})} =
\sum_{i=1}^{Q_1Q_5} \sum_{j=1, j\neq i}^{Q_1Q_5}
 \Sigma^{(\frac{1}{2}, \,  \frac{1}{2})}_{(ij)}
\ee
We can define the rest of the twist operators for the orbifold in a
similar manner. The conformal dimensions of these operators are
$(1/2,1/2)$. They transform as $(\bf 2 , \bf 2)$ under the $SU(2)_R
\times \widetilde{SU(2)}_R$ symmetry of the SCFT. They belong to the
bottom component of the short multiplet $(\bf 2, \bf 2)_S$. The
operator $\Sigma^{(\frac{1}{2}, \, \frac{1}{2})}$ is a chiral primary.
As before the $4$ top components of this short multiplet,
which we denote by
\bea
T^{(\frac{1}{2}, \, \frac{1}{2})}, \;\; T^{(\frac{1}{2}, \,
-\frac{1}{2})} \\ \nonumber
T^{(-\frac{1}{2}, \, \frac{1}{2})}, \;\;
T^{(-\frac{1}{2}, \, -\frac{1}{2})} 
\eea 
are given  by the leading pole in the following OPE's respectively
\bea 
\label{raising}
G^2(z)\widetilde{G}^2(\bar z)\Sigma^
{(\frac{1}{2}, \,  \frac{1}{2})} (w, \bar w), \;\;
G^2(z) \widetilde{G}^{1\dagger}(\bar z)\Sigma^{(\frac{1}{2}, \,  
\frac{1}{2})} (w, \bar w),  \\  \nonumber
 G^{1\dagger}(z) \widetilde{G}^2(\bar z)
 \Sigma^{(\frac{1}{2}, \, \frac{1}{2})} (w, \bar w), \;\;
G^{1\dagger}(z)\widetilde{G}^{1\dagger}
(\bar z)\Sigma^{(\frac{1}{2}, \,\frac{1}{2})} (w, \bar w)
\eea
These are the $4$ blow up modes of the $R^4/Z_2$ singularity
\cit{chp2:CveDix}\ and they have conformal weight $(1,1)$%
\footnote{Relevance of $Z_2$ twist operators
to the marginal deformations of the SCFT has earlier
been discussed in [\ref{chp2:HasWad1},\,\ref{chp2:DijVerVer}]}. 
They transform as
$(\bf 1 , \bf
1)$ under the $SU(2)_R \times \widetilde{SU(2)_R}$.  As before, since
these are top components of the short multiplet $(\bf 2 , \bf 2)_S$
they can be added to the free SCFT as perturbations without violating
the ${\cal N} = (4,4)$ supersymmetry of the SCFT. 
The various quantum numbers of
these operators are listed below.
\bea
\begin{array}{lccc}
\label{twist_operator}
\mbox{Operator} & (j^3, \widetilde{j}^3)_I & 
SU(2)_R\times \widetilde{SU(2)}_R &  (h, \bar{h}) \\
{\cal T}^1_{(1)}=T^{(\frac{1}{2}, \,  \frac{1}{2})} & 
( 0, 1) & (\bf 1, \bf 1) & (1,1) \\
{\cal T}^1_{(0)}= T^{(\frac{1}{2}, 
\,  -\frac{1}{2})}+ T^{(-\frac{1}{2}, \,  \frac{1}{2})} &  
(0,0) & (\bf 1, \bf 1) & (1,1) \\
{\cal T}^1_{(-1)}= T^{(-\frac{1}{2}, \,  -\frac{1}{2})} 
& (0,-1) & (\bf 1, \bf 1) & (1,1) \\
{\cal T}^0= T^{(-\frac{1}{2}, \, -\frac{1}{2})} - T^{(-\frac{1}{2}, 
\,  -\frac{1}{2})}
& (0, 0) & (\bf 1, \bf 1) & (1,1) 
\end{array}  
\eea
The first three operators of the above table can be organized as a
$(\bf 1, \bf 3)$ under $SU(2)_I\times\widetilde{SU(2)}_I$. We will
denote these $3$ operators as ${\cal T}^1$. The last
operator transforms as a scalar $(\bf 1, \bf 1)$ under 
$SU(2)_I\times\widetilde{SU(2)}_I$ and is denoted by ${\cal T}^0$. 
The simplest way of figuring out the $(j^3, \widetilde{j}^3)_I$
quantum numbers in the above
table is to note that
(a) the $\Sigma$-operators
of \eq{defSigma} are singlets under $SU(2)_I\times\widetilde{SU(2)}_I$,
as can be verified by computing the action on them of
the operators $I_1, I_2$ and $\widetilde{I}_1,
\widetilde{I}_2$, (b) the ${\cal T}$-operators are
obtained from $\Sigma$'s by the action of the supersymmetry
currents as in \eq{raising} and (c) the
quantum numbers of the supersymmetry
currents under $I_1, I_2$ and $\widetilde{I}_1,
\widetilde{I}_2$ are given by \eq{so4_on_g}. 

\subsection{Higher twists}
\label{chp2:high_twist}

We now show that the twist operators corresponding to any other
conjugacy class of $S(Q_1 Q_5)$ are irrelevant. Consider the class
with $N_1= Q_1Q_5-3, N_3 =1$ and the rest of $N_n=0$. A representative
element of this class is 
\be
\label{z3_element}
(X_1\rightarrow X_2, X_2 \rightarrow X_3, X_3 \rightarrow X_1), 
\; X_4\rightarrow X _4, \ldots ,
\; X_{Q_1Q_5}\rightarrow X_{Q_1Q_5}. 
\ee
To make the action of this group element transparent we diagonalize
the group action as follows. 
\bea
\left(
\begin{array}{c}
\phi_1 \\ \phi_2 \\ \phi_3
\end{array} \right)
=
\left(
\begin{array}{ccc}
1 & 1 & 1 \\
1 & \omega & \omega^2 \\
1 & \omega^2 & \omega^4 
\end{array}
\right)
\left(
\begin{array}{c}
X_1 \\ X_2 \\  X_3
\end{array}
\right)
\eea
where $\omega = \exp(2\pi i/3)$.
These new coordinates are identified under the group action 
\eq{z3_element} $\phi_1
\rightarrow \phi_1$, $\phi_2 \rightarrow \omega^2 \phi_2$  and
$\phi_3 \rightarrow \omega \phi_3$. 
These identifications are locally characteristic of the
orbifold
\be
R^4\times R^4/\omega \times R^4/\omega^2
\ee
The dimension of the supersymmetric twist operator which twists the
coordinates by a phase $e^{2\pi i k /N}$ in $2$ complex dimensions is
$h(k,N)= k/N$\cit{chp2:DixFriMarShe}. The twist operator which implements the
action of the group element \eq{z3_element} combines the
supersymmetric twist operators acting on $\phi_2$ and $\phi_3$
and therefore has total dimension  
\be
h =h(1,3) + h(2,3) = 
1/3+ 2/3 =1 
\ee 
It is
the superpartners of these which could be candidates for the blow up
modes. However, these have weight $3/2$, These operators are therefore
irrelevant.

For the class $N_1= Q_1Q_5 -k$ , $N_k =k$ and the rest of $N_n =0$,
the total dimension of the twist operator is 
\be
h = \sum_{i=1}^{k-1} h(i,k) = (k-1)/2
\ee 
Its superpartner has dimension $k/2$. Now it is easy to see that all
conjugacy classes other than the exchange of $2$ elements give
rise to irrelevant twist operators. Thus the orbifold ${\cal M}$ is
resolved by the $4$ blow up modes corresponding to the conjugacy class
represented by \eq{group_element}. We have thus
identified the $20$ marginal operators of the ${\cal N}=(4,4)$ SCFT on
$\widetilde{T}^4$. They are all top components of the $5(\bf 2 ,\bf
2)_S$ short multiplets. The $5(\bf 2, \bf 2)_S$ have $20$ 
operators of conformal dimensions $(h, \bar{h})=(1/2, 1/2)$. These are
relevant operators for the SCFT. It would be interesting to
investigate the role of these relevant operators.

\section{The chiral primaries of the ${\cal N}=(4,4)$ SCFT on ${\cal
M}$ }
\label{chp2:chprimary}

In this section we will explictly construct all the chiral primaries 
corresponding to single particle states 
of the SCFT on the orbifold ${\cal M}$. 
For this purpose we will have to 
construct the
twist operator corresponding to the conjugacy class $N_1 = Q_1Q_5 -k,
N_k =k$ and the rest of $N_n=0$. 

\subsection{The k-cycle twist operator}
\label{chp2:k_cycle}
We will extend the method of construction of 
the 2-cycle twist operator of
Section \ref{chp2:z_2twists} to the construction of the k-cyle twist
operator. Consider the conjugacy class given by $N_1 =Q_1 Q_5 -k, N_k
=k$ and the rest of $N_n=0$. A representative element of this class is
the following group action 
\be
\label{chp2:g_action}
(X_1 \rightarrow X_2 , \ldots , X_k\rightarrow X_1 ), X_{k+1}
\rightarrow
X_{k+1}, \ldots ,
X_{Q_1 Q_5} \rightarrow X_{Q_1 Q_5} .
\ee
We can diagonalize the group action as follows
\be
\left(
\begin{array}{c}
\phi_k \\
\phi_{k-1} \\
\phi_{k-2} \\
\vdots \\
\phi_1
\end{array}
\right)
=
\left(
\begin{array}{ccccc}
1 & 1 & 1 & \ldots & 1 \\ 
1 & \omega & \omega^2 & \ldots & \omega^{k-1} \\ 
1 & \omega^2 & \omega^4 & \ldots & \omega^{2(k-1)} \\ 
\vdots &\vdots & \vdots & \ldots & \vdots \\
1 & \omega^{k-1} & \omega^{(k-1)2} & \ldots & \omega^{(k-1)(k-1)}
\end{array}
\right)
\left(
\begin{array}{c}
X_1 \\
X_2 \\
X_3 \\
\vdots \\
X_k
\end{array}
\right)
\ee
where $\omega = e^{2\pi i /k}$. These new coordinates are identified
under the group action \eq{chp2:g_action} as
\be
\phi_1 \rightarrow \omega \phi_1 ,\;\; \phi_2 \rightarrow \omega^{2} \phi_2 ,
\;\; \phi_3 \rightarrow \omega^3 \phi_3 , \;\; \ldots ,\;\; \phi_{k-1}
\rightarrow \omega^{k-1} \phi_{k-1},\;\;
\phi_k \rightarrow \omega^k \phi_k
\ee
These identifications are locally characteristic of the orbifold
\be
R^4\times R^4/\omega \times R^4/\omega^2 \times \ldots \times 
R^4/\omega^{k-1}
\ee
The coordinate $\phi_m$ is twisted by the phase $\omega^{m}$ ( $m$ runs
from $1\ldots k $).
The bosonic twist operators corresponding to this twist are defined by
the following OPE's
\bea
\label{chp2:OPEs}
\del \phi^1_m (z) \sigma^1_m(w, \bar{w} ) = 
\frac{ \tau^1_m(w, \bar w ) }{ (z-w)^{1-m/k} }  \; &\;& \;
\del {\phi}^{1\dagger}_m (z) \sigma^1_m(w, \bar{w} ) = 
\frac{ \tau'^1_m(w, \bar w ) }{ (z-w)^{m/k} }  \\   \nonumber
\del \phi^2_m (z) \sigma^2_m(w, \bar{w} ) =
\frac{ \tau^2_m(w, \bar w ) }{ (z-w)^{1-m/k} }  \; &\;& \;
\del {\phi}^{2\dagger}_m (z) \sigma^2_m(w, \bar{w} ) =
\frac{ \tau'^2_m(w, \bar w ) }{ (z-w)^{m/k} }  \\  \nonumber
\bar{\del} \widetilde{\phi}^1_m (\bar{z}) \sigma^1_m(w, \bar{w} ) = 
\frac{ \widetilde\tau'^1_m(w, \bar w ) }{ (\bar z-\bar w)^{m/k} } 
\; &\;&
\;
\bar{\del} \widetilde{\phi}^{1\dagger}_m 
(\bar z) \sigma^1_m(w, \bar{w} ) = 
\frac{ \widetilde\tau^1_m(w ,\bar w ) }{ (\bar z-\bar w)^{1-m/k} } 
\\   \nonumber
\bar{\del} \widetilde{\phi}^2_m (\bar z) \sigma^2_m(w, \bar{w} ) =
\frac{ \widetilde\tau'^2_m(w, \bar w ) }{ (\bar z-\bar w)^{m/k} } 
\; &\;& \;
\bar{\del} \widetilde{\phi}^{2\dagger}_m (\bar z ) 
\sigma^2_m(w, \bar{w} ) =
\frac{ \widetilde\tau^2_m(w, \bar w ) }{ (\bar z-\bar w)^{1-m/k} }  
\eea
As in Section \ref{chp2:z_2twists} $\tau$'s are excited twist
operators. The fermionic twists are constructed from bosonized currents
defined by
\bea
\chi^1_m(z) = e^{iH^1_m(z)} \; &\;& \; 
\chi^{1\dagger}_m(z) = e^{-iH^1_m(z)} \\
\nonumber
\chi^2_m(z) = e^{iH^2_m(z)} \; &\;& \; 
\chi^{2\dagger}_m(z) = e^{-iH^2_m(z)} \\
\nonumber
\eea
Where the $\chi_m$'s are 
the superpartners of the bosons $\phi_m$'s. The twist operators
corresponding to the fermions $\chi_m$'s are given by 
$e^{\pm i mH_m/k}$.

We now assemble all these operators to construct the k-cycle twist
operator which is a chiral primary. 
The k-cycle twist operator is given by
\be
\Sigma_{(12\ldots k)}^{(k-1)/2} =
\prod_{m=1}^{k-1} \left[
\sigma^1_m (z, \bar{z})
\sigma^2_m (z, \bar{z})
e^{i m H_m^1(z)/k}
e^{-i m H_m^2(z)/k}
e^{i m \tilde{H}_m^1(\bar{z})/k}
e^{-i m \tilde{H}_m^2(\bar{z})/k}
\right]
\ee
The subscript $(12\ldots k)$ refers to the fact that these twist
operators were constructed for the representative group element 
\eq{chp2:g_action} which cyclically permutes the $1,\ldots , k$ labels
of the coordinates of $\tilde{T}^4$. The superscript $(k-1)/2$ stands
for the conformal dimension of this operator. As we saw in Section
\ref{chp2:high_twist} 
the conformal dimension of the twist operator for the conjugacy
class $N_1 = Q_1Q_5 -k, N_k =k$ and the rest of $N_n=0$ is
$(h, \bar{h}) = ((k-1)/2, (k-1)/2)$. 
The twist operator for the conjugacy
class under consideration is obtained by summing over the k-cycle
twist operators for all representative element of these class.
\be
\Sigma ^{(k-1)/2} (z, \bar{z} ) = \sum_{ \{ i_i, \ldots , i_k \} }
\Sigma_{i_1i_2\ldots i_k} (z, \bar{z} )
\ee
where the sum runs over all  $k$-tuples $\{ i_i \ldots , i_k \}$ such
that $i_i\neq i_2 \neq \ldots \neq i_k$. $i_m$ take values from $1$ to
$Q_1Q_5$. The operator $\Sigma^{(k-1)/2}$ is a chiral primary with
conformal dimension 
$(h, \bar{h}) = ((k-1)/2, (k-1)/2)$ and $(j_R^3, \tilde{j}_R^3)
=((k-1)/2, (k-1)/2)$. As the largest cycle is of length $Q_1Q_5$, the
maximal dimension of the k-cycle twist operator is 
$((Q_1Q_5-1)/2, (Q_1Q_5-1)/2)$. It belongs to the bottom component of the 
short multiplet $(\bf{k}, \bf{k})_S$.
The other components of the shortmultiplet $(\bf{k}, \bf{k})_S$ 
corresponding to the k-cycle twists can be generated by the
action of supersymmetry currents and the R-symmetry currents of the
${\cal N}=(4,4)$ theory on ${\cal M}$.

\subsection{The complete set of chiral primaries}

We have seen is Section \ref{chp2:res} there are five chiral
primaries corresponding to the shortmultiplet $5(\bf{2},
\bf{2})_S$. In this section we will construct the
complet set of chiral primaries from single particle states of the
SCFT on ${\cal M}$. It is known that the chiral primaries 
with weight $(h, \bar{h})$ of a
${\cal N}=(4,4)$ superconformal field theory on a manifold $K$
correspond to the elements of the cohomology ${\cal H}_{2h\,
2\bar{h}}(K)$ \cit{chp2:Wit_susy}. 
The chiral primaries are formed by the product of
the chiral primaries corresponding to the cohomology of the diagonal
$\tilde{T}^4$ denoted by $B^4$ 
(the sum of all copies of $\tilde{T}^4$)
and the various k-cycle chiral primaries constructed in
Section \ref{chp2:k_cycle}. We will list the chiral primaries below
\\
{\bf Chiral primaries with  $h-\bar{h} =0$}

All the k-cyle chiral primaries have $h-\bar{h}=0$. To construct
chiral primaries with $h-\bar{h} =0$ we need the four chiral
primaries which 
correspond to the cohomology ${\cal H}_{11}(B^4)$  
with weight $(1/2,1/2)$. They are given in \eq{chiral}.
Using this we can construct the following chiral primaries
\bea 
\Sigma^{(k-1)/2}(z, \bar{z})
\Psi^1_A(z) \widetilde{\Psi}^1_A(\bar{z})  \;&\;&\;  
\Sigma^{(k-1)/2}(z, \bar{z})
\Psi^1_A(z)\widetilde{\Psi}^{2\dagger}_A(\bar z) \\   \nonumber
\Sigma^{(k-1)/2}(z, \bar{z})
\Psi^{2\dagger}_A(z)\widetilde{\Psi}^1_A(\bar z )  \;&\;&\;
\Sigma^{(k-1)/2}(z, \bar{z})
\Psi^{2\dagger}_A(z)\widetilde{\Psi}^{2\dagger}_A(\bar z) 
\eea
where summation over $A$ is implied. These four operators have 
conformal dimension $(k/2, k/2)$. There is one more chiral primary
corresponding to the cohomology ${\cal H}_{22}(B^4)$ for which $h-\bar{h}=0$. 
It is given by 
\be
\Psi^1_A(z) \Psi^{2\dagger}_A(z)
\widetilde{\Psi}^1_A(\bar{z})  \widetilde{\Psi}^{2\dagger}_A(\bar z) 
\ee
where summation over all indices of $A$ is implied. 
This chiral primary corresponds to the top form of 
$B^4$. The cohomology ${\cal H}_{00}(B^4)$ gives rise to a 
chiral primaries of conformal dimension $(k/2,k/2)$. 
It is given by
\be
\Sigma^{(k-2)/2}(z, \bar{z})
\Psi^1_A(z) \Psi^{2\dagger}_A(z)
\widetilde{\Psi}^1_A(\bar{z})  \widetilde{\Psi}^{2\dagger}_A(\bar z) 
\ee
From the equation above we see that these chiral primaries exist only
of $k\geq 2$. Finally we have the chiral primary
$
\Sigma^{(k)/2}(z, \bar{z})
$
of conformal dimension $(k/2,k/2)$. Thus for $k\geq 2$ 
and $k\leq Q_1Q_5 -1$ there are $6$
chiral primaries of dimension $(k/2,k/2)$ 

The complete list of chiral primaries with $(h, \bar{h})$ 
with $h -\bar{h} =0$ 
corresponding
to single particle states are given by
\be
\begin{array}{lc}
(h, \bar{h}) & \mbox{Degeneracy} \\
   &  \\ 
(1/2, 1/2) &  5 \\
(1, 1)    &   6 \\
(3/2,3/2)     &   6 \\
\vdots    & \vdots \\
((Q_1Q_5 -1)/2 , (Q_1Q_5 -1)/2 ) & 6 \\
( (Q_1Q_5)2, (Q_1Q_1)/2 )  & 5 \\
( (Q_1Q_5 + 1)/2, (Q_1Q_5 + 1)/2 ) & 1 
\end{array}
\ee
In the above table we have ignored the vacuum with weight $(h,
\bar{h})=(0,0)$.

\noindent
{\bf Chiral primaries with  $h-\bar{h} =1/2$ }

The ciral primaries of $B^4$ which correspond to
the elements of the cohomology ${\cal H}_{10}(B^4)$ are given by
\be
\sum_{A=1}^{Q_1Q_5}
\Psi^1_A(z) \;\;\; \mbox{and} \;\;\;
\sum_{A=1}^{Q_1Q_5}
\Psi^{2\dagger}_A(z)
\ee
We can construct chiral primaries with weight $( (k+1)/2 , k/2) )$ by
taking the product of the above chiral primaries with the twist
operator $\Sigma^{k/2}(z, \bar{z})$. These give the following
chiral primaries 
\be
\Sigma^{k/2}(z, \bar{z}) 
\sum_{A=1}^{Q_1Q_5}
\Psi^1_A(z) \;\;\; \mbox{and} \;\;\;
\Sigma^{k/2}(z, \bar{z}) 
\sum_{A=1}^{Q_1Q_5}
\Psi^{2\dagger}_A(z)
\ee
The chiral primary of the diagonal $B^4$ which 
correspond to the
elements of the cohomology ${\cal H}_{21}(B^4)$  are 
\be
\Psi^1_A(z) \Psi^{2\dagger}_A(z)
\widetilde{\Psi}^1_A(\bar{z})  \;\; \mbox{and} \;\;
\Psi^1_A(z) \Psi^{2\dagger}_A(z)
\widetilde{\Psi}^{2\dagger}_A(\bar z) 
\ee
Here summation over all the three indices of $A$ is implied. 
From these the  one can construct chiral 
primaries with weight $((k+1)/2, k/2)$ are follows
\be
\Sigma^{(k-1)/2}(z, \bar{z}) 
\Psi^1_A(z) \Psi^{2\dagger}_A(z)
\widetilde{\Psi}^1_A(\bar{z})  \;\; \mbox{and} \;\;
\Sigma^{(k-1)/2}(z, \bar{z}) 
\Psi^1_A(z) \Psi^{2\dagger}_A(z)
\widetilde{\Psi}^{2\dagger}_A(\bar z) 
\ee
Therefore there are 4 chiral primaries with weight $((k+1)/2, k/2)$
for $1\leq k\leq (Q_1Q_5 -1)$ and 2 chiral primaries with weight
$((Q_1Q_5+1)/2, Q_1Q_5/2)$. There are also 2 chiral primaries with
weight $(1/2, 0)$.

\noindent
{\bf Chiral primaries with  $\bar{h}-h =1/2$ }

The procedure for constructing these chiral primaries are identical to
the  case $h -\bar{h} =1/2$. 
The four chiral primaries with weight $(k/2, (k+1)/2)$ are given by

\bea
 \Sigma^{k/2}(z, \bar{z}) 
\sum_{A=1}^{Q_1Q_5}\widetilde{\Psi}^1_A(\bar{z}) \;&\;&\;
\Sigma^{k/2}(z, \bar{z}) 
\sum_{A=1}^{Q_1Q_5}\widetilde{\Psi}^{2\dagger}_A(\bar z) \\ \nonumber
\Sigma^{(k-1)/2}(z, \bar{z}) 
\Psi^1_A(z) 
\widetilde{\Psi}^1_A(\bar{z})  \widetilde{\Psi}^{2\dagger}_A(\bar z) 
\;&\;&\;
\Sigma^{(k-1)/2}(z, \bar{z}) 
 \Psi^{2\dagger}_A(z)
\widetilde{\Psi}^1_A(\bar{z})  \widetilde{\Psi}^{2\dagger}_A(\bar z) 
\eea
 There are 4 chiral primaries with weight $(k/2, (k+1)/2)$
for $1\leq k\leq (Q_1Q_5 -1)$  2 chiral primaries with weight
$(Q_1Q_5/2, (Q_1Q_5+1)/2)$ and 2 chiral primaries with weight $(0,
1/2)$.

\noindent
{\bf Chiral primaries with  $h-\bar{h} =1$}

As in the previous cases let us first look at the
chiral primaries corresponding to the cohomology element 
${\cal H}_{20}(B^4)$. There is only one element which is given by
\be
\Psi^1_A(z) \Psi^{2\dagger}_A(z)
\ee
where summation over $A$ is implied.  
There is a single  chiral 
primary with weight $((k+2)/2, k/2)$  constructed out of the above
chiral primary is  
\be
\Sigma^{k/2}(z, \bar{z}) 
\Psi^1_A(z) \Psi^{2\dagger}_A(z)
\ee
Thus there is a one chiral primary with weight $((k+2)/2, k/2)$  
for $0\leq k\leq (Q_1Q_5-1)$

The operator product expansion of two chiral primaries will give rise
to other chiral primaries consistent with conservation laws.
There are known to form a ring. It will be interesting to understand
the structrure of this ring.

\noindent
{\bf Chiral primaries with  $\bar{h}-h =1$}

The construction of these is parallel to the case 
for $h-\bar{h}=1$. The single
chiral primary with weight $(k/2, (k+2)/2)$ for 
$0\leq k\leq (Q_1Q_5-1)$ is given by 
\be
\Sigma^{k/2}(z, \bar{z}) 
\widetilde{\Psi}^1_A(\bar{z})  \widetilde{\Psi}^{2\dagger}_A(\bar z) 
\ee

\section{Shortmultiplets of ${\cal N}= (4,4)$ SCFT on ${\cal M}$}
\label{chp2:sec_shortmultiplets}

Using the results of Section \ref{chp2:chprimary}
we will write the complete
set of shortmultiplets of single particle states of
the ${\cal N}=(4,4)$ SCFT onf ${\cal M}$. 
In Chapter 3 we will compare this set of short multiplets with that
obtained from supergravity. We will see in Chapter 3 
that supergravity is a good approximation in string theory only when
$Q_1\rightarrow \infty, Q_5\rightarrow \infty$. Therefore we write
down the list of shortmultiplets for 
$(\tilde{T}^4)^{(\infty)}/S(\infty)$. 
Basically this means that the list
of chiral primaries of the previous Section \ref{chp2:chprimary} 
does not terminate. 

We have seen that each chiral primary  of weight $(h, h')$ gives rise
gives rise to the short multiplet 
$({\bf 2h} +{\bf 1} , {\bf 2h^{'}} +{\bf 1}))_S$. Therefore the results of
Section \ref{chp2:chprimary} indicate that the list of
shormultiplets corresponding to the single particle states of
${\cal N}=(4,4)$ SCFT on 
$(\tilde{T}^4)^{(\infty)}/S(\infty)$. 
is given by
\bea
\label{chp2:eq.short}
&5 (\re 2 , \re 2 )_S + 6 \oplus_{\re m \geq \re 3 } (\re m , \re m
)_S \\ \nonumber 
&2(\re 1, \re 2)_S + 2(\re 2, \re 1)_S + (\re 1, \re 3)_S + 
(\re 3, \re 1)_S \\ \nonumber
&\oplus_{\re m \geq \re 2 } [\, (\re m , \re m + \re2
)_S + ( \re m + \re2 , \re m )_S + 4 ( \re m , \re m + \re 1 )_S + 4 (
\re m + \re 1 , \re m )_S \, ] 
\eea

\section{The location of the symmetric product}

The complete specification of the D1/D5 system includes
various moduli. Most of the studies of the D1/D5 system so far have been
focused on the situation with no moduli. It is known from supergravity
that the D1/D5 system with no moduli turned on is marginally stable
with respect to the decay of a subsystem consisting of $Q'_5$ D5 and
$Q'_1$ D1-branes. It has been observed recently \cit{chp2:SeiWit} that
such a decay in fact signals a singularity in the world volume gauge
theory associated with the origin of the Higgs branch.  The issue of
stability in supergravity in the context of the D1/D5 system has also
been discussed in [\ref{chp2:Dijkgraaf},\,\ref{chp2:LarMar}].

The singularity mentioned above leads to a singular conformal field
theory and hence to a breakdown of string perturbation
theory. However, generic values of the supergravity moduli which do
not involve fragmentation into constituents are described by
well-defined conformal field theories and therefore string
perturbation theory makes sense.  
We have seen in section \ref{chp2:res} that important 
singularity structure of
the ${\cal M}= (4,4)$ SCFT on the orbifold ${\cal M}$ is locally of the 
type $R^4/Z_2$. The resolution of this singularity gives rise to
marginal operators. An orbifold theory realized as a free 
field SCFT on $R^4/Z_2$
is nonsingular as all correlations fuctions are finite. 
The reason for this can
be understood from the linear sigma model description of the $R^4/Z_2$
singluarity discussed in section \ref{chp2:z_2sing}. 
We will see that the though
the $R^4/Z_2$ singularity is geometrically singular the 
SCFT is finite because it
corresponds to a non-zero theta term in the linear sigma model. The 
geometric resolution of
this singularity corresponds to adding Fayet-Iliopoulos 
terms to the D-term
equations of the linear sigma model. This deforms 
the $R^4/Z_2$ singularity to an
Eguchi-Hansen space. In the orbifold theory this deformation is caused
by the twist operator ${\cal T}^1$.
The Eguchi-Hansen space is 
asymptotically $R^4/Z_2$ but the
singularity at the origin is blown up to a 2-sphere. 
One can use the $SU(2)_R$ symmetry of the linear sigma model to rotate
the three 
Fayet-Iliopoulos terms to one term. This term 
corresponds to the radius of the blown up 2-sphere. 
The theta term of the linear
sigma model corresponds to $B$-flux through the 2-sphere. The change
of this $B$-flux is caused by deforming the orbifold SCFT by the twist
operator ${\cal T}^0$.
Thus SCFT realized as a free field theory on the orbifold $R^4/Z_2$ is 
regular even though the 2-sphere is squashed to zero size because of
the non-zero value of $B$-flux trapped in the squashed 2-sphere
\cit{chp2:Aspinwall}. 

The realization
of the SCFT of the D1/D5 system as ${\cal N}=(4,4)$ theory on ${\cal
M}$ implies that we are at a  point in the moduli space of the
D1/D5 system at which the orbifold is geometrically singular but
because of the non-zero value of the theta term the SCFT is regular
and {\em not at the singularity corresponding to
fragmentation}. In other words, the orbifold SCFT corresponds to a
bound state of $Q_1$ D1-branes and $Q_5$ D5-branes (henceforth denoted
as the $(Q_1,Q_5)$ bound state). Thus we use the free field
realization of ${\cal N}=(4,4)$ SCFT on the orbifold ${\cal M}$ and
its resolutions using the marginal operators of this theory to
describe the boundary SCFT corresponding to the D1/D5 system at
generic values of the moduli. 

The symmetric product orbifold ${\cal
M}$ as we have seen before in section \ref{chp4:instanton} is the
moduli space of $Q_1$ instantons of a $U(Q_5)$ gauge theory on $T^4$.
We argue that the orbifold ${\cal M}$ resolved by the twist
operators ${\cal T}^1$ is the moduli space of $Q_1$ instantons of a
$U(Q_5)$ gauge theory on a non-commutative torus $T^4$. We first look
at the case of instantons on $R^4$. The moduli space of $Q_1$
instantons of a $U(Q_5)$ gauge theory is given by the D-terms of a
 $1+1$ dimensional $U(Q_1)$ gauge theory with $(4,4)$ supersymmetry
 and $Q_5$ hypemultiplets. On deforming these D-term equations by
 Fayet-Iliopolous terms we obtain the moduli space of $Q_1$ instantons
 of a $U(Q_5)$ gauge theory on a non-commutative $R^4$
 \cit{chp2:ShwarzNekrasov}. 
 Now let us turn to instantons on $T^4$
 The $R^4/Z_2$ singularity of the orbifold ${\cal M}$ is resolved by
 adding Fayet-Iliopoulos terms to the D-term equations in the linear
 sigma model description of the $R^4/Z_2$ singularity. 
As we have seen in the orbifold theory this is caused by adding the
twist operators ${\cal T}^1$ to the free theory.
 The analogy with the case of instantons on $R^4$  gives us
 the hint that the resolved orbifold ${\cal M}$ 
 is the moduli space of $Q_1$ instantons
 of a $U(Q_5)$ gauge theory on a non-commutative torus $T^4$. It will
 be interesting to make this argument more precise.


\section{The linear sigma model}
\label{chp2:linear_sigma}
In this section we will analyze the gauge theory description of the
D1/D5 system.  We show that that the gauge theory  has four 
parameters which control the break up of the $(Q_1, Q_5)$ system to 
subsystems $(Q'_1, Q'_5)$ and $(Q''_1, Q''_5)$. These are the
Fayet-Iliopoulos D-terms and the theta term in the effective $U(1)$
$(4,4)$ linear sigma model of the relative coordinate between the
subsystems $(Q'_1, Q'_5)$ and $(Q''_1, Q''_5)$.
To motivate this 
we will review the linear  sigma model corresponding to the
of the $R^4/Z_2$ singularity. 

\subsection{The linear sigma model description of $R^4/Z_2$}
\label{chp2:z_2sing}

The linear sigma model is a $1+1$ 
dimensional $U(1)$ gauge theory with $(4,4)$ supersymmetry \cit{chp2:Wit}. 
It has $2$
hypermultiplets charged under the $U(1)$. The scalar fields of the 
hypermultiplets can be
organized as doublets under the $SU(2)_R$ symmetry of the $(4,4)$
theory as
\bea
\chi_1=
\left(
\begin{array}{c}
A_1 \\ B_1^\dagger
\end{array}
\right)  \;\mbox{and} \;
\chi_2=
\left(
\begin{array}{c}
A_2 \\ B_2^\dagger
\end{array}
\right) 
\eea
The $A$'s have charge $+1$ and the $B$'s have charge $-1$ under the
$U(1)$. The vector multiplet has $4$ real 
scalars $\varphi_i$, $i=1,\ldots ,4$.
They do not transform under the $SU(2)_R$. One can 
include $4$ parameters
in this theory consistent with (4,4) supersymmetry. They are the $3$
Fayet-Iliopoulos terms and the theta term. 

Let us first investigate the hypermultiplet moduli space of this
theory with the $3$ Fayet-Iliopoulos terms and the theta term set to
zero. The Higgs phase of this theory is obtained by setting $\phi_i$
and the D-terms to zero. The D-term equations are
\bea
\label{dtermr4}
|A_1|^2 + |A_2|^2 -|B_1|^2 -|B_2|^2 &=&0 \\  \nonumber
A_1B_1 + A_2B_2 &=&0
\eea
The hypermultiplet moduli space is the space of solutions of the above
equations modded out by the $U(1)$ gauge symmetry. Counting the number
of degrees of freedom indicate that this space is $4$ dimensional. To
obtain the explicit form of this space it is convenient to introduce
the following gauge invariant variables
\bea
\label{dtermdef1}
M=A_1B_2 \; &\;& \; N= A_2B_1 \\ \nonumber
P= A_1B_1 &=& - A_2B_2 \\
\eea
These variables are not independent. Setting the D-terms equal to zero
and modding out the resulting space by $U(1)$ is equivalent to the
equation
\be
\label{surface}
P^2 +MN=0.
\ee
This homogeneous equation is an equation of the space $R^4/Z_2$. To see
this the solution of the above equation can be parameterized by $2$
complex numbers $(\zeta , \eta)$ such that
\be
\label{dtermdef2}
P= i\zeta \eta \;\; M=\zeta^2 \;\; N= \eta^2
\ee
Thus the point $(\zeta, \eta)$ and $(-\zeta, -\eta)$ are the same
point in the space of solutions of \eq{surface}. We have
shown that the hypermultiplet moduli space is $R^4/Z_2$. 

The above singularity at the origin of the moduli space 
is a geometric singularity in the hypermultiplet
moduli space. We now argue that this singularity is a genuine
singularity of the SCFT that the linear sigma model flows to in the
infrared. At the origin of the classical moduli space the Coulomb
branch meets the Higgs branch. In addition to the potential due to the
D-terms the linear sigma model contains the following term in the
superpotential\footnote{These terms can be understood from
the coupling $A_\mu A^\mu \chi^* \chi$ in six dimensions,
and recognizing that under dimensional reduction to two
dimensions $\varphi_i$'s appear from the components of $A_\mu$
in the compact directions.}
\be
V = 
(|A_1|^2 + |A_2|^2 + |B_1|^2 + |B_2|^2)(\varphi_1^2 + 
\varphi_2^2 +\varphi_3^2 +\varphi_4^2)
\ee
Thus at the origin of the hypermultiplet moduli space a flat direction
for the Coulomb branch opens up. The ground state at this point is not
normalizable due to the non-compactness of the Coulomb branch. This
renders the infrared SCFT singular.

This singularity can be avoided in two distinct ways. If one turns on
the Fayet-Iliopoulos D-terms, the D-term equations are modified to
\cit{chp2:Wit} 
\bea |A_1|^2 + |A_2|^2 -|B_1|^2 -|B_2|^2 &=&r_3 \\
\nonumber A_1B_1 + A_2B_2 &=&r_1+ir_2 
\eea 
Where $r_1,r_2,r_3$ are the
$3$ Fayet-Iliopoulos D-terms transforming as the adjoint of the
$SU(2)_R$. Now the origin is no more a solution of these equations and
the non-compactness of the Coulomb branch is avoided. In this case
wave-functions will have compact support all over the hypermultiplet
moduli space. This ensures that the infrared SCFT is
non-singular. Turning on the Fayet-Iliopoulos D-terms thus correspond
to the geometric resolution of the singularity. The resolved
space is known  to be [\ref{chp2:Wit},\,\ref{chp2:Aspinwall}] 
described by 
an Eguchi-Hanson metric in which $r_{1,2,3}$ parameterize
a shrinking two-cycle.

The second way to avoid the singularity in the SCFT is to turn on the
theta angle $\theta$. This induces a constant electric field in the
vacuum. This electric field is screened at any other point than the
origin in the hypermultiplet moduli space as the $U(1)$ gauge field is
massive with a mass proportional to the vacuum expectation value of the
hypers. At the origin the $U(1)$ field is not screened and thus it
contributes to the energy density of the vacuum. This energy is
proportional to $\theta^2$. Thus turning on the theta term lifts the
flat directions of the Coulomb branch. This ensures that the
corresponding infrared SCFT is well defined though the hypermultiplet
moduli space remains geometrically singular. In terms of the
Eguchi-Hanson space, the $\theta$-term corresponds to a flux of the
antisymmetric tensor through the two-cycle mentioned above.

The $(4,4)$ SCFT on $R^4/Z_2$ at the orbifold point is well defined.
Since the orbifold has a geometric singularity but the SCFT is
non-singular it must correspond to the linear sigma model with a
finite value of $\theta$ and the Fayet-Iliopoulos D-terms set to
zero.  Deformations of the $R^4/Z_2$ orbifold by its $4$ blow up modes
correspond to changes in the Fayet-Iliopoulos D-terms and theta term
of the linear sigma model%
\footnote{If we identify the $SU(2)_R$ of the linear sigma-model
with $\widetilde{SU(2)}_I$ of the orbifold SCFT, then the
Fayet-Iliopoulos parameters will correspond to ${\cal T}^1$
and the $\theta$-term to ${\cal T}^0$. 
This is consistent with Witten's observation \cit{chp2:Wit_cft}
that $SO(4)_E$ symmetry  of the linear sigma-model (one
that rotates the $\phi_i$'s) corresponds to the $SU(2)_R$
of the orbifold SCFT.} 
The global description of the moduli of a
${\cal N} =(4,4)$ SCFT on a resolved $R^4/Z_2$ orbifold is provided by
the linear sigma model.  In conclusion let us describe this linear
sigma model in terms of the gauge theory of D-branes. The theory
described above arises on a single D1-brane in presence of $2$
D5-branes. The singularity at the point $r_1, r_2, r_3, \theta
=0$ is due to noncompactness of the flat direction of the Coulomb
branch.  Thus it corresponds to the physical situation of the D1-brane
leaving the D5-branes.

\subsection{The gauge theory description of the moduli of the D1/D5
system}
\label{chp2:gt}

As we have seen in Section 2 the resolutions of the ${\cal N}=(4,4)$
SCFT on ${\cal M}$ is described by 4 marginal operators which were
identified in the last subsection with the Fayet-Iliopoulos D-terms
and the theta term of the linear sigma model description of the
$R^4/Z_2$ singularity. We want to now indicate how these four
parameters would make their appearance in the gauge theory description
of the full D1/D5 system.

Motivated by the
D-brane description of the $R^4/Z_2$ singularity we look for the
degrees of freedom characterizing the break up of $(Q_1, Q_5)$ system
to $(Q'_1, Q'_5)$ and $(Q''_1, Q''_5)$. Physically the relevant degree
of freedom describing this process is the relative coordinate between
the centre of mass of the $(Q'_1, Q'_5)$ system and the $(Q''_1,
Q''_5)$. We will describe the effective theory of this degree of
freedom below.

For the bound state $(Q_1, Q_5)$ the hypermultiplets the $\chi_{a,
\bar{b}}$ are charged under the relative $U(1)$ of $U(Q_1)\times
U(Q_5)$, that is under the gauge field $A_\mu = Tr_{U(Q_1)}( A^{a \bar
a}_\mu) - Tr_{U(Q_5)}(A^{b \bar b})$.  The gauge multiplet
corresponding to the relative $U(1)$ corresponds to the degree of
freedom of the relative coordinate between the centre of mass of the
collection of $Q_1$ D1-branes and $Q_5$ D5-branes. At a generic point
of the Higgs phase, all the $\chi_{a\bar{b}}$'s have expectation
values, thus making this degree of freedom becomes massive. This is
consistent with the fact that we are looking at the bound state $(Q_1,
Q_5)$.

Consider the break up of the $(Q_1, Q_5)$ bound state to 
the bound states $(Q'_1, Q'_5)$ and $(Q''_1, Q''_5)$. To find out the
charges of the hypermultiplets under the various $U(1)$, we will
organize the hypers as
\bea
\left(
\begin{array}{cc}
\chi_{a'b'} &  \chi_{a'b''}    \\ 
\chi_{a''b'} & \chi_{a''b''}
\end{array}
\right), \quad
\left(
\begin{array}{cc}
Y_{i(a'\bar{a}')}^{(1)} & Y_{i(a'\bar{a}'')}^{(1)}  \\
Y_{i(a''\bar{a}')}^{(1)} & Y_{i(a''\bar{a}'')}^{(1)}
\end{array}
\right)
\quad \mbox{and} \quad
\left(
\begin{array}{cc}
Y_{i(b'\bar{b}')}^{(5)} & Y_{i(b'\bar{b}'')}^{(5)}  \\
Y_{i(b''\bar{b}')}^{(5)} & Y_{i(b''\bar{b}'')}^{(5)}
\end{array}
\right)
\eea
where $a',\bar{a}'$ runs from $1,\ldots , Q'_1$, $b',\bar{b}'$  
from $1,\dots ,Q'_5$,
$a''\bar{a}''$ from $1, \ldots Q''_1$ and 
$b'',\bar{b}''$ from $1\ldots ,Q''_5$. 
We organize the scalars of the vector multiplet corresponding to the
gauge group $U(Q_1)$ and $U(Q_5)$ as
\bea
\left(
\begin{array}{cc}
\phi_{i}^{^(1)a'\bar{a}'} & \phi_{i}^{(1)a'\bar{a}''} \\
\phi_{i}^{^(1)a''\bar{a}'}& \phi_{i}^{(2)a'\bar{a}''}
\end{array}
\right)
\quad
\mbox{and}
\quad
\left(
\begin{array}{cc}
\phi_{i}^{^(5)b'\bar{b}'} & \phi_{i}^{(5)b'\bar{b}''} \\
\phi_{i}^{^(5)b''\bar{b}'}& \phi_{i}^{(5)b'\bar{b}''}
\end{array}
\right)
\eea
where $i=1,2,3,4$. 

Let us call the the $U(1)$ gauge fields (traces) of 
$\, U(Q'_1),\, U(Q'_5),
U(Q''_1),\, U(Q''_5)\, $ as $\, A'_1,\,  A'_5, \, A''_1, \, A''_5\, $ 
respectively.  We
will also use the notation $A'_\pm \equiv A'_1 \pm A'_5$ and $A''_\pm
\equiv A''_1 \pm A''_5$.

As we are interested in the bound states $(Q'_1, Q'_5)$ and $(Q''_1,
Q''_5)$, in what follows we will work with a specific classical
background in which we give {\em vev}'s to the block-diagonal hypers
$\chi_{a'b'},\chi_{a''b''}, Y^{(1)}_{i( a' \bar a')},Y^{(5)}_{i(b'
\bar b')},Y^{(1)}_{i( a'' \bar a'')}$ and $Y^{(5)}_{i(b'' \bar b'')}$.
These {\em vev}'s are chosen so that the classical background
satisfies the D-term equations \cit{chp2:HasWad}.

The {\em vev}'s of the $\chi$'s  render the fields
$A'_-$ and $A''_-$ massive with a mass proportional to {\em vev}'s. In
the low energy effective Lagrangian these gauge fields can therefore
be neglected.  In the following we will focus on the $U(1)$ gauge
field $A_r=1/2(A'_+ - A''_+)$ which does not get mass from the above
{\em vev}'s. The gauge multiplet corresponding to $A_r$ contains four
real scalars denoted below by $\varphi_i$. These represent the
relative coordinate between the centre of mass of the $(Q'_1, Q'_5)$
and the $(Q''_1, Q''_5)$ bound states. We will be interested in the
question whether the $\varphi_i$'s remain massless or otherwise. The
massless case would correspond to a non-compact Coulomb branch and
eventual singularity of the SCFT.

In order to address the above question we need to find the low energy
degrees of freedom which couple to the gauge multiplet corresponding
to $A_r$.  

The fields charged under $A_r$ are the hypermultiplets 
$\, \chi_{a'\bar
b''},\;  \chi_{a''\bar b'}, \; Y^{(1)}_{i(a'\bar a'')}, \;
Y^{(1)}_{i(a''\bar a')},$ \\
$Y^{(5)}_{i(b'\bar b'')},\,  Y^{(5)}_{i(b''\bar b')}\, $ 
and the vector multiplets
$\phi^{(1)a'\bar a''}_{i}, \phi^{(1){a''\bar a'}}_i, 
\phi^{(5)b'\bar
b''}_{i}, \phi^{(5)b''\bar
b'}_i$. In order to 
find out which of these are massless, we
look at the following terms in the Lagrangian of
 $U(Q_1)\times U(Q_5)$ gauge theory:
\bea
\label{lagrangian}
L &=& L_1 + L_2 + L_3 + L_4\\  \nonumber
L_1&=& 
\chi_{a_1\bar{b}_1}^*\phi_i^{(1)a_2\bar{a}_1*}
\phi_i^{(1)a_2\bar{a}_3}\chi_{a_3\bar{b_1}} \\ \nonumber
L_2&=&
\chi_{a_1\bar{b_1}}^*\phi_i^{(5)b_1\bar{b}_2*}
\phi_i^{(5)b_3\bar{b}_2}\chi_{a_1\bar{b}_3} \\  \nonumber
L_3&=&Tr([Y^{(1)}_i,Y^{(1)}_j][Y^{(1)}_i, Y^{(1)}_j]) \\  \nonumber
L_4&=&Tr([Y^{(5)}_i,Y^{(5)}_j][Y^{(5)}_i, Y^{(5)}_j]) \\   \nonumber
\eea
where the  $a_i$'s run from $1,\ldots,Q_1$ 
and the $b_i$'s run form $1,\ldots, Q_5$. 
The terms $L_1$ and $L_2$ originate from terms of
the type $|A_M\chi|^2$ where $A_M \equiv (A_\mu, \phi_i)$
is the $(4,4)$ vector multiplet in two dimensions. The terms
$L_3$ and $L_4$ arise from commutators of gauge fields in
compactified directions.  

The fields $Y$ are in general
massive. The reason is that the traces $y^{\prime (1)}_{i}
\equiv Y_{i(a'\bar{a}')}^{(1)}$, representing
the centre-of-mass position in the $T^4$ of $Q'_1$
D1-branes, and  $y^{\prime\prime (1)}_{i}
\equiv Y_{i(a''\bar{a}'')}^{(1)}$, representing
the centre-of-mass position in the $T^4$ of $Q''_1$
D1-branes, are neutral and will have {\em vev}'s which
are generically separated (the centres of mass can be
separated in the torus even when they are on top
of each other in physical space). 
 The mass of $Y_{i(a'\bar{a}'')}^{(1)} , 
Y_{i(a''\bar{a}')}^{(1)}$ can be read off from the term $L_3$ in 
\eq{lagrangian}, to be  proportional to
 $ (y^{\prime (1)} - y^{\prime\prime (1)})^2$
Similarly the
mass of $Y_{i(b'\bar{b}'')}^{(5)} , Y_{i(b'\bar{b}'')}^{(5)} $ is
proportional to $(y^{\prime (5)} -y^{\prime\prime (5)})^2 $ 
(as can be read off from the term $L_4$ in \eq{lagrangian})
where $y^{\prime (5)}$ and
$y^{\prime\prime (5) }$ are the centers of mass of the 
$Q'_5$ D5-branes and $Q''_5$
D5-branes along the direction of the dual four torus $\hat{T}^4$. 
(At special points when their centres of mass coincide, these fields
become massless. The analysis for these cases can also be carried out
by incorporating these fields in \eq{dterm}-\eq{resolved}, with no
change in the conclusion) The fields $\phi_{i}^{(1)a'\bar{a}''},
\phi_{i}^{^(1)a''\bar{a}'}$ are also massive. Their masses can be read
off from the $L_1$ in \eq{lagrangian}.
Specifically they arise from the following terms
\be
\chi^*_{a''_1\bar{b}''}\phi_i^{(1)a'\bar{a}''_1*}
\phi_i^{(1)a'\bar{a}''_2}
\chi_{a''_2\bar{b}''}+
\chi^*_{a'_1\bar{b}'}\phi_i^{(1)a''\bar{a}'_1*}\phi_i^{(1)a''\bar{a}'_2}
\chi_{a'_2\bar{b}'} 
\ee ~
where $a'_i$ run from $1,\dots Q'_1$ and
$a''_i$ run form $1,\ldots Q''_1$.  These terms show that
their masses are proportional to the expectation values of the hypers
$\chi_{a'b'}$ and $\chi_{a''b''}$. Similarly the terms of $L_2$ in
\eq{lagrangian} 
\be
\chi^*_{a''\bar{b}''_1}\phi_i^{(5)b''_1{b}'*}\phi_i^{(5)b''_2\bar{b}'}
\chi_{a''\bar{b}''_2}+
\chi^*_{a'\bar{b}'_1}\phi_i^{(5)b'_1\bar{b}''*}\phi_i^{(5)b'_2\bar{b}''}
\chi_{a'\bar{b}'_2} 
\ee 
show that the fields $\phi_{i}^{(5)b'\bar{b}''}
\phi_{i}^{^(5)b''\bar{b}'}$ are massive with masses proportional to the
expectation values of the hypers $\chi_{a'\bar{a}''}$ and
$\chi_{a''\bar{b}''}$. In the above equation $b'_i$ take values
from $1,\ldots , Q'_5$ and $b''_i$ take values from $1,\ldots
, Q''_5$. Note that these masses remain non-zero even in the limit
when the $(Q'_1,Q'_5)$ and $(Q''_1,Q''_5)$ are on the verge of
separating.

Thus the relevant degrees of freedom describing the splitting process
is a $1+1$ dimensional $U(1)$ gauge theory of $A_r$ with $(4,4)$
supersymmetry.  The matter content of this theory consists of
hypermultiplets $\chi_{a'\bar{b}''}$ with charge $+1$ and
$\chi_{a''\bar{b}'}$ with charge $-1$.

Let us now describe the dynamics of the splitting process. This is
given by analyzing the hypermultiplet moduli space of the effective
theory described above with the help of the D-term equations:
\bea
\label{dterm}
A_{a'\bar{b}''}A^*_{a'\bar{b}''}-A_{a''\bar{b'}}A^*_{a''\bar{b}'} - 
B_{b''\bar{a}'}B^*_{b''\bar{a}'}
+B_{b'\bar{a''}}B^*_{b'\bar{a}''} = 0 \\ \nonumber
A_{a'\bar{b}''}B_{b''\bar{a}'} -A_{a''\bar{b}'}B_{b'\bar{a}''} =0
\eea
In the above equations the sum over $a', b', a'' , b''$ is understood.
These equations are generalized version of \eq{dtermr4} 
discussed for the
$R^4/Z_2$ singularity in Section 4.1. 
At the origin of the Higgs branch where the
classical moduli space meets the Coulomb branch this linear sigma model
would flow to an infrared conformal field theory which is singular. 
The reason for this is the same as for the
$R^4/Z_2$ case. The linear sigma model contains the following term in
the superpotential
\be
\label{potential}
V= (A_{a'\bar{b}''}A^*_{a'\bar{b}''} + 
A_{a''\bar{b}'}A^*_{a''\bar{b}'} + B_{b''\bar{a}'}B^*_{b''\bar{a}'} +
B_{b'\bar{a}''}B^*_{b'\bar{a}''})
(\varphi_1^2 + \varphi_2^2 + \varphi_3^2 +\varphi_4^2)
\ee
As in the discussion of the $R^4/Z_2$ case, at the origin of the
hypermultiplet moduli space the flat direction of the Coulomb branch
leads to a ground state which is not normalizable.  This singularity
can be avoided by deforming the D-term equations by
the Fayet-Iliopoulos terms:
\bea
\label{resolved}
A_{a'\bar{b}''}A^*_{a'\bar{b}''}-A_{a''\bar{b}'}A^*_{a''\bar{b}'} -
B_{b''\bar{a}'}B^*_{b''\bar{a}'} +B_{b'\bar{a}''}B^*_{b'\bar{a}''} =
r_3 \\ \nonumber A_{a'\bar{b}''}B_{b''\bar{a}'}
-A_{a''\bar{b}'}B_{b'\bar{a}''} = r_1 + ir_3 
\eea
We note here that
the Fayet-Iliopoulos terms break the relative $U(1)$ under discussion
and the gauge field becomes massive.  The reason is that the D-terms
with the Fayet-Iliopoulos do not permit all $A,B$'s in the above
equation to simultaneously vanish.  At least one of them must be
non-zero. As these $A, B$'s are charged under the $U(1)$, the non-zero
of value of $A,B$ gives mass to the vector multiplet. This can be seen
from the potential \eq{potential}. The scalars of the vector multiplet
becomes massive with the mass propotional to the {\em vev}'s of $A,B$.
Thus the relative $U(1)$ is broken.

The singularity associated with the non-compact Coulomb branch can also
be avoided by turning on the $\theta$ term, the mechanism being
similar to the one discussed in the previous subsection. If any of the
$3$ Fayet-Iliopoulos D-terms or the $\theta$ term is turned on, the
flat directions of the Coulomb branch are lifted, leading to
normalizable ground state is of the Higgs branch. This prevents the
breaking up of the $(Q_1,Q_5)$ system to subsystems. Thus we see that
the $4$ parameters which resolve the singularity of the ${\cal
N}=(4,4)$ SCFT on ${\cal M}$ make their appearance in the gauge theory
as the Fayet-Iliopoulos terms and the theta term.

It would be interesting to extract the singularity structure of the
the gauge theory of the D1/D5 system through mappings similar to
\eq{dtermdef1}- \eq{dtermdef2}\footnote{The singularity structure for
a $U(1)$ theory coupled to $N$ hypermultiplets has been obtained in
\cit{chp2:GASY}}.

\subsection{The case $(Q_1,Q_5)\to (Q_1-1,Q_5)+
(1,0)$: splitting of 1 D1-brane}

It is illuminating to consider the special case
in which  1 D1-brane splits off from the bound state
$(Q_1,Q_5)$. The effective dynamics is again 
described in terms of a $U(1)$ gauge theory associated
with the relative separation between the single D1-brane
and the bound state $(Q_1-1,Q_5)$. The massless
hypermultiplets charged under this $U(1)$ correspond
to open strings joining the single D1-brane with
the D5-branes and are denoted by
\be
\chi_{b'} = \left( \begin{array}{c}
A_{b'} \\
B^\dagger_{b'}
\end{array}
\right)
\ee
The D-term equations, with the
Fayet-Iliopoulos terms, become in this case
 \be
\label{dterm1}
\sum_{b'=1}^{Q_5} \left( |A_{b'}|^2 - |B_{b'}|^2 \right)=
r_3, \quad  \sum_{b'=1}^{Q_5} A_{b'} B_{b'} = r_1 + ir_2
\ee
while the potential is
\be
\label{potential1}
V= \left[ \sum_{b'=1}^{Q_5} \left( |A_{b'}|^2 + |B_{b'}|^2 \right)
\right] (\varphi_1^2 + \varphi_2^2 +\varphi_3^2 +\varphi_4^2 )
\ee
In this simple case it is easy to see that the presences of the
Fayet-Iliopoulos terms in \eq{dterm1} ensures that all $A,B$'s do not
vanish simultaneously. The {\em vev}'s of $A,B$ gives mass to the
$\varphi$'s. Thus the relative $U(1)$ is broken when the
Fayet-Iliopoulos term is not zero.
The D-term equations above agree with those in
\cit{chp2:SeiWit} which discusses the splitting
of a single D1-brane. It is important to emphasize
that the potential and the D-term equations describe
an {\em effective dynamics} in the classical background 
corresponding to the $(Q_1-1, Q_5)$ bound state. This
corresponds to the description in \cit{chp2:SeiWit}
of the splitting process in an AdS$_3$ background
which represents a mean field of the above bound state. 

\subsection{The dynamics of the splitting at the singularity of the
Higgs branch}
\label{chp2:sing}

We have seen in Section \ref{chp2:gt} that the effective theory
describing the dynamics of the splitting of the $(Q_1, Q_5)$ system to
substems $(Q'_1, Q'_5)$ and $(Q''_1, Q''_5$) is $(4,4)$, $U(1)$
super Yang-Mills coupled to $Q'_1Q''_5 + Q'_5 Q''_1)$ hypermultiplets.
The SCFT which this gauge theory flows in the infra-red is singular
if the Fayet-Iliopoulos terms and the theta term is set to zero. The
description of the superconformal theory of the
Higgs branch of a $U(1)$ gauge theory with $(4,4)$
supersymmetry and $N$ hypermultiplets 
near the singularity was recently found in
\cit{chp2:AhaBer}. It was expressed in the Coulomb variables. 
It consists of a bosonic $SU(2)$ 
Wess-Zumino-Witten model at level $N-2$, four free fermions and a
linear dilaton with background charge given by
\be
Q= \sqrt{\frac{2}{N}} (N-1)
\ee
The central charge of this SCFT is $6(N-1)$. Using this result for the
$U(1)$ theory describing the splitting we get a background charge for
the linear dilaton given by
\be
Q=\sqrt{\frac{1}{Q'_1Q''_5 + Q'_5Q''_1}}(Q'_1Q''_5 + Q'_5Q''_1 -1)
\ee

\section*{References}
\begin{enumerate}
\bibi{chp2:DavManWad2} J.R. David, G. Mandal and S.R. Wadia, ``D1/D5
moduli in SCFT and gauge theory and Hawking radiation,''
hep-th/9907075.
\bibi{chp2:Malthesis} J. Maldacena, ``D-branes and near extremal black
holes at low energies,'' Phys. Rev. {\bf D55} (1997) 7645.
\bibi{chp2:HasWad} S.F. Hassan and S.R. Wadia, ``Gauge theory
description of D-brane black holes: emergence of the effective SCFT
and Hawking radiation,'' Nucl. Phys. {\bf B526} (1998) 311,
hep-th/971213.
\bibi{chp2:Vafa1} C. Vafa, ``Instantons on D-branes,'' Nucl. Phys.
{\bf B463} (1996) 435, hep-th/9512078.
\bibi{chp2:Douglas} M. R. Douglas, ``Branes within branes,'' 
hep-th/9512077.
\bibi{chp2:Dijkgraaf} R. Dijkgraaf, ``Instanton strings and
hyperkaehler geometry,'' Nucl. Phys. {\bf B543} (1999) 545,
hep-th/9810210.
\bibi{chp2:Vafa2} C. Vafa, ``Gas of D-branes and Hagedorn Density of
BPS States,'' Nucl. Phys. {\bf B463} (1996) 415, hep-th/9511088.
\bibi{chp2:Sen} A. Sen, ``U-Duality and Intersecting D-branes,''
 Phys. Rev. {\bf D53} (1996) 2874, hep-th/9511026. 
\bibi{chp2:Yu} M. Yu, ``The unitary representations of the ${\cal N}=4$
$SU(2)$ extendend superconformal algebras,'' Nucl. Phys. {\bf B294}
(1987) 890.
\bibi{chp2:VafWit}
L. Dixon, J. Harvey, C. Vafa and E. Witten, ``Strings
on orbifolds,'' Nucl.
Phys. {\bf B261} (1985) 620;
Strings on orbifolds. 2, Nucl. Phys. {\bf B274} (1986) 285.
\bibi{chp2:DijMooVerVer}
R. Dijkgraaf, G. Moore, E. Verlinde and H.
Verlinde, ``Elliptic genera of symmetric products and second quantized
strings,'' Commun. Math. Phys. {\bf 185} (1997) 197, hep-th/9608096.
\bibi{chp2:DixFriMarShe}
L. Dixon, D. Friedan, E. Martinec and S.
Shenker, ``The conformal field theory of orbifolds,'' Nucl. Phys. {\bf
B282} (1987) 13.
\bibi{chp2:Wit_susy} E. Witten, ``Constraints on supersymmetry
breaking,'' Nucl. Phys. {\bf B202} 253.
\bibi{chp2:CveDix}
M. Cvetic, ``Effective Lagrangian of the
(blown up) orbifolds,'' in {\em Superstrings, Unified Theories And
Cosmology} ed. G. Furlan et. al., (1987).
\bibi{chp2:HasWad1}
S.F. Hassan and S.R. Wadia, ``D-brane black holes:
large-N limit and the effective string description,'' Phys.Lett.
{\bf B402} (1997) 43, hep-th/9703163.
\bibi{chp2:DijVerVer}
R. Dijkgraaf, E. Verlinde and H. Verlinde,
``5D black holes and matrix strings,'' Nucl.Phys.
{\bf B506} (1997) 121, hep-th/9704018.
\bibi{chp2:SeiWit}
N. Seiberg and E. Witten, ``The D1/D5 system and
singular CFT,'' hep-th/9903224.
\bibi{chp2:LarMar}
F. Larsen and E. Martinec,
``U(1) charges and moduli in the D1/D5 system,''  JHEP
{\bf 9906} (1999) 019, hep-th/9905064.
\bibi{chp2:Aspinwall}
P.S. Aspinwall, ``Enhanced gauge symmetries and
$K_3$ surfaces,'' Phys. Lett. {\bf B357} (1995) 329, hep-th/9507012.
\bibi{chp2:ShwarzNekrasov} N. Nekrasov and A. Schwarz, ``Instantons on
noncommutative $R^4$ and $(2,0)$ superconformal six dimensional
theory,'' Commun. Math. Phys. {\bf 198} (1998) 689, hep-th/9802068.
\bibi{chp2:Wit}
E. Witten, ``Some comments on string dynamics,''
hep-th/9507121, in Strings 95, {\em Future Perspectives In String
Theory}, eds. I. Bars et. al.
\bibi{chp2:Wit_cft}
E. Witten, ``On the conformal field theory of the
Higgs branch,'' JHEP {\bf 9707}:003 (1997), hep-th/9707093.
\bibi{chp2:GASY} A. Dhar, G. Mandal, S.R. Wadia and K.P.
Yogendran,``D1/D5 system with B-field, noncommutative geometry and the
CFT of the Higgs branch,'' hep-th/9910194.
\bibi{chp2:AhaBer} O. Aharony and M. Berooz, ``IR dynamics of $d=2,
{\cal N} =(4,4)$ gauge theories and DLCQ of `Little String
Theories','' hep-th/9909101.
\end{enumerate}

%% file: chapter3.tex
\chapter{Coupling with the bulk fields}
\markright{Chapter 3. Coupling with the bulk fields}
\vspace{-1cm}

This chapter focuses on finding the coupling of the microscopic theory
discussed in Chapter 2 with fields of the supergravity. The 
supergravity field $\phi$ couples with the 
operator  ${\cal O}(z,\bar{z} )$ of 
${\cal N}=(4,4)$ SCFT on ${\cal M}$ in the form of an interaction
\be
\label{chp3:coupling}
S_{\mbox{int}} = 
\mu\int d^2 z \, \phi (z, \bar{z}) {\cal O} (z, \bar{z})
\ee
$\mu$ denotes the strength of the coupling. 
We fix the operator ${\cal O} (z, \bar{z})$ corresponding to the
supergravity field $\phi$  by appealing to symmetries. We have seen
that in Chapter 2 the marginal operators of the 
SCFT can be classified according to
the quantum numbers of the supergroup $SU(1,1|2)\times SU(1,1|2)$
together with the $SO(4)_I$ quantum numbers.  In this chapter we 
will see
that these symmetries are also present in the near horizon limit of
the supergravity solution.  We will classify the supergravity
fields according to these symmetries. In particular we will focus on
all the scalars present in the supergravity background \cit{chp3:DavManWad1}. 
The coupling
in \eq{chp3:coupling} is 
fixed by demanding that the operator ${\cal O}(z,
\bar{z})$ and the supergravity field $\phi$ have the same quantum
number under the various symmetries \cit{chp3:DavManWad2}.

\section{Near horizon symmetry}

In this section we will review the near horizon symmetries of the
D1/D5 system.
The solution of the type IIB supergravity equations in the string
metric is given by \cit{chp3:CalMal}
\bea
\label{chp3:d1_d5}
ds^2 &=& f_1^{-\frac{1}{2}} f_5^{-\frac{1}{2}} (-dt^2 + dx_5^2)
+ f_1^{\frac{1}{2}} f_5^{\frac{1}{2}} (dx_1^2 + \cdots + dx_4^2)
\\ \nonumber
& & + f_1^{\frac{1}{2}} f_5^{-\frac{1}{2}}
(dx_6^2 + \cdots + dx_9^2),
\\ \nonumber
e^{-2 \phi} &=& \frac{1}{g_s^2} f_5 f_1^{-1} , \\ \nonumber
B_{05} &=& \frac{1}{2} (f_1^{-1} -1), \\ \nonumber
H_{abc} &=& (dB^{'})_{abc}
=\frac{1}{2}\epsilon_{abcd}\partial_{d} f_5, \;\;\;\;
a, b, c, d = 1, 2, 3, 4
\eea
where $f_1$ and $f_5$ are given by
\be
f_1 = 1 + \frac{16 \pi ^4 g_s \alpha '^3 Q_1}{V_4 r^2}, \;\;\;\;
f_5= 1+ \frac{ g_s \alpha' Q_5 }{r^2},\;
\ee
Here $r^2 = x_1^2 + x_2^2 + x_3^2 + x_4^2$ denotes the 
distance measured in the transverse direction to all the D-branes.

The near horizon scaling limit is obtained by 
\bea
\alpha' \rightarrow 0, \;&\;&\; \frac{r}{\alpha '} 
\equiv U=  \mbox{fixed} \\  \nonumber
v \equiv \frac{V_4}{16\pi^4\alpha^{\prime 2}} = \mbox{fixed}, \;&\;&\;
g_6 = \frac{g_s}{\sqrt{v}} = \mbox{fixed}
\eea
In this limit the metric in \eq{chp3:d1_d5} becomes
\be
\label{chp3:near_horizon}
ds^2 = \alpha ' \left[ \frac{U^2}{g_6\sqrt{Q_1Q_5}} (-dx_0^2 + dx_5^2)
+ g_6\sqrt{Q_1 Q_5} \frac{dU^2}{U^2} + g_6\sqrt{Q_1Q_5} d\Omega_3^2
\right] + \sqrt{\frac{Q_1}{vQ_5}}(dx_6^2 + \ldots dx_9^2)
\ee
Thus the near horizon geometry is that of $AdS_3\times S^3\times T^4$.
Our notation for coordinates here is as follows: $AdS_3:
(x_0, x_5, r); S^3:(\chi, \theta, \phi); T^4:(x_6, x_7, x_8,
x_9)$. $r, \chi,\theta,\phi$ are spherical polar coordinates for the
directions $x_1,x_2,x_3,x_4$. 
The radius of $S^3$ 
and the anti-de Sitter space 
is $\sqrt{\alpha'} (g_6^2 Q_1Q_5)^{1/4}$. 
Let us examine the symmetries of this near horizon geometry. 
The bosonic symmetries arise from the isometries of $AdS_3$ and $S^3$.
The
isometries of the $AdS_3$ space form the non-compact group $SO(2,2)$,
while the isometries of $S^3$ form the group  $SO(4)_E= SU(2)_E\times
\widetilde{SU(2)}_E$. 
Though the
compactification on $T^4$ breaks the $SO(4)$ rotations of the
coordinates $x_6, \ldots ,x_9$ we can still 
use this symmetry to classify
supergravity fields. We will call this symmetry $SO(4)_I$.
The D1/D5 system preserves eight out of the 32 supersymmetries
of the type IIB theory. In the near horizon limit the number of
supersymmetries gets enhanced from eight to sixteen \cit{chp3:Town}. 
These symmetries
fix the form of the effective anti-de Sitter supergravity theory near
the horizon. The bosonic symmetries $SO(2,2)\times SO(4)_E = (SL(2
R)\times SU(2)) \times {(SL(2, R) \times SU(2))}$ 
form the bosonic
symmetries of the anti-de Sitter supergravity in three-dimensions. 
Simple anti-de Sitter supergroups in three-dimensions were classified
in \cit{chp3:towngun}. It can be seen that the only simple 
supergroups whose bosonic part is $SL(2,R)\times SU(2)$ are
$Osp(3|2, R)$ and $SU(1,1|2)$. 
The former contains the bosonic
subgroup $O(3)\times SL(2,R)$. 
The supercharges of the supergroup
$Osp(3|2, R)$ transform as the vector representation of the group
$O(3)$, while the supercharges of the supergroup $SU(1,1|2)$ transform
as ${\bf 2}$ of the group $SU(2)$. 
The unbroken supercharges of the D1/D5 system transform in 
the spinor representation of $SO(4)_E$ and therefore they 
transform as ${\bf 2}$ of $SU(2)$. This rules out $Osp(3|2, R)$.
Therefore the near horizon anti-de Sitter supergravity is based on the
supergroup $SU(1,1|2)\times SU(1,1|2)$ with matter fields. The pure
anti-de Sitter supergravity based on the 
super group $SU(1,1|2)\times SU(1,1|2)$ 
was constructed in \cit{chp3:Jus} using the fact that it
is a Chern-Simons theory .

\section{Classification of the supergravity modes}
\label{chp3:classification}

In this section we analyze the spectrum of Type IIB supergravity
compactified on $AdS_3 \times S^3 \times T^4$. 
From \eq{chp3:near_horizon} we see the volume of $T^4$ is 
 $16\pi^4 \alpha^{' 2} Q_1/Q_5$. Therefore we ignore 
Kaluza-Klein modes on the $T^4$. The radius of the $S^3$ is
$\sqrt{\alpha^{\prime}}(g_6Q_1Q_5)^{1/4}$.
This 
is large when 
\be
g_sQ_1>>1 \;\;\mbox{and}\;\; g_sQ_5>>1
\ee
These inequalities are true in the limit \eq{l_1}.
Therefore we are justified in using supergravity. Kaluza-Klein
reduction of type IIB supergravity to six dimensions leaves six
dimensional $(2,2)$ supergravity.
We show that the Kaluza-Klein spectrum of the six dimensional
theory on $AdS_3\times S^3$ can be completely organized as short
multiplets of the supergroup $SU(1,1|2)\times SU(1,1|2)$.  We will
follow the method developed by \cit{chp3:deB}. 

The massless spectrum of $(2,2)$ six-dimensional supergravity
consists
of: 
a graviton, 8 gravitinos, 5 two-forms, 
16 gauge fields, 40 fermions and 25 scalars.
Since these are massless, 
the physical degrees of freedom 
fall into various representations $R_4$
of the little group $SO(4)_{L}$ of $R^{(5,1)}$. 
For example, the graviton transform as a  $(\bf 3 ,\bf 3)$ under the
little group $SO(4)_{L}= SU(2)_L\times \widetilde{SU(2)}_L$.
On further compactifying
$R^{(5,1)}$ into $AdS_3 \times S^3$, each representation $R_4$
decomposes into various representations $R_3$ of $SO(3)$, the local
Lorentz group of the $S^3$. This $SO(3)\simeq SU(2)$ 
is the diagonal $SU(2)$ of $SU(2)_L\times \widetilde{SU(2)}_L$. 
For example, the graviton decomposes as $\bf{1} + \bf{3} +\bf{5}$ 
under the $SO(3)$, the local Lorentz group of $S^3$.
The dependence of each of these fields on
the angles of $S^3$ leads to decomposition in terms of Kaluza-Klein
modes on the $S^3$ which transforms according to some representation of
the isometry group $SO(4)$ of $S^3$. Only those representations of
$SO(4)$ occur in these decompositions which contain the representation
$R_3$ of $S^3$. To be more explicit, consider the field 
$\phi_{R_{SO(3)}} (x_0, x_5, r, \theta,\phi,\chi)$ 
which
transforms as some representation $R_{SO(3)}$ of the local Lorentz group
of $S^3$. The Kaluza-Klein expansion of this field on $S^3$ is given
by
\be
\phi_{R_{SO(3)}} (x_0, x_5, r, \theta,\phi,\chi ) =
\sum_{R_{SO(4)}} \tilde{\phi}_{R_{SO(4)}}(x_0, x_5, r)
Y^{R_{SO(4)}}_{R_{SO(3)}}(\theta, \phi, \chi).
\ee
Here
$Y^{R_{SO(4)}}_{R_{SO(3)}}(\theta, \phi, \chi)$ stands for the spherical
harmonics on $S^3$.  In the above expansion the only representation of
$R_{SO(4)}$ allowed are the ones which contain $R_{SO(3)}$. For
example, $\phi (x_0, x_5, r, \theta, \phi, \chi)$ which is a scalar under the 
local Lorentz group of $S^3$ can be expanded as
\be
 \phi (x_0, x_5, r, \theta, \phi, \chi) 
= \sum_{\bf{m}, \bf{m'};\; \bf{m}= \bf{m'}}
\tilde{\phi}_{\bf{m} \bf{m'}} (x_0, x_5, r) 
Y^{(\bf{m}, \bf{m'})} (\theta, \phi, \chi)
\ee
Once the complete set of Kaluza-Klein modes are obtained we will
organize them into short multiplets of the supergroup $SU(1,1|2)\times
SU(1,1|2)$.

Let us now consider all the massless field of $(2,2)$ supergravity in
six-dimensions individually.
The graviton transforms as $(\re{3},\re{3})$  of the little group
in 6 dimensions. The Kaluza-Klein harmonics of this field 
according to the rules discussed above are
\bea
\label{eq.gravi}
&(\re{1}, \re{1} ) + 2 (\re{2},\re{2}) + (\re{3},\re{1}) +
(\re{1},\re{3})  \\ \nonumber
&+ 3 \oplus_{\re{m}\geq \re{3}} (\re{m},\re{m}) + 2
\oplus_{\re{m}\geq \re{2}} [\, (\re{m} + \re 2,\re{m} ) + 
(\re{m} , \re{m} + \re2 ) \,]  \\ \nonumber
&+ \oplus_{\re{m}\geq \re 1} [\, (\re{m} + \re 4 ,\re{m} ) +
(\re{m}, \re{m} + \re 4 )  \, ]
\eea
The little group representations of the 8 gravitinos is 
$4(\re 2 , \re 3) + 4(\re 3 ,\re 2) $. Their Kaluza-Klein 
harmonics are
\bea
& 8 [\, (\re 1, \re 2 ) + (\re 2 , \re 1) \,] + 
16 \oplus_{\re m  \geq  \re 2 } 
[\, (\re m + \re 1 , \re m ) + ( \re m , \re m + \re 1 ) \, ]
\\  \nonumber
&+ 8 \oplus_{\re m \geq \re 1 }[ \, ( \re m + \re 3 , \re m ) +
(\re m , \re m + \re 3 ) \, ]
\eea
The Kaluza-Klein harmonics of the 5 two-forms transforming
in  $(\re 1 ,\re 3 ) + (\re 3 , \re 1 )$ of the little group are
\be
10 \oplus_{\re m \geq \re 2} (\re m, \re m ) + 10 \oplus_{\re m \geq
\re 1 } [\, (\re m +\re 2 , \re m ) + ( \re m , \re m + \re 2 ) \, ]
\ee
The Kaluza-Klein harmonics of the 16 gauge fields, $(\re 2 , \re 2)$
 are 
\be
\label{eq.gauge}
16 (\re 1 , \re 1 ) + 32  \oplus_{\re m \geq \re 2 } (\re m , \re m )
+ 16 \oplus_{\re m \geq \re 1 } [\, (\re m , \re m + \re 2 ) + ( \re
m + \re 2 , \re m ) \, ]
\ee
The 40 fermions $ 20 (\re 2 , \re 1) + 20 ( \re 1 , \re 2 )$ give
rise to the following harmonics
\be
40 \oplus_{ \re m \geq \re 1 } [\, ( \re m , \re m + \re 1 ) + (
\re m + \re 1 , \re m ) \, ]
\ee 
The 25 scalars $( \re 1 , \re 1 )$  give rise to the harmonics
\be
25 \oplus_{\re m \geq \re 1} (\re m , \re m )
\ee
Putting all this together the complete Kaluza-Klein 
spectrum of type IIB on
$AdS_3 \times S^3 \times T^4$ yields 
\bea
\label{eq.spec}
&42 ( \re 1 , \re 1 ) + 69 (\re 2 , \re 2 ) + 48 [\, (\re 1 , \re
2 ) + ( \re 2 , \re 1 ) \, ] + 27 [ \, (\re 1 , \re 3 ) + ( \re 3
, \re 1 ) \, ]  \\    \nonumber
&70 \oplus_{\re m \geq \re 3 } ( \re m , \re m ) + 56 \oplus_{\re
m \geq \re 2 } [\, (\re m , \re m + \re 1  ) + ( \re m + \re 1 ,
\re m ) \,] \\ \nonumber
&+ 28 \oplus_{\re m \geq \re 2 } [\, ( \re m , \re m + \re 2 ) +
( \re m + \re 2 , \re m ) \, ] + 8 \oplus_{\re m \geq \re 1 } [\,
(\re m , \re m + \re 3 ) + (\re m + \re 3 , \re m ) \, ] \\
\nonumber
&+ \oplus_{\re m \geq \re 1} [ \, ( \re m , \re m + \re 4 ) + ( \re m
+ \re 4 , \re m ) \, ]
\eea
We now organize  the above Kaluza-Klein modes into 
short representations of $SU (1,1 | 2) \times SU ( 1,1 |2 ) $
\cit{chp3:deB}. The short multiplet of $ SU(1,1 | 2)$ consists of
the following states 
\be
\begin{array}{ccc}
  j      & L_0   & \mbox{Degeneracy}     \\    
\hline 
\vspace{-.2ex}
h     & h       & 2h +1    \\    
h-1/2~~   & ~~h + 1/2 & 2(2h)     \\    
h-1   &h+1 & 2h -1  
\end{array}
\ee 
In the above table $j$ labels the representation of $SU(2)$ which
is identified as one of the $SU(2)$'s of the isometry
group of $S^3$. $L_0$ denotes the conformal weight of the state. 
We denote the short multiplet of $SU(1,1 |2) \times SU(1,1 |2)$ as 
$(\re{2h} + \re{1} , \re{2h'} + \re 1 )_S$. On organizing the
Kaluza-Klein spectrum into short multiplets we get the following
set
\bea
\label{eq.short}
&5 (\re 2 , \re 2 )_S + 6 \oplus_{\re m \geq \re 3 } (\re m , \re m
)_S \\ \nonumber &\oplus_{\re m \geq \re 2 } [\, (\re m , \re m + \re2
)_S + ( \re m + \re2 , \re m )_S + 4 ( \re m , \re m + \re 1 )_S + 4 (
\re m + \re 1 , \re m )_S \, ] 
\eea 
Equation \eq{eq.spec} shows that
there are $42 (\re 1 ,\re 1)$ $SO(4)$ representations in the
supergravity Kaluza Klein spectrum. We know that one of these arises from the
$s$-wave of $g_{55}$ from equation \eq{eq.gravi}.  This is one of the
fixed scalars. $16 (\re 1 ,\re 1)$ comes from the $s$-waves of the
$16$ gauge fields (the components along $x^5$) as seen in equation
\eq{eq.gauge}.  The remaining $25$ comes from the $25$ scalars of the
six dimensional theory. We would like to see where these $42 (\re 1
,\re 1)$ fit in the short multiplets of $SU(1,1|2)\times
SU(1,1|2)$. From equation \eq{eq.short} one can read that $20$ of them
are in the $ 5 (\re 2 , \re 2 )_S$ with $(j=0, L_0 =1 ;\, j=0, L_0 =
1)$. 6 of them are in in $6 (\re 3 ,\re 3 )_S$ with $(j=0, L_0=2;\,
j=0, L_0=2)$. These correspond to the fixed scalars. Finally, the
remaining $16$ of them belong to $4(\re 2 , \re 3 )_S + 4(\re 3 , \re
2 )_S $. $8$ of them have $(j=0, L_0=1 ;\, j=0, L_0=2 )$ and $8$ of
them have $(j=0, L_0=2 ;\, j=0, L_0=1 )$. These scalars can be
recognized as the intermediate scalars.

\subsection{Comparison of supergravity shortmultiplets with SCFT}

In Section \ref{chp2:sec_shortmultiplets} we have listed the complete
set of shortmultiplets corresponding to single particle states of the
${\cal N}=(4,4)$ SCFT on the orbifold ${\cal M}$. Comparing equation
\eq{chp2:eq.short} and the list of shortmultiplets of single particle
states obtained from supergravity in \eq{eq.short} we find that they
are identical except for the presence of the following 
additional shortmultiplets  in the SCFT
\be
2(\re 1, \re 2)_S + 2(\re 2, \re 1)_S + (\re 1, \re 3)_S + (\re 3, \re
1)_S
\ee
These correspond to non-propagating degrees of freedom in the
supergravity \cit{chp3:deB}. 
Therefore they are not present in the list of
shortmultiplets obtained form supergravity \eq{eq.short}. 

\section{The supergravity moduli}

In this section we will analyze in detail the massless
scalars in the near horizon geometry of the D1/D5 system.
Type IIB supergravity
compactified on $T^4$ has 25 scalars. There are $10$ scalars $h_{ij}$
which arise from compactification of the metric.  $i, j, k \ldots $
stands for the directions of $T^4$. There are $6$ scalars $b_{ij}$
which arise from the Neveu-Schwarz $B$-field and similarly there are
$6$ scalars $b'_{ij}$ from the Ramond-Ramond $B'$-field. The remaining
3 scalars are the ten-dimensional dilaton 
$\phi_{10}$, the Ramond-Ramond scalar $\chi$ and
the Ramond-Ramond $4$-form $C_{6789}$. These scalars parameterize the
coset $SO(5,5)/(SO(5)\times SO(5))$.  The near horizon limit of the
D1/D5 system is $AdS_3\times S^3\times T^4$ \cit{chp3:MalStr}. 
In this geometry $5$ of
the $25$ scalars become massive.  They are the $h_{ii}$ (the trace of
the metric of $T^4$ which is proportional to the volume of $T^4$), the
$3$ components of the anti-self dual part of the Neveu-Schwarz
$B$-field $b_{ij}^-$ and a linear combination of the Ramond-Ramond
scalar and the $4$-form \cit{chp3:SeiWit}.  The massless scalars in the
near horizon geometry parameterize the coset $SO(5,4)/(SO(5)\times
SO(4))$ \cit{chp3:GivKutSei}.

As we have seen 
the near horizon symmetries form the supergroup $SU(1,1|2)\times
SU(1,1|2)$. 
We have classified  all
the massless supergravity fields of type IIB supergravity on
$AdS_3\times S^3\times T^4$ ignoring the Kaluza-Klein modes on $T^4$
according to the short multiplets of the supergroup $SU(1,1|2)\times
SU(1,1|2)$. The isometries of the anti-de Sitter space allow us to
relate the quantum number  $L_0 +\bar{L}_0$  to the mass of the scalar
field through the  relation.
\be
h+\bar{h}= 1+ \sqrt{1+m^2}
\ee
Here $m$ is the mass of the scalar in units of the radius of
$AdS_3$ and $(h,\bar h)$ is the eigenvalue of $L_0, \bar{L}_0$ under
the classification of the scalar in shortmultiplets of
$SU(1,1|2)\times SU(1,1|2)$. 
Thus the massless fields of the 
near horizon geometry of the D1/D5 system fall into the top
component of the $5(\bf 2, \bf 2 )_S$ short multiplet.  We further
classify these fields according to the representations of the
$SO(4)_I$, the rotations of the $x_6,x_7,x_8,x_9$ directions. As we have
mentioned before this 
is not a symmetry of the supergravity as it is compactified on $T^4$,
but it can be used to classify states. The quantum number of the
massless supergravity fields are listed below.
\bea
\label{sugra_fields}
\begin{array}{lccc}
\mbox{Field} & SU(2)_I\times \widetilde{SU(2)_I} 
& SU(2)_E\times \widetilde{SU(2)}_E &
\mbox{Mass} \\
h_{ij } -\frac{1}{4}
\delta_{ij} h_{kk} & (\bf 3, \bf 3) & (\bf 1, \bf 1) & 0 \\
b'_{ij} & (\bf 3, \bf 1) + (\bf 1, \bf 3) &(\bf 1, \bf 1) &  0 \\
\phi_6 &  (\bf 1, \bf 1) &(\bf 1, \bf 1) &  0 \\
a_1 \chi + a_2 C_{6789}&  (\bf 1, \bf 1) &(\bf 1, \bf 1) &  0 \\
b^+_{ij} &  (\bf 1, \bf 3) &(\bf 1, \bf 1) & 0 
\end{array}
\eea
The linear combination appearing on the fourth line is the
one that remains massless in the near-horizon limit.
$\phi_6$ refers to the six-dimensional dilaton.
The $SU(2)_E \times \widetilde{SU(2)}_E$ stands for the $SO(4)$
isometries of the $S^3$.
All the above fields are s-waves of scalars in the near horizon
geometry. 

\section{AdS$_3$/CFT$_2$ correspondence}

Maldacena's conjecture 
[\ref{chp3:Maldacena},\,\ref{chp3:Witten98},\,\ref{chp3:GubKlePol},\,\ref{chp3:MalStr98}] 
for the case of the D1/D5 system states
that string theory on $AdS_3\times S^3\times T^4$ is dual to the 
$1+1$ dimensional 
conformal field theory of the Higgs branch of gauge theory of the
D1/D5 system. There is evidence for this conjecture from symmetries.
To describe the D1/D5 system at a generic point in
the moduli space we can use the ${\cal N}=(4,4)$ SCFT on the orbifold
${\cal M}$ to describe  the Higgs branch of the gauge theory of the
D1/D5 system as we have argued in Chapter 2. 
The volume of $T^4$ is of the order of string length and radius of
$S^3$ is large, therefore we can pass over from string theory on
$AdS_3\times S^3\times T^4$ to six-dimensional $(2,2)$ supergravity
on $AdS_3\times S^3$. We will compare symmetries in the supergravity
limit.  The identification of the isometries  of the near horizon
geometry with that of the symmetries of the SCFT are given in the
following table.
\\

\begin{tabular}{ll}
Symmetries of the Bulk & Symmetries of SCFT \\
\hline
 & \\
(a) Isometries of $AdS_3$ 
&The global part of the Virasoro group \\
 $SO(2,2)\simeq SL(2, R) \times \widetilde{SL(2,R)}$ &
$SL(2,R)\times \widetilde{SL(2,R)}$ \\
 & \\
(b) Isometries of $S^3$ &
R-symmetry of the SCFT \\
$SO(4)_E\simeq SU(2)\times SU(2)$ &
$SU(2)_R\times\widetilde{SU(2)}_R$ \\
 & \\
(c) Sixteen near horizon symmetries & Global supercharges  of 
${\cal N}= (4,4) $ SCFT \\
& \\
(d) $SO(4)_I$ of $T^4$ & $SO(4)_I$ of $\tilde{T}^4$ \\
\hline
\end{tabular} \\

To summarize the $SU(1,1|2)\times SU(1,1|2)$ symmetry of the near
horizon geometry is identified with the global part of the ${\cal
N}=(4,4)$ SCFT on the orbifold ${\cal M}$ together with the
identification of the $SO(4)_I$ algebra of $T^4$ and $\tilde{T}^4$.

\section{Supergravity moduli and the marginal operators}

We would like to match the twenty supergravity moduli appearing
in \eq{sugra_fields} with the twenty marginal operators 
appearing in \eq{untwist_operator} and \eq{twist_operator} 
by comparing their symmetry properties under the AdS/CFT correspondence
\cit{chp3:DavManWad}. 

\noindent
The symmetries, or equivalently quantum numbers, to be compared under
the  AdS/CFT correspondence
are as follows:

(a) The isometries of the supergravity are identified with the global
symmetries of the superconformal field theory. For the $AdS_3$ case
the symmetries form the supergroup $SU(1,1|2)\times SU(1,1|2)$.  
The identification of this supergroup with the
global part of the ${\cal N}= (4,4)$ superalgebra leads
to the following mass-dimension relation
\be
\label{chp3:massdim}
h+\bar{h}= 1+ \sqrt{1+m^2}  
\ee
where $m$ is the mass of the bulk field and 
{\em $(h,\bar h)$
are the dimensions of the SCFT operator}. Since in 
our case the SCFT operators are marginal and the supergravity
fields are massless, the mass-dimension relation is obviously
satisfied. 

(b) The $SU(2)_E\times \widetilde{SU(2)}_E$
quantum number of the bulk supergravity field corresponds to the
$SU(2)_R\times \widetilde{SU(2)}_R$ 
quantum number of the boundary operator.
By an inspection of column three of the tables in
\eq{untwist_operator}, \eq{twist_operator} and \eq{sugra_fields}, 
we see that these quantum numbers
also match.
 
(c) The location of the bulk fields and the boundary operators as
components of the short multiplet can be found by the supersymmetry
properties of the bulk fields and the boundary operators.
Noting the fact that all the twenty bulk fields as
well as all the marginal operators mentioned above correspond to top
components of short multiplets, this property also matches.  

(d) The above symmetries alone do not distinguish between the twenty
operators or the twenty bulk fields. To further distinguish these
operators and the fields we identify the $SO(4)_I$ symmetry of the
directions $x_6,x_7,x_8,x_9$ 
with the $SO(4)_I$ of the SCFT. At the level
of classification of states this identification is reasonable though
these are not actual symmetries. Using the quantum numbers under this
group we obtain the following matching of the boundary operators and
the supergravity moduli.
\bea
\begin{array}{llc}
\mbox{Operator} & \mbox{Field}  &  
SU(2)_I\times \widetilde{SU(2)}_I \\
\del x^{ \{ i }_A(z) \bar{\del}x^{ j\} }_A (\bar z) -1/4\delta^{ij}
\del x^{k}_A \bar{\del}x^k_A
& h_{ij} -1/4\delta_{ij} h_{kk} & (\bf 3, \bf 3 ) \\
\del x^{[i}_A(z) \bar{\del}x^{j]}_A (\bar z) 
& b'_{ij}
& (\bf 3, \bf 1 ) +
(\bf 1, \bf 3)  \\ 
\del x^i_A(z) \bar{\del}x^i_A (\bar z) 
& \phi
&( \bf 1, \bf 1 ) \\
{\cal T}^1 & b^+_{ij} & (\bf 1, \bf 3 ) \\
{\cal T}^0 & a_1\chi + a_2C_{6789} & (\bf 1, \bf 1 )
\end{array}
\eea
Note that both the representations ${\bf (1,3)}$ and ${\bf (1,1)}$ occur
twice in the above table. This could give rise to a two-fold ambiguity
in identifying either ${\bf (1,3)}$ or ${\bf (1,1)}$ operators with
their corresponding bulk fields. The way we have resolved it here is
as follows.  The operators ${\cal T}^1$ and ${\cal T}^0$ correspond to
blow up modes of the orbifold, and since these are related to the
Fayet-Iliopoulos terms and the $\theta$-term in the gauge theory (see
Section \ref{chp2:linear_sigma}), 
tuning these operators one can reach the singular SCFT
\cit{chp3:SeiWit} that corresponds to fragmentation of the D1/D5
system. In supergravity, similarly, it is only the moduli $b^+_{ij}$
and $a_1\chi + a_2C_{6789}$ which affect the stability of the D1/D5
system 
[\ref{chp3:SeiWit},\,\ref{chp3:LarMar},\,\ref{chp3:Dijkgraaf}]. 
As a result, it
is $b^+_{ij}$ (and not $b^{\prime +}_{ij}$) which
should correspond to the operator ${\cal T}^1$ and similarly
$a_1\chi + a_2C_{6789}$ should correspond to ${\cal T}^0$. Another
reason for this identification is as follows. $b^+_{ij}$ and $a_1\chi
+a_2C_{6789}$ are odd under world sheet parity while $b^{\prime
+}_{ij}$ and $\phi$ are even under world sheet parity. 
Assign a $Z_2$ quantum number $-1$ to the twisted sectors and $+1$ to
the untwisted sectors in the SCFT. If under the AdS/CFT correspondence
one can identify these $Z_2$ quantum numbers in the boundary SCFT and
the bulk then the corespondence we have made is further justified.

Thus, we arrive at a one-to-one correspondence between
operators of the SCFT and the supergravity moduli.

%

\section{Fixed scalars}
\label{chp3:sec_fixed}

Out of the 25 scalars mentioned earlier which form part of the
spectrum of IIB supergravity on $T^4$, five become massive when
further compactified on $AdS_3 \times S^3$. There is an important
additional scalar field which appears after this compactification:
$h_{55}$. Let us remind ourselves the 
notation used for the coordinates: $AdS_3:
(x_0, x_5, r), S^3:(\chi, \theta, \phi); T^4:(x_6, x_7, x_8,
x_9)$. $r, \chi,\theta,\phi$ are spherical polar coordinates for the
directions $x_1,x_2,x_3,x_4$. In terms of the D-brane wrappings, the
D5 branes are wrapped along the directions 
$x_5, x_6, x_7, x_8, x_9$ 
and D1 branes are aligned 
along $x_5$. The field $h_{55}$ is scalar in the sense
that it is a scalar under the local Lorentz group $SO(3)$ of $S^3$.

In what follows we will specifically consider the three scalars
$\phi_{10}, h_{ii}$ and $h_{55}$. The equations of motion of these
fields in supergravity are coupled and have been discussed in detail
in the literature [\ref{chp3:CalGubKleTse},\,\ref{chp3:KraKle}]. 
It turns out that the six-dimensional dilaton 
$\phi_6 = \phi_{10} - h_{ii}/4$
which is a linear combination of $h_{ii}$ and $\phi_{10}$ remains massless; 
it is
part of the twenty massless (minimal) scalars previously discussed.
The two other linear combinations $\lambda$ and $\nu$ 
 are defined as
\bea
\lambda &=& \frac{h_{55}}{2} -\frac{\phi_{10}}{2} + \frac{h_{ii}}{8} \\
\nonumber
\nu &=& \frac{h_{ii}}{8}
\eea
$\lambda$ and $\nu$ satisfy coupled
differential equations. These are examples of fixed
scalars. The pick up masses in the 
background geometry of the D1/D5 system. 
In the near horizon limit the equation of motions of $\lambda$ and
$\nu$ become decoupled. They give rise to the massive Klein-Gordon
equation in $AdS_3$.  The
near horizon mass of $\lambda$ and $\nu$ is $m^2= 8$ in units of the
radius of $AdS_3$.

Understanding the absorption and emission properties of fixed scalars
is an important problem, because the D-brane computation and
semiclassical black hole calculation of these properties appear to be
at variance \cit{chp3:KraKle}.  The discrepancy essentially originates
from the `expected' couplings of $\lambda$ and $\nu$ to SCFT operators
with $(h, \bar h)=(1,3)$ and $(3,1)$.  These SCFT operators lead to
qualitatively different greybody factors from what the fixed scalars
exhibit semiclassically.
The semiclassical greybody
factors are in agreement with D-brane computations if the couplings
were only to (2,2) operators. 

The coupling to $(1,3)$ and $(3,1)$ operators is guessed from
qualitative reasoning based on the Dirac-Born-Infeld action. Since we
now have a method of  deducing the couplings in \eq{chp3:coupling} 
based on near-horizon symmetries, 
let us use it to the case of the fixed
scalars.

(a) By the mass dimension relation \eq{chp3:massdim} 
we see that the fixed
scalars $\lambda$ and $\nu$ correspond to operators with weights
$h +\bar{h}=4$.

(b) The fixed scalars have $SU(2)_E\times \widetilde{SU(2)}_E$ quantum
numbers $(\bf{1}, \bf{1})$. 

As all the supergravity fields are
classified according to the shortmultiplets of $SU(1,1|2)\times
SU(1,1|2)$ we can find the field corresponding  to these quantum
numbers among the shortmultiplets.
%
Searching through the shortmultiplets (see below \eq{eq.short}),
we find that the fixed scalars belong to the
short multiplet $( \re 3 , \re 3 )_S$ of
$SU(1,1|2) \times SU(1,1|2)$. They occur as top component of 
$( \re 3 , \re 3 )_S$.  There are six fixed scalars in all.
We conclude that the operators with
$(h, \bar{h}) = (1,3)$ or
$(h, \bar{h}) = (3,1)$  (which were inferred by the DBI method) are ruled out 
by the analysis of symmetries.

In summary, since the $(1,3)$ and $(3,1)$ operators are ruled out
by our analysis, the discrepancy between the D-brane calculation and
the semiclassical calculation of absorption and emission rates
disappears. It is important to note here that couplings guessed from
reasoning based on Dirac-Born-Infeld action turns out to be incorrect.

\section{Intermediate Scalars}

We just  remark that the classification presented in Section
\ref{chp3:classification}
 correctly account for all sixteen intermediate scalars, and predict
that they should couple to SCFT operators with $(h,\bar h) =(1,2)$
belonging to the short multiplet $({\bf 2}, {\bf 3})_S$ or operators
with $(h,\bar h) =(2,1)$ belonging to the short multiplet $({\bf 3},
{\bf 2})_S$ (see below equation \eq{eq.short}). 
This agrees with the `phenomenological' prediction made
earlier in the literature \cit{chp3:KleRajTse}.

\section{Supergravity from gauge theory}

We have seen in this chapter there is a detailed map of the
Kaluza-Klein spectrum of super gravity in
the near horizon geometry of the D1/D5 system to the shortmultiplets 
of the $(4,4)$ SCFT on ${\cal M}$. In this section we  show that
the gauge theory description of the
the break up the $(Q_1, Q_5)$ system to subsystems $(Q'_1, Q'_5)$ and
$Q''_1, Q''_5)$ agrees with the description of this process in
supergravity. 
In Chapter 2 we have shown that
for the $(Q_1, Q_5)$ system to break up into subsystems have to we set
the Fayet-Iliopoulos terms and the theta term in the gauge theory to
zero. This corresponds to the D1/D5 system  with the
$b^{+}_{ij}$ and $a_1\chi + a_2C_{6789}$ moduli set to zero.
From now on we will denote the subsystem $(Q'_1, Q'_5)$
as $(Q_1, Q_5)$ and the subsytem $Q''_1, Q''_5)$ as $(q_1, q_5)$ for
convenience.
We saw in Section \ref{chp2:sing} 
that in the infrared the description of the
splitting process consisted of a linear dilaton with background charge
given by
\be
\label{chp3:charge}
Q_{{\rm SCFT}}= 
\sqrt{\frac{2}{Q_1q_5 + Q_5q_1}} \left( Q_1q_5 + Q_5q_1 -1\right)
\ee
We show that the splitting process in supergravity is also controlled
by a linear dilaton theory with the background charge given in
\eq{chp3:charge}. 

Lets us first consider the case when $q_5=0$ \cit{chp3:SeiWit}. 
The Ads/CFT
correspondence tells us that we need to consider $q_1$ D1-branes in
the background of $AdS_3\times S^3\times T^4$. 
(We will work in  the Euclidean $AdS_3$ coordinates.)
The radius of $S^3$ and
the anti-de Sitter space is given by $r_0=\sqrt{\alpha^{\prime}}
(g_6^2Q_1Q_5)^{1/4}$. For the supergravity to be valid we need to
consider the limit \eq{l_1}. In the gauge theory the linear dilaton
corresponded to the distance between the centre of mass of the two
subsytems. Therefore we should focus on the distance between the
boundary of $AdS_3\times S^3\times T^4$ and the set of $q_1$
D1-branes. We are interested in the infrared description of the
splitting process. By the  
UV/IR correspondence the D1-branes should be close
to the boundary of the $AdS_3\times S^3\times T^4$ to obtain the
infrared description of the splitting process in the supergravity. We
assume that the D1-branes are fixed at a particular point on the $S^3$
and the $T^4$. The action of $q_1$ D1-branes in the background of
$AdS_3$ and the Ramond-Ramond two-form $B_{05}$ is given by the DBI
action. We can use the DBI action for multiple D1-branes as we are
interested only in the dynamics of the centre of mass of thee
collection of $q_1$ D1-branes. The DBI action of $q_1$ D1-branes is
given by
\be
\label{chp3:dbi}
S= \frac{q_1}{2\pi g_s\alpha^{\prime}} \int d^2 \sigma
e^{-\phi}\sqrt{\mbox{ det }( g_{\alpha\beta}^{{\rm ind}})} -
\frac{q_1}{2\pi
g_s\alpha^{\prime}} \int B
\ee
where $\sigma$ stands for the world volume coordinates and
$\alpha,\beta$ label these coordinates. $g_{\alpha\beta}^{{\rm ind}}$ 
is the
induced metric on the world volume. $B$ is the Ramond-Ramond 2-form
potential. We chose a gauge in which the world volume coordinates are
the coordinates of the boundary of the $AdS_3$. Let the metric on the
boundary be $g_{\alpha\beta}(\sigma)$. 
One can extend the metric $g_{\alpha\beta}(\sigma)$ on the boundary to
the interior of $AdS_3$ in the neighbourhood of the boundary
\cit{chp3:SeiWit}. This is
given by
\be
\label{chp3:bcmetric}
ds^2= \frac{r_0^2}{t^2}\left( dt^2 + \hat{g}_{\alpha\beta}
(\sigma,t)d\sigma^{\alpha}d\sigma^{\beta} \right)
\ee
with
\be
\hat{g}_{\alpha\beta}(\sigma, 0) = g_{\alpha\beta}(\sigma), \;\;\;
\hat{g}_{\alpha\beta}(\sigma, t) = g_{\alpha\beta}(\sigma ) -
t^2P_{\alpha\beta} + O(t^3)+\ldots
\ee
Here $g_{\alpha\beta}P^{\alpha\beta}=R/2$. $R$ is the world sheet
curvature. The global coordinates of Euclidean $AdS_3$ is given by
\be
ds^2= r_0^2(d\phi^2 + \sinh^2\phi d\Omega^2)
\ee
where $d\Omega^2$ is the round metric on $S^2$. 
Near the boundary the metric is given by
\be
ds^2 = r_0^2(d\phi^2 +\frac{e^{2\phi}}{4} d\Omega^2)
\ee
Motivated by this we
use $\phi$ defined as 
$t=2e^{-\phi}$ to measure the distance from the boundary of
$AdS_3$. Substituting the metric in \eq{chp3:bcmetric} and the near
horizon value of the Ramond-Ramond 2-form and the
dilaton  in \eq{chp3:dbi} we obtain
the following effective action of the D1-branes near the boundary.
\bea
\label{chp3:d1eff}
S&=& \frac{q_1r_0^2 }{4\pi g_s \alpha^{\prime}}
\sqrt{\frac{Q_5 v}{Q_1}}
\int \sqrt{g} \left(
\partial_{\alpha}\phi\partial^{\alpha}\phi + \phi R -\frac{1}{2}R
 + O(e^{-2\phi}) \right) \\ \nonumber
&=& \frac{q_1Q_5}{4 \pi} \int \sqrt{g} \left( 
\partial_{\alpha}\phi\partial^{\alpha}\phi + \phi R -\frac{1}{2}R
+ O(e^{-2\phi})\right)
\eea

Now consider the case when $q_1=0$. The $q_5$ D5-branes are wrapped on
$T^4$. Therefore the world volume of the D5-branes is of the form
$M_2\times T^4$ where $M_2$ is any 2-manifold. The D5-branes are
located at a point on the $S^3$.
We ignore the fluctuations on $T^4$ as we are
interested in the dynamics on $AdS_3$. The DBI action of $q_5$
D5-branes is given by
\be
\frac{q_5}{32\pi^5g_s\alpha^{\prime 3} }
\left( \int d^6\sigma 
e^{-\phi} \sqrt{\mbox{det}(g_{\alpha\beta}^{{\rm ind}})}
-\int C^{6} \right)
\ee
where $C^6$ is the Ramond-Ramond 6-form potential coupling to 
the D5-brane. 
Performing a similar calculation for the D5-branes and substituting
the near horizon values of the 6-from Ramond-Ramond potential, the
dilaton and the volume of $T^4$ one obtains the following effective
actions for the D5-branes
\bea
\label{chp3:d5eff}
S&=& \frac{q_5r_0^2}{4\pi g_2\alpha^{\prime}}\sqrt{\frac{Q_1 v}{Q_5}}
\int \sqrt{g} \left(
\partial_{\alpha}\phi\partial^{\alpha}\phi + \phi R -\frac{1}{2}R
+ O(e^{-2\phi})\right) \\ \nonumber
&=& \frac{q_5Q_1}{4 \pi} \int \sqrt{g} \left( 
\partial_{\alpha}\phi\partial^{\alpha}\phi + \phi R -\frac{1}{2}R
+ O(e^{-2\phi})\right)
\eea

For the case when $q_1\neq 0$ and $q_5\neq 0$ and we just add the 
contribution from \eq{chp3:d1eff} and \eq{chp3:d5eff} to obtain the
effective action of the $(q_1, q_5)$ string in $AdS_3$. The reason we
can do this is because there is no force between the test D1 and
D5-branes \footnote{I thank G. Horowitz and R. Myers for a discussion
which helped to clarify this point.}. Thus to the leading order in
$\phi$ the total effective action
of the $(q_1, q_5)$ string near the boundary is given by
\be
S= \frac{(q_1Q_5 + q_5Q_1)}{4\pi} \int 
\sqrt{g} \left( 
\partial_{\alpha}\phi\partial^{\alpha}\phi + \phi R -\frac{1}{2}R
\right)
\ee
Rescaling $\phi$ so that the the normalization of the kinetic energy
term is canonical one obtains a linear dilaton action with a back
ground charge given by
\be
Q_{{\rm SUGRA}} = \sqrt{2(q_1Q_5 + q_5Q_1)}
\ee
We see that the $Q_{{\rm SUGRA}}= Q_{{\rm SCFT}}$ for large $Q_1$
and $Q_5$. Thus we are able to derive the dynamics of the break up of
the $(Q_1, Q_5)$ system to subsystems from the gauge theory.

\section*{References}

\begin{enumerate}
\bibi{chp3:DavManWad1} J.R. David, G. Mandal and S.R. Wadia, ``D1/D5
moduli in SCFT and gauge theory and Hawking radiation,''
hep-th/9907075.
\bibi{chp3:DavManWad2} J.R. David, G. Mandal and S.R. Wadia, 
``Absorption and Hawking radiation of minimal and fixed
scalars, and $AdS/CFT$ correspondence,''
Nucl. Phys. {\bf B544} (1999) 590, hep-th/9808168.
\bibi{chp3:CalMal} C. G. Callan and J. Maldacena, ``D-brane
 approach to black hole quantum mechanics,'' Nucl. Phys. {\bf
B472} (1996) 591, hep-th/9602043.
\bibi{chp3:Town} P. Claus, R. Kallosh, J. Kumar, P. Townsend and A. Van 
Proeyen, ``Conformal theory of M2, D3, M5 and D1+D5 branes,'' JHEP {\bf 9806} 
(1998) 004, hep-th/9801206.
\bibi{chp3:towngun} M. G\"{u}naydin, G. Seirra and P.K. Townsend, ``The 
unitary supermultiplets of d=3 Anti-deSitter and d=2 conformal superalgebras,
'' Nucl. Phys. {\bf B274} (1986) 429.
\bibi{chp3:Jus} J. David, ``Anti-de Sitter gravity associated with the 
supergroup $SU(1,1|2) \times SU(1,1|2)$,'' Mod. Phys. Lett. {\bf A14} (1999)
 1143, hep-th/9904068.
\bibi{chp3:deB} J.de Boer, ``Six-dimensional supergravity on
$S^3\times AdS_3$ and 2d conformal field theory,'' hep-th/9806104.
\bibi{chp3:MalStr}J. Maldacena and A. Strominger, 
``Black hole greybody factors and D-brane spectroscopy,'' Phys. Rev.
{\bf D55} (1997) 861, hep-th/9609026; ``Universal low-energy dynamics
for rotating black holes,'' Phys. Rev. {\bf D56} (1997) 4975,
hep-th/9702015. 
\bibi{chp3:SeiWit}N. Seiberg and E. Witten, ``The D1/D5 System and
singular CFT,'' hep-th/9903224. 
\bibi{chp3:GivKutSei} A. Giveon, D. Kutasov and N. Seiberg, ``Comments on
string theory on $AdS_3$,'' Adv. Theor. Math. Phys. {\bf 2} (1998)
733, hep-th/9806194.
\bibi{chp3:DavManWad}J.R. David, G. Mandal and S.R. Wadia, ``Absorption
and Hawking radiation of minimal and fixed scalars, and AdS/CFT
correspondence,'' Nucl. Phys. {\bf B544} (1999) 590, hep-th/9808168.
\bibi{chp3:Maldacena}J. Maldacena, ``The large N limit of superconformal
field theories and supergravity,'' Adv. Theor. Math. Phys. {\bf 2}
(1997) 231, hep-th/9711200.
\bibi{chp3:Witten98}E. Witten, ``Anti-de Sitter space and
holography,'' Adv. Theor. Math. Phys. {\bf 2} (1998) 253, 
hep-th/9802150.
\bibi{chp3:GubKlePol}S. S. Gubser, I. R. Klebanov and A. M. Polyakov,
 ``Gauge theory correlators from non-critical string theory,'' 
hep-th/9802109. 
\bibi{chp3:MalStr98}J. Maldacena and A. Strominger, ``$AdS_3$ black holes
and a stringy exclusion principle,'' JHEP {\bf 9812}:005 (1998),
hep-th/9804085.
\bibi{chp3:LarMar}F. Larsen and E. Martinec,
``U(1) charges and moduli in the D1-D5 system,''  JHEP
{\bf 9906} (1999) 019, hep-th/9905064.
\bibi{chp3:Dijkgraaf}R. Dijkgraaf, ``Instanton strings and
HyperKaehler geometry,'' Nucl. Phys. {\bf B543} (1999) 545,
hep-th/9810210.
\bibi{chp3:CalGubKleTse}C. G. Callan, S. S. Gubser, I. R.
Klebanov and A. A. Tseytlin, ``Absorption of fixed scalars and the
D-brane approach to black holes,'' Nucl. Phys. {\bf B489} (1997)
65, hep-th/9610172.
\bibi{chp3:KraKle}I. R. Klebanov and  M. Krasnitz, ``Fixed scalar
greybody factors in five and four dimensions,'' Phys. Rev. {\bf
D55} (1997) 3250, hep-th/9612051; ``Testing effective string
models of black holes with fixed scalars,'' Phys. Rev. {\bf D56}
(1997) 2173, hep-th/9703216.
\bibi{chp3:KleRajTse}I. R. Klebanov, A. Rajaraman, A. A. Tseytlin, 
``Intermediate scalars and the effective string model of black
holes,'' Nucl. Phys. {\bf B503} (1997) 157, hep-th/9704112.
\end{enumerate}

%% file: chapter4.tex
\chapter{ The Hawking Process}
\markright{Chapter 4. The Hawking Process}
\vspace{-1cm}
\def\om{\omega}
\def\tM{{\widetilde{\cal M}}}
\def\o{{\cal O}}
\def\z{{\vec z}}

In this chapter we put the results of Chapter 2 and Chapter 3 together and 
work towards a precise understanding of Hawking
radiation from the D1/D5 black hole starting from the microscopic SCFT 
\cit{chp4:DavManWad1}.
We first review the 
supergravity calculation of Hawking radiation of minimal scalars
from the D1/D5 black hole. We are interested in the D1/D5 black
hole with various moduli turned on. We show that in the supergravity
calculation the Hawking radiation is independent of the moduli.
Then we discuss the SCFT relevant to the D1/D5 black hole and its
relationship with the SCFT of the D1/D5 system. 
We review the formulation of the absorption cross-section 
calculation from the SCFT as an evaluation of the thermal Green's
function of the operators ${\cal O} (z,\bar{z})$ corresponding to the
supergravity field ${\phi}$. 
We are able to show that the Hawking radiation calculated from the
SCFT agrees precisely with that of the supergravity calculation if we
fix the strength of the coupling $\mu$ in \eq{interaction} 
using the AdS/CFT correspondence.  
We investigate the dependence of the Hawking radiation 
of minimal scalars calculated from
the SCFT and show that it is independent of the moduli as expected
from supergravity.

\section{Supergravity calculation of absorption/Hawking radiation
in presence of moduli}
\label{chp4:sugra_abs_moduli}

We recall that the D1/D5 black hole solution in the absence
of moduli is [\ref{chp4:StrVaf},\,\ref{chp4:CalMal}] 
obtained from the D1/D5 system by
further compactifying $x^5$ on a circle of radius $R_5$ and adding left(right)
moving Kaluza-Klein momenta along $x^5$.  
The corresponding supergravity solution is given in \eq{chp1:nonextremal}.

The absorption cross-section of minimal scalars in the absence of
moduli is given by [\ref{chp4:DhaManWad},\,\ref{chp4:MalStr96}]
\be
\label{classabs}
\sigma_{abs} = 2 \pi^2 r_1^2 r_5^2 \frac{\pi \om}{2} 
\frac{\exp(\om/T_H)-1}{(\exp(\om/2T_R)-1) (\exp(\om/2T_L)-1)}
\ee
where $T_H$ is given by
\be
\frac{2}{T_H} = \frac{1}{T_L} + \frac{1}{T_R}
\ee
The quantities $T_L, T_R, r_1, r_5$ are as defined in \eq{temp} and 
\eq{chp1:nonextremal}.
We will now show that the absorption cross-section remains unchanged
even when the moduli are turned on.

{}From the equations of motion of type IIB supergravity
\cit{chp4:CalGubKleTse}, we can explicitly see that the five-dimensional
Einstein metric $ds_{5,Ein}^2$ is not changed by turning on the
sixteen moduli corresponding to the metric $G_{ij}$ on $T^4$ and the
Ramond-Ramond 2-form potential $B$. As regards the four blowing up
moduli, the invariance of $ds_{5,Ein}^2$ can be seen from the fact
that turning on these moduli corresponds to $SO(4,5)$ transformation
(which is a part of a U-duality transformation) and from the fact that the
Einstein metric does not change under U-duality.  
Now we know that the minimal scalars $\phi^i$ all
satisfy the wave-equation
\be 
D_\mu \del^\mu \phi^i =0 
\ee 
where the Laplacian is with respect to the Einstein metric in five
dimensions. Since it is  only this  wave equation that
determines the absorption cross-section completely, we see that
$\sigma_{abs}$ is the same as before.

It is straightforward to see that the Hawking rate, given by
\be
\label{classdecay}
\Gamma_H = \sigma_{abs} (e^{\om/T_H}-1)^{-1} 
\frac{d^4k}{(2 \pi)^4}
\ee
is also not changed when moduli are turned on.

\section{Near horizon geometry of the D1/D5 black hole}

In order to find out the SCFT relevant for the D1/D5 black 
hole let us
review the near horizon geometry of the D1/D5 black hole. The near
horizon scaling limit is given by \cit{chp4:MalStr98}
\be
\label{chp4:Maldalim}
\alpha^{\prime}\rightarrow 0, \;\;\;  r\rightarrow 0,\;\;\;
r_0\rightarrow 0
\ee
with
\bea
 U\equiv \frac{r}{\alpha^{\prime}} =\mbox{fixed} \;&\;&\; 
 U_0\equiv \frac{r_0}{\alpha^{\prime}} =\mbox{fixed}  \\ \nonumber
 v\equiv \frac{V_4}{16\pi^4\alpha^{\prime 2}} =\mbox{fixed} \;&\;&\;
 g_6= \frac{g_s}{\sqrt{v}} =\mbox{fixed} \;\;\; R_5 = \mbox{fixed}
 \eea
In this limit the metric of the D1/D5 black hole \eq{chp1:nonextremal}
reduces to the following
\bea
\label{chp4:Maldametric}
ds^2 &=& \frac{\alpha^{\prime} U^2}{l^2}(-dx_0^2 + dx_5^2)  
+ \frac{\alpha^{\prime} U_0^2}{l^2} (\cosh \sigma dt + \sinh\sigma
dx_5^2)^2 
+\frac{\alpha^{\prime} l^2}{U^2- U_0^2} dU^2  \\ \nonumber
&+& \alpha^{\prime}l^2 d\Omega_3^2 +
\sqrt{\frac{Q_1}{vQ_5}}(dx_6^2+ \ldots dx_9^2) 
\eea
where $l^2= g_6\sqrt{Q_1Q_5}$. It can be seen that this geometry is
that of $BTZ\times T^4\times S^3$  
using the co-ordinate redefinitions given
below. (Here $BTZ$ refers to the black hole in three-dimensional anti
de-Sitter space discovered by \cit{chp4:BTZ}.)
\bea
\tilde{r}^2&=& (U^2 + U_0^2\sinh^2\sigma)\frac{R_5^2}{l^2} \\ \nonumber
r_+&=& \frac{R_5 U_0\cosh\sigma}{l} \\   \nonumber
r_-&=& \frac{R_5 U_0\sinh\sigma}{l} \\  \nonumber
\phi= \frac{x_5}{R_5} \;&\;&\; t = \frac{lx_0}{R_5}
\eea
The metric in these new coordinates is given by
\bea
\label{chp4:btz}
ds^2&=& -\frac{\alpha^{\prime}(\tilde{r}^2-r_+^2)(\tilde{r}-r_-^2)}
{l^2\tilde{r}^2} dt^2
+ \frac{\alpha^{\prime} \tilde{r}^2l^2}{(\tilde{r}-r_+^2)
(\tilde{r}-r_-^2)} dr^2
+\alpha^{\prime}\tilde{r}^2\left( d\phi + \frac{r_+r_-}
{\tilde{r}^2l}dt\right)^2 \\ \nonumber
&+&\alpha^{\prime}l^2 d\Omega_3^2 + \sqrt{\frac{Q_1}{vQ_5}}(dx_6^2
+\ldots + dx_9^2)
\eea
Our co-ordinate definitions are as follows: $t, \phi, \tilde{r}$ refer
to BTZ co-ordinates, $\Omega_3$ stands for the $S^3$ and $x_6, x_7,
x_8, x_9$ stand for the co-ordinates of $T^4$.
The mass $M$ and the angular momentum $J$
of the BTZ black hole are given by
\be
\label{chp4:parameters}
M = \frac{r_{+}^2 + r_{-}^2}{l^2}  \;\;\;\;
J = \frac{\sqrt{\alpha '}2r_+ r_-}{l}
\ee
The mass $M$ and the angular momentum $J$ for the BTZ black hole are
related to the parameters of the D1/D5 black hole by
\bea
\label{chp4:parmetersd1_d5}
\frac{M}{2} &=& L_0 + \bar{L}_0 = \frac{N_L + N_R}{Q_1Q_5} \\
\nonumber
\frac{J}{2\sqrt{\alpha '} l} &=& 
L_0 -\bar{L}_0 = \frac{N_L - N_R}{Q_1Q_5}
\eea
where $N_L$, $N_R$ are defined in \eq{temp} and $L_0$, $\bar{L}_0$ 
are the levels of the SCFT.
The extremal limit is given by $r_{+}=r_{-}$. From
\eq{chp4:parameters} and \eq{chp4:parmetersd1_d5} 
we see that in the extremal limit $N_R=0$ as
expected for the D1/D5 black hole.

In order to find the relevant SCFT corresponding to the D1/D5 black
hole let us look at the limit $r_+=r_-=0$. Substituting these values
in \eq{chp4:btz} we find the metric is given by
\be
\label{chp4;zero_btz}
\frac{ds^2}{\alpha^{\prime}} = -\frac{\tilde{r}^2}{l^2}dx_0^2 +
\frac{l^2}{\tilde{r}^2}dr^2 + \tilde{r}^2d\phi^2
\ee
In the above equation we have suppressed the $T^4$ and the $S^3$ part of
the metric. By comparison with \eq{chp3:near_horizon} one can see that 
this metric is locally $AdS_3$ except for the global identification 
$\phi\sim\phi + 2\pi$. Let us compare this metric with that of $AdS_3$
in this global co-ordinates. In this co-ordinate system the metric of
the $AdS_3$ is given by
\be
\label{chp4:ads_btz}
\frac{ds^2}{\alpha '}= -(\frac{\tilde{r}^2+l^2}{l^2} ) dx_0^2 + 
\frac{l^2}{\tilde{r}^2 +l^2} dt^2 + \tilde{r}^2d\phi^2
\ee
The geometry of the zero mass $BTZ$ black hole
and that of the $AdS_3$ (although
identical locally) has an important difference in the boundary
conditions for the fermions. For the case of $AdS_3$ the fermions are
anti-periodic in $\phi$ and for the zero mass $BTZ$ black hole
they are periodic
in $\phi$. One can easily see that the constant time slice of the metric in
 \eq{chp4:ads_btz} has the topology of a disk. This forces the fermions to be
anti-periodic in $\phi$ for $AdS_3$. For the case of the zero mass $BTZ$ 
black hole in \eq{chp4:ads_btz}, the constant time slice has a singularity at
 $\tilde{r}=0$. 
Therefore the fermions can be both periodic or anti-periodic.
An analysis of the killing spinors in the background of the 
zero mass $BTZ$ \cit{chp4:CouHen} shows that the fermions have to be
periodic. Thus the SCFT at the boundary of the zero mass $BTZ$ is in the 
Ramond sector while for the $AdS_3$ case it is in the
Neveu-Schwarz sector. 

From the above discussion we find that the SCFT relevant for 
the D1/D5 black hole with Kaluza-Klein momentum $N=0$
is the Ramond vacuum of the ${\cal N}=(4,4)$
SCFT on the orbifold ${\cal M}$. The microscopic states 
corresponding to the general D1/D5 black hole are states with
$L_0\neq 0$ and $\bar{L}_0\neq 0$ excited over the Ramond vaccum of
the ${\cal N}= (4,4)$ SCFT on the orbifold ${\cal M}$.

\section{The coupling with the bulk fields in the Ramond sector}
\label{chp4:ramond_sector}

In Chapter 3 the coupling of the supergravity fields $\phi$ with the 
operators ${\cal O}$ of the ${\cal N}=(4,4)$ SCFT
on the ${\cal M}$ in \eq{chp3:coupling} 
was determined using the Neveu-Schwarz sector. In
this section we argue that these couplings will not change in the
Ramond sector. Interaction terms in SCFT Lagrangian  do not 
depend on whether one is in the Ramond sector or in the
Neveu-Schwarz sector. It is only the quantities like partition function
which change on going from the
Neveu-Schwarz sector to the Ramond sector. 
The scaling dimension of an operator is given by operator product
expansions(OPEs) with the stress energy tensor. Since OPEs are  local relations , they do not change on
going from the Neveu-Schwarz sector to the Ramond sector. 
Therefore the operators
corresponding to the minimal scalars determined in Chapter 3 will not
change for the D1/D5 black hole. 

\section{Determination of the strength of the coupling $\mu$}
\label{chp4:constant}

Before we perform the calculation of Hawking radiation/absorption
cross-section from the SCFT corresponding to the D1/D5 black hole it
is important to determine the strength of the coupling $\mu$ in
\eq{interaction}. In this section we will determine $\mu$ for the case
of minimal scalars $h_{ij}.$%
\footnote{From now on $h_{ij}$ will denote the 
traceless part of the metric
fluctuations of $T^4$.}
In Chapter 2 we have identified the SCFT operator
corresponding to these fields of the supergravity. The SCFT operator
is given by  
\be
{\cal O}^{ij}(z, \bar{z})=
\del x^{ \{ i }_A(z, \bar{z})
\bar{\del} x^{ j\} }_A (z, \bar{z} ) -\frac{1}{4} \delta^{ij} 
\del x^k_A \bar{\del} x^k_A (z, \bar{z})
\ee

Let us suppose the background metric of the torus $T^4$  is 
$g_{ij} = \delta_{ij} $. 
The interaction Lagrangian of the SCFT with the fluctuation $h_{ij}$
is given by
\be
\label{chp4:interaction} 
S_{\rm int} = \mu  T_{\rm eff} \int d^2 z\; 
\left[ h_{ij} \del x^i_A \bar{\del} x^j_A \right]
\ee
The effective string tension $T_{\rm eff}$ of the conformal
field theory , which also appears in the free part of the action 
\be
\label{chp4:free}
S_0 = T_{\rm eff} \int d^2 z\; \left[\del_z
x^i_A \del_{\bar z} x_{i,A} + {\rm fermions} \right]
\ee
has been discussed in [\ref{chp4:CalGubKleTse},\,\ref{chp4:MALTH},\,
\ref{chp4:HasWad}]. The specific
value of $T_{\rm eff}$ is not important for the calculation of the $S$-matrix
for absorption or emission, since the factor 
just determines the normalization of the two-point function of the
operator ${\cal O}^{ij}(z, \bar{z})$. 
In this section we will argue that the constant $\mu = 1$.

A direct string theory computation would of course provide
the constant $\mu$ as well (albeit at weak coupling).  This
would be analogous to fixing the normalization of the
Dirac-Born-Infeld action for a single D-brane by comparing with 
one-loop open string diagram \cit{chp4:Pol}. However, for a large number
and more than one type of D-branes it is a difficult proposition
and we will not attempt to pursue it here. Fortunately, 
the method of symmetries using the AdS/CFT employed 
for determining the operator ${\cal O}$
helps us determine the value of $\mu$ as well. For the latter,
however, we need to use the more quantitative version
[\ref{chp4:Witads},\,\ref{chp4:GubKlePol}] of the Maldacena conjecture.  
We will see below
that for this quantitative conjecture to be true for the two-point
function (which can be calculated independently from the 
${\cal N}=(4,4)$
SCFT and from supergravity) we need $\mu=1$.

We have seen that the above normalization leads to precise
equality between the absorption cross-sections (and consequently
Hawking radiation rates) computed from the moduli space of the D1/D5
system and from semiclassical gravity. This method of fixing the
normalization can perhaps be criticized on the ground that it borrows
from supergravity and does not rely entirely on the SCFT.  However, we
would like to emphasize two things:\\ 
(a) We have fixed $\mu=1$ by
comparing with supergravity around AdS$_3$ background which does 
{\sl not} have a black hole.  On the other hand, 
the supergravity calculation of absorption
cross-section and Hawking flux is performed around a black hole
background represented in the near-horizon limit by the BTZ black
hole. From the viewpoint of semiclassical gravity  these two 
backgrounds are rather different. The fact that normalizing
$\mu$ with respect to the former background leads to the
correctly normalized absorption cross-section around the
black hole background is a rather remarkable prediction.\\
(b)  Similar issues are involved in fixing the coupling
constant between the electron and the electromagnetic field in the
semiclassical theory of radiation in terms of the physical electric
charge, and in similarly fixing the gravitational coupling of extended
objects in terms of Newton's constant.  These issues too are decided
by comparing two-point functions of currents with Coulomb's or
Newton's laws respectively. In the present case the quantitative
version of the AdS/CFT conjecture 
[\ref{chp4:Witads},\,\ref{chp4:GubKlePol}]
 provides the
counterpart of Newton's law or Coulomb's law at strong coupling.
Without this the best result one can achieve is that the Hawking
radiation rates computed from D1/D5 branes and from semiclassical
gravity are {\sl proportional}.

We should remark that fixing the normalization by the use of
Dirac-Born-Infeld action, as has been done previously, is not
satisfactory since the DBI action is meant for single D-branes and
extending it to a system of multiple D1/D5 branes does not always give
the right results as we have seen in Section \ref{chp3:sec_fixed}. The
method of equivalence principle to fix the normalization is not very
general and cannot be applied to the case of non-minimal scalars, for
example. 

Let us now compare the two-point function for the minimal scalar
$h_{ij}$ determined from the AdS/CFT correspondence and the SCFT to
determine the normalization constant $\mu$.
We will discuss the more quantitative version of the
AdS/CFT conjecture [\ref{chp4:GubKlePol},\,\ref{chp4:Witads}] 
to compare the 2-point
correlation function of $\o_{ij}$ from supergravity and SCFT.

The relation between the correlators  are as
follows. Let the supergravity Lagrangian be
\bea 
\label{lag}
L &=& \int d^3 x_1 d^3 x_2 b_{ij,i'j'}(x_1,x_2) h_{ij}(x_1) 
h_{i'j'}(x_2)\nonumber\\
 &+& \int d^3 x_1 d^3 x_2 d^3
x_3 c_{ij,i'j',i''j''}(x_1, x_2, x_3) h_{ij}(x_1) h_{i'j'}(x_2)
h_{i''j''}(x_3) + \ldots\nonumber\\ 
\eea 
where we have only exhibited terms quadratic and cubic in the 
$h_{ij}$'s. The coefficient $b$ determines the propagator and
the coefficient $c$ is the tree-level 3-point vertex in supergravity. 

The 2-point function of the $\o_{ij}$'s (at large $g_sQ_1,g_sQ_5$)
is given by [\ref{chp4:GubKlePol},\,\ref{chp4:Witads}], 
assuming $S_{\rm int}$
given by \eq{chp4:interaction}
\bea
\label{2pt}
& \langle \o_{ij}(z_1) \o_{i'j'}(z_2)\rangle
\\    \nonumber 
& = 2(\mu T_{\rm eff})^{-2}
\int d^3 x_1 d^3 x_2 \left[ b_{ij,i'j'}(x_1, x_2) 
K(x_1|z_1) K(x_2|z_2)\right],
\eea
where $K$ is the boundary-to-bulk Green's function for massless
scalars \cit{chp4:Witads}.
\be
\label{green}
K(x|z)= \frac{1}{\pi} \left[ \frac{x_0}{(x_0^2 + (|z_x - 
z|^2)}\right]^2
\ee
We use complex $z$ for coordinates of the SCFT, and
$x = (x_0, z_x)$ for the Poincar\'{e} coordinates of bulk theory.

\subsection{Evaluation of the tree-level vertices in supergravity}

We begin with the bosonic sector of Type IIB supergravity. The
Lagrangian is (we follow the conventions of \cit{chp4:BACHAS}) 
\bea  
I &=& I_{\rm NS} + I_{\rm RR}\nonumber\\
I_{\rm NS} &=& 
-\frac{1}{2k_{10}^2} \int d^{10} x \sqrt{-G}\left[ e^{-2\phi} \left( R -
4(d\phi)^2 + \frac{1}{12} (dB_{NS})^2 \right)  \right] \nonumber\\
I_{\rm RR} &=&  -\frac{1}{2k_{10}^2} \int d^{10} x \sqrt{-G}  
\left (\sum_{n=3, 7, ...}\frac{1}{2n!}
(H^{n})^2 \right)
\eea 
with $k_{10}^2= 64\pi^7 g^2_s \alpha^{\prime 4}$. 
We use $\hat{M}, \hat{N} \ldots $ to denote $10$ dimensional indices,
$i,j, \ldots$ to denote coordinates on the torus $T^4$, $M ,N
\ldots $ to denote the remaining $6$ dimensions and $\mu, \nu,
\ldots $ to denote coordinates on the AdS$_3$.  
We have separately indicated the terms
depending on Neveu-Schwarz Neveu-Schwarz and Ramond-Ramond backgrounds.

Our aim will  be to obtain the Lagrangian of the minimally coupled
scalars corresponding to the fluctuations of the metric of the $T^4$
in the D1/D5-brane system. We will find the Lagrangian up to
cubic order in the near horizon limit. Let us first focus
on $I_{\rm NS}$. 
We substitute the  values of the background fields of the D1/D5 system
in the Type
IIB Lagrangian with the  following change in the metric
\be
\label{changemetric}
f_1^{\frac{1}{2}} f_5^{-\frac{1}{2}}\delta_{ij} \rightarrow 
f_1^{\frac{1}{2}} f_5^{-\frac{1}{2}}(\delta_{ij} + h_{ij}).
\ee
where $h_{ij}$ are the minimally coupled scalars with trace 
zero. These scalars are functions of the 6 dimensional
coordinates. Retaining the terms upto $O(h^3)$ and ignoring the traces, the 
Lagrangian can be written as
\be
I_{\rm NS}= -\frac{V_4}{2k_{10}^2}\int d^6x \sqrt{-G}
\frac{G^{M N}}{4} \left[ \partial_M h_{ij} \partial_N h_{ij} +
\partial_M (h_{ik}h_{kj} ) \partial_N h_{ij} \right]
\ee
In the above equation we have used the near horizon limit 
and $V_4$ is the volume of the $T^4$. The
metric $G_{MN}$ near the horizon is 
\be
ds^2 = \frac{r^2}{R^2} ( -dx_0^2 + dx_5
^2 ) + \frac{R^2}{r^2} dr^2 + R^2 d\Omega_3 ^2
\ee
We make a change of variables to the Poincar\'{e} coordinates 
by substituting
\bea
z_0 &=& \frac{R}{r} \\  \nonumber
z_1 &=& \frac{x_0}{R}  \\   \nonumber
z_2 &=& \frac{x_5}{R}  
\eea
The metric becomes 
\be
ds^2 = R^2 \frac{1}{z_0^2} ( dz_0^2 - dz_1^2 + dz_2^2) +
R^2 d\Omega_3 ^2 .
\ee
Here $R= \sqrt{\alpha^{\prime}}(g_6^2 Q_1Q_5)^{1/4}$ 
is the radius of curvature
of $AdS_3$ (also of the $S^3$).
For $s$-waves the minimal scalars do not
depend on the coordinates of the $S^3$. Finally, in Poincar\'{e} coordinates  
$I_{\rm NS}$ (correct to cubic order in $h$) can be written as
\be
I_{\rm NS}= -\frac{V_4}{8k_{10}^2} R^3 
V_{S^3} \int d^3z \sqrt{-g} g^{\mu\nu} \left[ \partial_\mu
h_{ij} \partial_\nu h_{ij} + \partial_\mu (h_{ik} h_{kj})
\partial_\nu h_{ij} \right],
\ee
where $V_{S^3}=2\pi^2$, the volume of a three-sphere of unit radius.

Now we would like to show that to all orders in h, $I_{RR}=0$ in the near 
horizon geometry.
The relevant terms in our case are 
\be
I_{RR}=-\frac{1}{4\times 3! k_{10}^2} \int d^{10} x \sqrt{-G} 
H_{\hat{M} \hat{N} \hat{O} } H^{\hat{M} \hat{N} \hat{O}}.
\ee
We substitute the values of $B$ due to the magnetic and electric
components of the Ramond-Ramond charges and the value of $G$. The
contribution from the electric part of $B'$, after
going to the  near-horizon limit and performing the integral over $S^3$ and
$T^4$ is 
\be
\frac{V_4}{4 k_{10}^2} R V_{S^3} \int d^3 z \sqrt{-g}
\sqrt{{\mbox{det}}(\delta_{ij} + h_{ij})}
\ee
The contribution of the magnetic part of $B'$ in the
same limit is 
\be
-\frac{V_4}{4 k_{10}^2} R V_{S^3} \int d^3 z \sqrt{-g}
\sqrt{{\mbox{det}}(\delta_{ij} + h_{ij})}
\ee
We note that the contribution of the electric and the magnetic
parts cancel giving no couplings for the minimal scalars
to the Ramond-Ramond background. Therefore the tree-level supergravity action 
correct to cubic order in $h$ is given by
\be
\label{chp4:effective}
I= -\frac{Q_1Q_5}{16\pi} \int d^3 z \left[ 
\del_\mu h_{ij} \del_\mu h_{ij}
+ \del_{\mu} (h_{ik} h_{kj}) \del_{\mu} h_{ij}\right]
\ee

The coefficient $Q_1Q_5/(16\pi)$ is U-duality invariant.
This is because it is a function of only the integers $Q_1$ and
$Q_5$.
This can be tested by computing the same coefficient from the
Neveu-Schwarz/fundamental string background which is related to the
D1/D5 system by S-duality. The Neveu-Schwarz/fundamental string
back ground also gives the same coefficient. U-duality transformations
which generate $B_{NS}$ backgrounds \cit{chp4:DavManWad99} 
also give rise to
the same coefficient.

\subsection{Two-point function}

The two-point function of the operator ${\cal O}_{ij}$ can be evaluated
by substituting the value of $b_{ij, i'j'}(x_1,x_2)$ obtained from 
\eq{chp4:effective} into \eq{2pt} and using the boundary-to-bulk
Green's function given in \eq{green}. On evaluating the integral in
\eq{2pt} using formulae given in \cit{chp4:FreMatMatRas}, we find that
\be
\label{chp4:coefficient}
\langle \o_{ij}(z) \o_{i'j'}(w) \rangle =
(\mu T_{\rm eff})^{-2}\delta_{ii'}\delta_{jj'} 
\frac{Q_1 Q_5}{16 \pi^2}\frac{1}
{|z - w|^4}
\ee  
This is exactly the value of the two-point function obtained
from the SCFT described by the free Lagrangian \eq{free} provided
we put $\mu=1$. 

We have compared the two-point function  obtained from the supergravity
corresponding to the near horizon geometry of the D1/D5 system with no
moduli to the orbifold SCFT. As we have argued before the orbifold
SCFT corresponds to the D1/D5 system with moduli. Thus naively this
comparison seems to be meaningless. On further examination we note that
the coefficient $b_{ij, i'j'}$ in \eq{2pt} was U-duality invariant.
Since the D1/D5 system with moduli can be obtained through U-duality
transformations we know that this coefficient will not change for the
D1/D5 system with moduli. It is only the value 
of this  coefficient which fixes $\mu$ to be $1$. Thus the comparison
we have made is valid.
It is  remarkable that even at strong coupling
the two-point function of $\o_{ij}$ can be computed 
from the free Lagrangian
\eq{chp4:free}.  This is consistent with the non-renormalization 
theorems involving the ${\cal N}=(4,4)$ SCFT.

The choice $\mu=1$ ensures that the perturbation \eq{chp4:interaction}
of \eq{chp4:free} is consistent with the perturbation implied in  
\eq{changemetric}. 
We will see in the next section  
that this choice leads to precise equality
between 
absorption
cross-sections (consequently Hawking radiation rates) calculated from
semiclassical gravity and from the D1/D5 branes. 
The overall
multiplicative constant $T_{\rm eff}$ will not be important for the
absorption cross-section calculation. This factor finally cancels off
in the calculation as we will see in Section \ref{chp4:abscft}.

\section{The black hole state}
\label{chp4:state}

As we have seen, the general non-extremal black hole will have Kaluza-Klein 
excitations along both the directions on the $S^1$. In the SCFT, it is 
represented by states with $L_0 \neq 0$ and $\bar{L}_0 \neq 0$ over the Ramond 
vacuum. The black hole is represented by a density matrix 
\be
\label{chp4:density-matrix}
\rho= \frac{1}{\Omega} \;\sum_{ \{i\} } |i\rangle\langle i|  
\ee
The states $|i\rangle$ 
belongs to the various twisted sectors of the orbifold theory.
They satisfy the constraint 
\be
L_0=\frac{N_L}{Q_1 Q_5} \;\;\;\; \bar{L}_0=\frac{N_R}{Q_1Q_5}
\ee
We have suppressed the index which labels the vacuum. $\Omega$ is the volume 
of the phase space in the micro-canonical ensemble. It can be seen that the 
maximally twisted sector of the orbifold gives rise to the dominant
contribution to the sum in \eq{chp4:density-matrix} over the various
twisted sectors. The maximally twisted sector is obtained by the
action of the twist operator $\sum^{(Q_1 Q_5-1)/2}$ on the Ramond
vacuum. From the OPE's in \eq{chp2:OPEs} we see that the twist
operator $\sum^{(Q_1 Q_5-1)/2}$ introduces a cut in the complex plane
such that 
\be
\label{chp4:bc}
X_A (e^{2\pi i}z,e^{-2 \pi i} \bar{z})=X_{A+1} (z,\bar{z})
\ee
Thus this changes the boundary conditions of the bosons and the fermions.
Again from the OPEs in \eq{chp2:OPEs} one infers that  the excitations like 
$\partial \phi_1 |\sum^{(Q_1 Q_5-1)/2}\rangle$
over the maximally twisted sector have modes in units of $1/Q_1Q_5$. A
simple way of understanding that the maximally twisted sector 
has modes in units of $1/(Q_1Q_5)$ is to note that the boundary conditions 
in \eq{chp4:bc} imply that $X_A(z,\bar{z})$ is periodic with a period of $2\pi
Q_1Q_5$. This forces the modes to be quantized in units of $1/(Q_1Q_5)$.

We now show that the maximally twisted sector can account for the
entire entropy of the black hole. The entropy of the D1/D5 black hole
given in \eq{mass_entropy} can be written as 
\be
S_{SUGRA} = 2 \pi \sqrt{N_L} + 2 \pi \sqrt{N_R}
\ee
Using Cardy's formula, the degeneracy of the states in the maximally
twisted sector with $L_0=N_L/Q_1 Q_5$ and $\bar{L}_0=N_R/Q_1Q_5$
is given by
\be
\Omega =e^{2 \pi \sqrt{N_L} + 2 \pi \sqrt{N_R}}
\ee
By the Boltzmann formula,
\be
S(\mbox{maximally twisted})=2 \pi \sqrt{N_L} + 2 \pi \sqrt{N_R}
\ee
Thus the maximally twisted sector entirely accounts for the D1/D5
black hole entropy. $N_L$ and $N_R$ are multiples of $Q_1Q_5$ due to
the orbifold projection. Therefore, the entropy can be written as
\be
S = 2 \pi \sqrt{N_L Q_1 Q_5} + 2 \pi \sqrt{N_R Q_1 Q_5}
\ee

With this understanding, we restrict the calculations of Hawking
radiation and absorption cross-section only to the maximally twisted
sector. The probability amplitude for the Hawking process is given by
\be
P=\frac{1}{\Omega} \sum_{f,i} |\langle f|S_{int}|i\rangle |^2
\ee
where $|f\rangle $ denotes the final states the black hole can decay into. We
have averaged over the initial states in the micro-canonical ensemble.

It is more convenient to work with the canonical ensemble. We now
discuss the method of determining the temperature of the canonical
ensemble. Consider the generating function
\be
\label{chp4:generating}
Z=\mbox{Tr}_{R}(e^{-\beta_L E_0} e^{-\beta_R \bar{E}_0})
\ee
where the trace is done over the Ramond states in the maximally
twisted sector. $E_0$ and $\bar{E}_0$ are energies of the left and the
right moving modes.
\be
E_0=\frac{L_0}{R_5}, \;\;\;\; \bar{E}_0=\frac{\bar{L}_0}{R_5}
\ee
From the generating function Z in \eq{chp4:generating} we see that the
coefficient of $e^{-(\beta_L N_L)/(Q_1Q_5R_5)}$ and $e^{-(\beta_R
N_R)/(Q_1Q_5R_5)}$ is the degeneracy of the states with
$L_0=N_L/Q_1Q_5$ and $\bar{L}_0=N_R/Q_1Q_5$
corresponding to the D1/D5 black hole. A simple way to satisfy this
constraint is to choose $\beta_L$ and $\beta_R$ such that Z is peaked
at this value of $L_0$ and $\bar{L}_0$.

Evaluating the trace one obtains
\be
Z=\prod_{n=1}^{\infty} \left( \frac{1+e^{-(\beta_L
n)/(Q_1Q_5R_5)}}{1-e^{-(\beta_L n)/(Q_1Q_5R_5)}} \right)^4   \;\; 
\left(\frac{1+e^{-(\beta_R n)/(Q_1Q_5R_5)}}{1-e^{-(\beta_R
n)/(Q_1Q_5R_5)}} \right)^4
\ee
Then
\bea
\ln Z= 4 \left[\sum_{n=1}^{\infty} \ln(1+e^{-\beta_L n/Q_1Q_5R_5}) -
\sum_{n=1}^{\infty} \ln (1-e^{-\beta_L n/Q_1Q_5R_5}) \right] \\
+ 4 \left[\sum_{n=1}^{\infty} \ln(1+e^{-\beta_R n/Q_1Q_5R_5}) -
\sum_{n=1}^{\infty} \ln (1-e^{-\beta_R n/Q_1Q_5R_5}) \right] 
\eea
We can evaluate the sum by approximating it by an integral given by
\be
\label{chp4:lnZ}
\ln Z=4 Q_1 Q_5 R_5 \int_0^\infty dx\;\; \ln 
\left(\frac{ 1+e^{-\beta_L x}}
{ 1-e^{-\beta_L x}} \right)+
\ln\left(\frac{ 1+e^{-\beta_R x}}{ 1-e^{-\beta_R x}}\right) 
\ee
From the partition function in \eq{chp4:generating} we see that 
\be
-\frac{\partial \ln Z}{\partial \beta_L} =\frac{<N_L>}{Q_1 Q_5 R_5}
\;\; \mbox{and} \;\;
-\frac{\partial \ln Z}{\partial \beta_R} =\frac{<N_R>}{Q_1 Q_5 R_5}
\ee
where $<\cdot>$ indicates the average value of $N_L$ and $N_R$. As the
distribution is peaked at $N_L$ and $N_R$ we assume that $<N_L>=N_L$
and $<N_R>=N_R$. Using \eq{chp4:lnZ} we obtain
\be
\frac{Q_1Q_5R_5 \pi^2}{\beta_L^2}= \frac{N_L}{Q_1Q_5R_5}
\;\;\; \mbox{and} \;\;\;
\frac{Q_1Q_5R_5 \pi^2}{\beta_R^2}= \frac{N_R}{Q_1Q_5R_5}
\ee
Thus
\be
T_L=\frac{1}{\beta_L}=\frac{\sqrt{N_L}}{\pi R_5 Q_1Q_5}
\;\;\; \mbox{and} \;\;\;
T_R=\frac{1}{\beta_R}=\frac{\sqrt{N_R}}{\pi R_5 Q_1Q_5}
\ee
Above we have introduced a left temperature $T_L$ and a right
temperature $T_R$ corresponding to the left and the right moving
excitations of the SCFT to pass over to the canonical ensemble.

\section{Absorption cross-section as thermal Green's function}
\label{chp4:abs_green}

Let us now relate the 
absorption cross-section of a supergravity
fluctuation $\delta \phi$ to the thermal Green's function of the
corresponding operator of the ${\cal N} =(4,4)$  SCFT on the orbifold
${\cal M}$ \cit{chp4:Gubser}.
The notation $\delta\phi$
implies that we are considering the supergravity field to be of the
form
\be
\phi = \phi_0 + \mu \delta \bar{\phi}
\ee 
where $\phi_0$ represents the background value and $\mu$ is the
strength of the coupling. 
\bea
S &=& S_0 +  \int d^2 z [\phi_0 + \mu  \delta \bar{\phi}] {\cal O}(z,\bar z)
\nonumber \\
&=& S_{\phi_0} + S_{int}
\nonumber \\
\eea
where
\be
S_{\phi_0} = S_0 + \int d^2 z \;\;  \phi_0 {\cal O}(z, \bar z)
\ee
\be
S_{int} = \mu \int d^2 z \;\; \delta  \bar{\phi} {\cal O}(z, \bar z)
\ee
${\cal O}$ is the operator corresponding to supergravity field $\phi$.
$S_0$ is the Lagrangian of the SCFT which includes the deformations
due to various backgrounds in the supergravity. For example, the free
Lagrangian in \eq{chp4:free} corresponds to the case when 
the field $a_1\chi + a_2 C_{6789}$ in \eq{sugra_fields} is turned on.

We  calculate the absorption of  a quanta $\delta\bar\phi =
\kappa_5 e^{-ipx}$  corresponding to the operator ${\cal O}$ 
using the Fermi's Golden Rule. $\kappa_5$ is related to the
five-dimensional Newton's  constant $G_5$ and the ten-dimensional
Newton's constant $G_{10}$ as 
\be
\label{chp4:kappa}
\kappa_5^2 = 8\pi G_5 = \frac{8\pi G_{10}}{V_4 2\pi R_5} = 
\frac{64 \pi^7 g_s^2 \alpha^{\prime 4}} {V_4 2\pi R_5}
\ee
We see that $\kappa_5$ is proportional to $\alpha^{'2}$. In the
Maldacena limit \eq{chp4:Maldalim} the coupling of the bulk
fluctuation to the SCFT drops out. We retain this term for our
calculation as Hawking radiation absorption cross-section is an
$O(1/N)$ effect. We will see below that the absorption cross-section
turns out to be proportional $g_6 Q_1 Q_5 \alpha^{'2}$. From
\eq{chp4:Maldametric} we see that this is the fourth power of the
radius of $S^3$. The radius of $S^3$ is large for the supergravity to
be valid. Thus although in the Maldacena limit \eq{chp4:Maldalim} the
coupling of the bulk fluctuation to the SCFT drops out, the factors of
$Q_1 Q_5$ picked up in the calculation ensure that this coupling
survives for the comparison with supergravity. Furthermore, in analogy
with $AdS_5/CFT_4$ correspondence the number of colours N of the
${\cal N}=4$ SuperYang Mills in four dimensions corresponds to
$Q_1Q_5$. As the absorption cross-section is proportional to $Q_1Q_5$,
this is a $O(1/N)$ effect.

In this computation of the absorption cross-section
the black hole is represented by a canonical
ensemble at a given temperature.
The above interaction gives
the thermally averaged transition probability ${\cal P}$ as
\be
\label{chp4:green}
{\cal P} = \sum_{i,f} \frac{e^{-\beta\cdotp p_i}}{Z} P_{i\rightarrow f} 
= \mu^2\kappa_5^2 Lt
\sum_{i,f} \frac{e^{-\beta\cdotp p_i}}{Z} (2\pi)^2 \delta^2(p+p_i
-p_f) |\langle f| {\cal O} (0,0) |i\rangle |^2
\ee
Here $i$ and $f$ refer to initial and final states respectively.
$p_i, p_f$ refers to the initial and final momenta of these states. 
$L = 2\pi R_5$ denotes
the length of the string and $t$ is the time of interaction. As we have seen 
in Section \ref{chp4:state},
the inverse temperature $\beta$ has two components $\beta_L$ and
$\beta_R$. The relation of  these temperatures to the 
parameters of the D1/D5 black hole is
\be
\beta_L= \frac{1}{T_L} \;\;\mbox{and} \;\; \beta_R= \frac{1}{T_R}
\ee
The left moving momenta $p_+$ and the right moving momenta $p_-$ are
in a thermal bath with inverse temperatures $\beta_L$ and 
$\beta_R$ respectively. $\beta\cdotp p$ is defined as
$
\beta\cdotp p = \beta_L p_+ + \beta_R p_-
$. $Z$ stands for the partition function of the thermal ensemble. 

The Green's function in Euclidean time is given by
\be
{\cal G} (-i\tau, x) = \langle{\cal O}^{\dagger} (-i\tau, x) {\cal
O}(0,0) \rangle =
\mbox{Tr} (\rho T_{\tau} \{ {\cal O}^{\dagger} (-i\tau , x) {\cal
O}(0,0) \} )
\ee
where $\rho=e^{-\beta\cdotp \hat{p}}/Z$.  Time ordering is defined as
$T_{\tau}$ with respect to $-\mbox{Imaginary} (t)$. This definition
coincides with radial ordering on mapping the co-ordinate 
$(\tau , x)$ from the cylinder to the plane.
The advantage of doing this is that the integral
\be
\label{chp4:green1}
\int dt\;dx\; e^{ip \cdotp x} {\cal G}(t-i\epsilon , x) =
\sum_{i,f} \frac{e^{-\beta\cdotp p_i}}{Z} (2\pi)^2 \delta^2(p+p_i
-p_f) |\langle f| {\cal O} (0,0) |i\rangle |^2
\ee
The Green's function 
${\cal G}$ is determined by the two-point function of the operator
${\cal O}$. This is in turn determined by 
conformal dimension $(h, \bar{h})$  of the operator ${\cal O}$ 
and the normalization of the two-point function.

As we have to subtract out the emission probability we get the
cross-section as
\be
\sigma_{abs} {\cal F} t = {\cal P} (1-e^{-\beta \cdotp p} )
\ee
where ${\cal F}$ is the flux and ${\cal P}$ is given by
\eq{chp4:green}. Substituting the value of ${\cal P}$ from
\eq{chp4:green1} we get
\be
\sigma_{abs} =\frac{\mu^2\kappa_5^2 L}{{\cal F}} \int dt\;dx
({\cal G} (t-i\epsilon , x) - {\cal G} (t+i\epsilon, x) )
\ee
In the above equation we have related the evaluation of the
absorption cross-section to the evaluation of the thermal Green's
function. Evaluating the integral one obtains
\bea
\label{chp4:cross-section}
\sigma_{abs}&=& \frac{\mu^2\kappa_5^2 L {\cal C_O}}{{\cal F}}
\frac{ (2\pi T_L)^{2h -1}  (2\pi T_R)^{2\bar{h} -1} }
{ \Gamma(2h) \Gamma(2\bar{h}) }
\frac{ e^{\beta\cdotp p/2} - (-1)^{2h + 2\bar{h}} e^{-\beta\cdotp p/2} 
}{2} \\ \nonumber
&\;&\left| 
\Gamma (h + i\frac{p_{+}}{2\pi T_L} )
\Gamma (\bar{h} + i\frac{p_{-}}{2\pi T_R} )
\right|^2
\eea
where ${\cal C_O}$ is the coefficient of the leading order
term in the OPE of the two-point  function of operator ${\cal O}$. 

\section{Absorption cross-section of minimal scalars from the D1/D5
SCFT}
\label{chp4:abscft}

In the previous section we related the thermal Green's function of the
SCFT operator to the absorption cross-section. We will apply the
results of the previous section for the case of the minimal scalars.
We will consider the case of the minimal scalars corresponding to the
fluctuation of the metric of $T^4$. 
Let the background metric of the torus be $\delta_{ij}$.
Consider the minimal scalar $h_{67}$. 
We know the SCFT operator corresponding to this  has
conformal dimension $(1,1)$. From Section \ref{chp4:constant} we know
that $\mu =1$. The interaction Lagrangian is given by
\be
S_{int} = 2 T_{\rm eff} \int d^2 z h_{67} \del x^6_A (z, \bar{z}) 
\bar{\del}
x^7_A (z, \bar{z}) 
\ee
where we have set $\mu=1$. The factor of $2$ arises because of the
symmetric property of $h_{67}$. $S_0$ is given by 
\be
S_{0} =T_{\rm eff} \int d^2 z \del x^i_A (z, \bar{z}) \bar{\del}
x^j_A(z, \bar{z})
\ee
Comparing with the previous section the operator ${\cal O} = 2T_{\rm
eff} \del x^6_A (z, \bar{z}) \bar{\del} x^7_A(z, \bar{z})$. 
For the absorption of a quanta of energy $\omega$ using
\eq{chp4:cross-section} we obtain
\be
\label{chp4:scftabs}
\sigma_{abs}=
2 \pi^2 r_1^2 r_5^2 \frac{\pi \om}{2} 
\frac{\exp(\om/T_H)-1}{(\exp(\om/2T_R)-1) (\exp(\om/2T_L)-1)}
\ee
where we have $L=2\pi R_5$, ${\cal F}= \omega$, \eq{chp4:kappa} for $\kappa_5$ 
and \eq{chp4:coefficient} for ${\cal C_O}$. Comparing the
 absorption cross-section of the minimal scalars obtained from
 supergravity in \eq{classabs} with \eq{chp4:scftabs} we find that
 \be
 \sigma_{abs}(\mbox{ SCFT}) = 
 \sigma_{abs}({\mbox{ Supergravity}})
 \ee
Thus the SCFT calculation and the supergravity calculation of the
absorption cross-section agree exactly.

It is important to note that we have used the ${\cal N}=(4,4)$ SCFT
realized as a free SCFT on the orbifold ${\cal M}$ as the background
Lagrangian $S_0$. As we have said before, this SCFT is non-singular
and therefore cannot correspond to the case of the D1/D5 system with
no moduli. In Section \ref{chp4:sugra_abs_moduli} 
we have argued that the supergravity
calculation of the absorption cross-section is independent of moduli.
Therefore it makes sense to compare it with the SCFT result for the
case with moduli turned on. In the next section we will 
show that the SCFT calculation is also independent of the moduli.

Another point worth mentioning is that the method followed in the thesis
for the calculation of the absorption-crossection from the SCFT can be
easily extended for the case of minimal scalars corresponding to the
four blow up modes. This is not  possible if one uses the long
string model as the microscopic theory for the D1/D5 black hole. The
simple reason being that these operators are not present in the long
string model.

\section{Independence of Hawking radiation on D1/D5 moduli}
\label{chp4:marginal}

In this section we will study the independence of the Hawking radiation on D1/D5 moduli.
In Chapter 2 we have listed the twenty (1,1) operators $ {\cal O}_i(z,\bar z) $
in the SCFT based on the symmetric product orbifold ${\cal M}$ which
is dual to the D1/D5 system.  
Turning on  various moduli $\phi^i$ of supergravity
corresponds to perturbing the SCFT
\be
\label{CFTperturbation}
S= S_0 +  \sum_i \int d^2 z \;\;\bar \phi^i {\cal O}_i(z,\bar z) 
\ee
where $\bar \phi^i$ denote the near-horizon limits of the
various moduli fields $\phi^i$. 
We note here that $S_0$ corresponds to the free SCFT based on the
symmetric product orbifold ${\cal M}$. Since this SCFT is non-singular
(all correlation functions are finite), it does not correspond to the
marginally stable BPS solution originally found in
[\ref{chp4:StrVaf},\,\ref{chp4:CalMal}].  
Instead, it corresponds to a five-dimensional
black hole solution in supergravity with suitable ``blow-up'' moduli
turned on.

Let us now  calculate the
absorption cross-section of a supergravity
fluctuation $\delta \phi_i$ to the thermal Green's function of the
corresponding operator of the SCFT.
The notation $\delta\phi_i$
implies that we are considering the supergravity field to be of the
form
\be
\phi^i = \phi^i_0 + \mu \delta \phi^i
\ee 
where $\phi^i_0$ represents the background value and $\mu$ is the
strength of the coupling. 
\bea
\label{CFTperturbation1}
S &=& S_0 +  \int d^2 z [\bar \phi^i_0 + \mu  \delta \bar 
\phi^i] {\cal O}_i(z,\bar z)
\nonumber \\
&=& S_{\phi_0} + S_{int}
\nonumber \\
\eea
where
\be
\label{S_phi_0}
S_{\phi_0} = S_0 + \int d^2 z \;\; \bar \phi_0^i {\cal O}_i(z, \bar z)
\ee
\be
\label{S_int}
S_{int} = \mu \int d^2 z \;\; \delta \bar \phi^i {\cal O}_i(z, \bar z)
\ee

As we have seen in the Section \ref{chp4:abs_green}
the absorption cross-section of the 
supergravity fluctuation
$\delta\phi^i$ involves 
essentially
the two-point function of the operator ${\cal O}_i$ calculated with respect
to the SCFT action $S_{\phi_0}$. Since ${\cal O}_i$ is a marginal operator,
its two-point function is completely determined apart from a constant.
Regarding the marginality of the operators ${\cal O}_i$, it is easy to
establish it upto one-loop order by direct computation
($c_{ijk}=0$). The fact that these operators are exactly marginal can
be argued as follows. The twenty operators ${\cal O}_i$ arise as top
components of five chiral primaries. It is known that the number of
chiral primaries with $(j_R, \tilde j_R)=(m,n)$ is the Hodge number
$h_{2m,2n}$ of the target space ${\cal M}$ of the SCFT. Since this
number is a topological invariant, it should be the same at all points
of the moduli space of deformations.

We showed in Section \ref{chp4:constant} that if the
operator ${\cal O}_i$ corresponding to $h_{ij}$ 
is canonically normalized (OPE has residue 1) and if
$\delta \phi_i$ is canonically normalized in supergravity, then the
normalization of $S_{int}$ as in \eq{S_int} ensures that
$\sigma_{abs}$ from SCFT agrees with the supergravity result.  The
crucial point now is the following: once we fix the normalization of
$S_{int}$ at a given point in moduli space, at some other point it may
acquire a constant ($\not=1$) in front of the integral when ${\cal O}_i$ and
$\delta\phi_i$ are canonically normalized at the new point. This would
imply that $\sigma_{abs}$ will get multiplied by this constant, in
turn implying disagreement with supergravity.  We need to show that
this does not happen.
 
To start with a simple example, let us first restrict to the moduli
$g_{ij}$ of the torus $\widetilde{T^4}$. We have
\be
\label{full_action}
S = \int d^2 z \, \del x^i \bar \del x^j g_{ij}
\ee
The factor of string tension has been absorbed in
the definition of $x^i$.

In Section \ref{chp4:constant} we had $g_{ij}
= \delta_{ij} +   h_{ij}$, leading to
\bea
S &=& S_0 + S_{int} \nonumber\\
S_0 &=&  \int d^2 z \;\;\del x^i \bar \del x^j \delta_{ij}
\nonumber\\
S_{int} &=&   
\int d^2 z \;\;\del x^i \bar \del x^j h_{ij}
\nonumber \\
\eea
In the above equation we have set $\mu =1$.
As we have remarked above, this $S_{int}$ gives rise to the
correctly normalized $\sigma_{abs}$. 

Now, if we expand around some other metric 
\be
g_{ij} = g_{0ij} +  h_{ij}
\ee
then the above action \eq{full_action} implies 
\bea
S &=& S_{g_0} + S_{int} \nonumber\\
S_{g_0} &=&  \int d^2 z \;\; \del x^i \bar \del x^j g_{0ij}
\nonumber\\
S_{int} &=& 
\int d^2 z \;\; \del x^i \bar \del x^j h_{ij}
\nonumber \\
\eea
Now the point is that neither $h_{ij}$ nor the operator $ {\cal O}^{ij} =
\del X^i \bar \del X^j $ in $S_{int}$ is canonically normalized at $g_{ij}=
g_{0ij}$. When we do use the canonically normalized operators, do we
pick up an additional constant in front?

Note that 
\be
\label{zamol_example}
\langle {\cal O}^{ij} {\cal O}^{kl} \rangle_{g_0}=g_0^{ik} g_0^{jl}|z-w|^{-4}
\ee
and
\be
\label{sugra_propagator_example}
\langle h_{ij} (x) h_{kl} (y) \rangle_{g_0} =
g_{0,ik} g_{0,jl} {\cal D}(x,y)
\ee
where $ {\cal D} (x,y)$ is the massless scalar propagator 

This shows that

Statement (1): 
{\em The two-point functions of $ {\cal O}^{ij} $ and $h_{ij}$
pick up inverse factors }.

As a result, $S_{int}$ remains correctly normalized when re-written
in terms of the canonically normalized $h$ and $ {\cal O} $ and no additional
constant is picked up.

The above result is in fact valid in the full twenty dimensional
moduli space $\tM$ because Statement (1) above remains true
generally. 

To see this, let us first rephrase our result for the special case of
the metric moduli \eq{full_action} in a more geometric way.  The
$g_{ij}$'s can be regarded as some of the coordinates of the moduli
space $\tM$ (known to be a coset $SO(4,5)/(SO(4)\times SO(5))$).  The
infinitesimal perturbations $h_{ij}, h_{kl}$ can be thought of as
defining tangent vectors at the point $g_{0,ij}$ (namely the vectors
$\del/\del g_{ij}, \del/\del g_{kl}$). The (residue of the) two-point
function given by \eq{zamol_example}\ defines the inner product
between these two tangent vectors according to the Zamolodchikov
metric [\ref{chp4:Zamolodchikov},\,\ref{chp4:Cecotti}].
 
The fact that the moduli space $\tM$ of the ${\cal N}=(4,4)$
SCFT on ${\cal M}$ is the coset $SO(4,5)/(SO(4) \times SO(5))$ is
argued in \cit{chp4:Cecotti}. If the superconformal theory has 
${\cal N}= (4,4)$ supersymmety and if the dimension of the moduli space is $d$
then it is shown in \cit{chp4:Cecotti} that the moduli space of the SCFT is given
by
\be
\label{chp4:m_space}
\frac{SO(4,d/4)}{SO(4)\times SO(d/4)}
\ee
The outline of the argument is a follows.
An ${\cal N}=(4,4)$ SCFT
has superconformal $SU(2)_R \times \widetilde{SU(2)_R}$ 
symmetry. We have
seen that the bottom component of the short multiplet which contains
the marginal operator $\bf{(2,2)_S}$ transforms as a $\bf{(2,2)}$
under $SU(2)_R \times \widetilde{SU(2)_R}$. The top component which
corresponds to the moduli transforms as a $\bf{(1,1)}$ under the
R-symmetry. 
The holonomy group of the 
Zamolodchikov metric 
should leave invariant the action of $SU(2)_R \times
\widetilde{SU(2)_R}$. Then the holonomy group should have a form 
\be
\label{chp4:moduli}
K \subset SU(2) \times SU(2) \times \tilde{K} \subset SO(d)
\ee
Then \eq{chp4:moduli}
together with ${\cal N}=(4,4)$ supersymmetry and the left-right
symmetry of the two $SU(2)_R$'s of the SCFT fixes the moduli space
to be uniquely that given in \eq{chp4:m_space}.
We have found in chapter  2 
that there are 20 marginal operators for the ${\cal
N}=(4,4)$ SCFT on the orbifold ${\cal M}$. 
Therefore the dimension of the moduli
space is $20$. 
Thus $\tM$ is given by
\be
\tM=\frac{SO(4,5)}{SO(4)\times SO(5)}
\ee

Consider, on the other hand, the propagator (inverse two-point
function) of $h_{ij}, h_{kl}$ in supergravity. The moduli space action
of low energy fluctuations is nothing but the supergravity action
evaluated around the classical solutions $g_{0,ij}$.  The kinetic term
of such a moduli space action defines the metric of moduli space. The
statement (1) above is a simple reflection of the fact that the
Zamolodchikov metric defines the metric on moduli space, and hence 

Statement (2):  {\em The propagator of supergravity fluctuations, viewed
as a matrix, is the inverse of the two-point functions in the SCFT.}

The last statement is of course not specific to the moduli $g_{ij}$
and is true of all the moduli.  We find, therefore, that fixing the
normalization of $S_{int}$ \eq{S_int} at any one point $\phi_0$
ensures that the normalization remains correct at any other point
$\phi'_0$ by virtue of Statement (2). We should note in passing that
Statement (2) is consistent with, and could have been derived from
AdS/CFT correspondence as applied to the two-point function.

Thus, we find that $\sigma_{abs}$ is independent of the
moduli, in agreement with the result from supergravity.

\section{Entropy and area}

Here we make a brief mention of the fact that 
the correspondence
between Bekenstein-Hawking entropy and the SCFT entropy remains true
in the presence of all the twenty moduli. The reason is that in
supergravity the Einstein metric remains unchanged (see Section 4.1)
and therefore the area of the event horizon remains the same (this can
be explicitly verified using the supergravity solution in
\cit{chp4:DavManWad99}).  
In the SCFT, since the operators corresponding to
the above moduli are all exactly marginal (Section \ref{chp4:marginal}) 
therefore the
central charge remains unchanged and hence, by Cardy's formula, the
entropy does not change, in agreement with the Bekenstein-Hawking
formula.

\section*{References}
\begin{enumerate}
\bibi{chp4:DavManWad1} J.R. David, G. Mandal and S.R. Wadia, 
``Absorption and Hawking radiation of minimal and fixed
scalars, and $AdS/CFT$ correspondence,''  
Nucl. Phys. {\bf B544} (1999) 590, hep-th/9808168.
\bibi{chp4:StrVaf} A. Strominger and C. Vafa, ``Microscopic origin
of the Bekenstein-Hawking entropy,'' Phys. Lett. {\bf B379}
(1996) 99, hep-th/9601029.
\bibi{chp4:CalMal} C. G. Callan and J. Maldacena, ``D-brane
 approach to black hole quantum mechanics,'' Nucl. Phys. {\bf
B472} (1996) 591, hep-th/9602043.
\bibi{chp4:DhaManWad}A. Dhar, G. Mandal and S.R. Wadia, ``Absorption vs
decay of black holes in string theory and T-symmetry,'' Phys.
Lett. {\bf B388} (1996) 51, hep-th/9605234.
\bibi{chp4:MalStr96}J. Maldacena and A. Strominger, ``Black hole
greybody factors and D-brane spectroscopy,'' Phys. Rev. {\bf D55}
(1997) 861, hep-th/9609026.
\bibi{chp4:CalGubKleTse}C. G. Callan, S. S. Gubser, I. R.
Klebanov and A. A. Tseytlin, ``Absorption of fixed scalars and the
D-brane approach to black holes,'' Nucl. Phys. {\bf B489} (1997)
65, hep-th/9610172.
\bibi{chp4:MalStr98} J. Maldacena and A. Strominger, ``$AdS_3$ black holes
and a stringy exclusion principle,'' JHEP {\bf 9812}:005 (1998),
hep-th/9804085.
\bibi{chp4:BTZ}M. Banados, C. Teitelboim and J. Zanelli,
``Black hole in
three-dimensional spacetime,'' Phys. Rev. Lett {\bf 69} (1992) 1849. 
\bibi{chp4:CouHen} O. Coussaert and M. Henneaux, ``Supersymmetry of the
$2+1$ black holes,'' 
Phys. Rev. Lett. {\bf 72} (1994) 183, hep-th/9310194. 
\bibi{chp4:DMW}A. Dhar, G. Mandal and S. R. Wadia, ``Absorption vs
decay of black holes in string theory and T-symmetry,'' Phys.
Lett. {\bf B388} (1996) 51, hep-th/9605234.
\bibi{chp4:DM}S. R. Das, and S. D. Mathur, ``Comparing decay rates
for black holes and D-branes,'' Nucl. Phys. {\bf B478} (1996)
561, hep-th/9606185; ``Interactions involving D-branes,'' Nucl.
Phys. {\bf B482} (1996) 153, hep-th/9607149.
\bibi{chp4:MALTH}Maldacena, ``Black holes in string theory,'' 
Ph.D. Thesis, hep-th/9607235.
\bibi{chp4:HasWad}S. F. Hassan and  S.R. Wadia, ``Gauge theory
description of D-brane black holes: emergence of the effective
SCFT and Hawking radiation,'' Nucl. Phys. {\bf B526} (1998) 311,
hep-th/9712213.
\bibi{chp4:MooDVV}R. Dijkgraaf, G. Moore, E. Verlinde and H. Verlinde,
``Elliptic genera of symmetric products and second quantized
strings,'' Commun. Math. Phys. {\bf 185} (1997) 197,
hep-th/9608096.
\bibi{chp4:DVV}R. Dijkgraff, E. Verlinde and H. Verlinde, ``Matrix
string theory,'' Nucl. Phys. {\bf B500} (1997) 43,
hep-th/9703030.
\bibi{chp4:Pol}J. Polchinski, S. Chaudhuri and C. V. Johnson,
``Notes on D-Branes,'' hep-th/9602052; J. Polchinski, 
``TASI Lectures on D-Branes,'' hep-th/9611050. 
\bibi{chp4:Witads}E.  Witten, ``Anti-de Sitter space and holography,'' 
hep-th/9802150.
\bibi{chp4:GubKlePol} S. Gubser, I. R. Klebanov and A. M. Polyakov,
 ``Gauge theory correlators from non-critical string theory,'' 
hep-th/9802109. 
\bibi{chp4:BACHAS}C. P. Bachas, ``Lectures on D-branes,'' hep-th/9806199.
\bibi{chp4:DavManWad99} A. Dhar, G. Mandal, S.R. Wadia and K.P.
Yogendran, ``D1/D5 system with B-field, noncommutative geometry and
the CFT of the Higgs branch,'' hep-th/9910194.
\bibi{chp4:Gubser} S.S. Gubser, ``Absorption of photons and fermions by 
black holes in four dimensions,'' Phys. Rev. {\bf D56} (1997) 7854,  
hep-th/9706100.
\bibi{chp4:FreMatMatRas} Z. Freedman, S. D. Mathur, A. Matusis
and L. Rastelli, ``Correlation functions in the CFT$_d/AdS_{d+1}$
correspondence,'' hep-th/9804058.
\bibi{chp4:Zamolodchikov} A.A. Zamolodchikov, 
``Irreversibility of the flux of the
renormalization group in a 2-D field theory,'' JETP Lett. 
{\bf 43}(1986) 730; Sov. J. Nucl. Phys. 64(1987) 1090. 
\bibi{chp4:Cecotti} S. Cecotti, ``${\cal N}=2$ 
Landau-Ginzburg vs. Calabi-Yau sigma models: nonperturbative
aspects,'',  Int. J. Mod. Phys. {\bf A6} (1991) 1749.

\end{enumerate}

%% file: chapter5.tex
\chapter{Concluding remarks and discussions}
\markright{Chapter 5. Concluding remarks and discussions}

In this thesis we have made precise the 
microscopic understanding of
Hawking Radiation from the D1/D5 black hole. 
Our investigations have the following conclusions.

\noindent
(a) We presented an explicit construction of all the marginal
operators in the SCFT of the D1/D5 system based on the orbifold 
${\cal M}$. 
These are twenty in number, four of which are constructed using
$Z_2$ twist operators and correspond to blowing up modes of the
orbifold.

\noindent
(b) We classified the the twenty near-horizon moduli of supergravity
on AdS$_3 \times S^3 \times T^4$ according to representations of
$SU(1,1|2) \times SU(1,1|2) \times SO(4)_I$.

\noindent
(c) We established one-to-one correspondence between the supergravity
moduli and the marginal operators by inventing a new $SO(4)$ symmetry
in the SCFT which we identified with the $SO(4)_I$ of supergravity.

\noindent
(d) We have explictly constructed all the chiral primaries of 
the ${\cal N} =
(4,4)$ SCFT on ${\cal M}$.

\noindent
(e) We analyzed gauge theory dynamics of the D1/D5 system relevant for
the splitting of the bound state $(Q_1,Q_5) \to (Q'_1,Q'_5) +
(Q''_1,Q''_5)$.

\noindent
(f) We have settled a problem in the context of fixed
scalars by showing that consistency with near-horizon symmetry demands
that they cannot couple to (1,3) or (3,1) operators. They can only
couple to (2,2) operators. This removes earlier discrepancies between
D-brane calculations and semiclassical calculations of absorption and
emission.

\noindent
(g) The black hole is represented by a density matrix of a 
microcanonical ensemble consisting of states of the various twisted
sectors of the ${\cal N}= (4,4)$ SCFT on the orbifold ${\cal M}$.
The dominant contribution to the
density matrix is
from the maximally twisted sector of the 
orbifold ${\cal M}$. The maximally twisted sector can be obtained 
by applying the $Q_1Q_5$-cycle twist operator over the Ramond vacuum.
The coupling
of the bulk field to the orbifold SCFT
was deduced using the method of
symmetries. These symmetries include the symmetries of the
near-horizon geometry, as emphasized by the AdS/CFT correspondence.
We calculate the absorption cross-section by relating it to the
thermal Greens funtion of the SCFT. The thermal nature of Hawking
radiation is because we have averaged over several microstates.

\noindent
(h) We determine the normalization of the interaction Lagrangian which
couples CFT operators to bulk modes by using the quantitative version
of the AdS/CFT correspondence, where we compare two-point functions
computed from CFT and from supergravity around AdS$_3$ background.
The normalization fixed this way remarkably leads to precise equality of
absorption cross-sections (consequently Hawking radiation rates)
computed from CFT and from supergravity around the black hole
background.

\noindent
(i) We showed in supergravity as well as in SCFT that the absorption
cross-section for minimal scalars is the same for all values of the
moduli, therefore establishing the agreement between SCFT and
supergravity all over the moduli space.

We will now discuss some directions for future research. 

\noindent
(a) It is important to  determine the absorption cross-section of the
fixed scalars from the orbifold SCFT.
We have deterimed all the chiral primaries of the shortmultiplet
$(\bf{2}, \bf{2})_S$ which contains the operators corresponding to
the fixed scalars as its top component. We can use these chiral
primaries to dermine 
all the
operators with conformal dimensions $(h,\bar{h})= (2,2)$ 
corresponding to the fixed scalars. It would be interesting to see if
the absorption cross-section calculated using these operators
reproduce the semi-classical result.

\noindent
(b) One can find the description of Hawking radiation of scalars
carrying angular momentum within the framework of the SCFT on ${\cal
M}$. 

\noindent
(d) It is very crucial to find the supergravity solutions of the D1/D5
system with moduli turned on. 
It is only then we can explicitly check
the prediction made in this thesis
that Hawking radiation for minimal scalars 
does not depend on the moduli for the D1/D5 black hole.

\noindent
(e) Recent developments has suggested that constant $B_{NS}$ moduli
is associated with non-commutative geometry. It is important to
investigate the D1/D5 system with $B_{NS}$ moduli turned on. The
microscopic theory of the D1/D5 system with $B_{NS}$ moduli 
encodes information about the moduli space of instantons on a 
non-commutative torus $T^4$.